\documentclass[twocolumn]{aastex63}

\submitjournal{ApJ}

\shorttitle{LSB Galaxies in DES}
\shortauthors{Tanoglidis {\it et al.}}

\reportnum{DES-2020-0528}
\reportnum{FERMILAB-PUB-20-228-AE}

\usepackage{graphicx}
\usepackage{natbib}
\defcitealias{DES:2018}{DES Collaboration 2018}
\defcitealias{2013A&A...558A..33A}{Astropy Collaboration 2013}

\usepackage{xcolor}
\usepackage{hyperref}
\usepackage{amsmath}
\usepackage{multirow}
\usepackage{enumitem}
\usepackage{accents}
\usepackage{subfigure}
\hypersetup{
  colorlinks      = true,
  linkcolor       = {black},
  linkbordercolor = {white},
  citecolor       = {black},
  citebordercolor = {white},
  urlcolor        = {blue},
  urlbordercolor  = {white},
}
\usepackage[all]{hypcap}                                      
\usepackage{comment}
\usepackage{lineno}

\usepackage{soul} 
\usepackage{amsmath}
\usepackage{amssymb}
\usepackage{xspace}
\usepackage{xifthen}

\newcommand{\CHECK}[1]{{\textcolor{black}{#1}}}

\newcommand{\NEW}[1]{\textcolor{blue}{#1}}

\mathchardef\mhyphen="2D

\newcommand{\roughly}{\ensuremath{ {\sim}\,} }

\newlength{\dhatheight}

\newcommand{\code}[1]{\texttt{#1}\xspace}

\newcommand{\unit}[1]{\ensuremath{\mathrm{\,#1}}\xspace}
\newcommand{\Gyr}{\unit{Gyr}}

\newcommand{\degree}{\ensuremath{{}^{\circ}}\xspace}

\newcommand{\pc}{\unit{pc}}
\newcommand{\kpc}{\unit{kpc}}
\newcommand{\Mpc}{\unit{Mpc}}

\newcommand{\pix}{\unit{pix}}

\newcommand{\magasec}{\unit{mag}\unit{arcsec}}
\newcommand{\magasecsq}{\unit{mag}\unit{arcsec^{-2}}}

\newcommand{\mumeaneff}{\ensuremath{\bar{\mu}_{\rm eff}}\xspace}

\newcommand{\Reff}{\ensuremath{R_{\rm eff}}\xspace}

\newcommand{\secref}[1]{Section~\ref{sec:#1}}
\newcommand{\appref}[1]{Appendix~\ref{app:#1}}
\newcommand{\tabref}[1]{Table~\ref{tab:#1}}

\newcommand{\figref}[1]{Figure~\ref{fig:#1}}

\newcommand{\bandvar}[2][]{%
  \ifthenelse{\isempty{#1}}{\var{#2}}{\var{#2\_#1}}%
}

\newcommand{\LCDM}{\ensuremath{\rm \Lambda CDM}\xspace}

\newcommand{\galfit}{\code{galfit}}
\newcommand{\galfitm}{\code{galfitm}}

\newcommand{\SExtractor}{\code{SourceExtractor}}

\newcommand{\var}[1]{\ensuremath{\texttt{\MakeUppercase{#1}}}\xspace}

\providecommand\physrep{\ref@jnl{Phys.~Rep.}}%
\providecommand\apjs{\ref@jnl{ApJS}}%
\providecommand{\jcap}{\ref@jnl{JCAP}}%

\newcommand{\NLSBG}{\CHECK{23,790}\xspace}

\begin{document}

\title{Shadows in the Dark: Low-Surface-Brightness Galaxies Discovered in the Dark Energy Survey}


\author[0000-0002-4631-4529]{D.~Tanoglidis}
\affiliation{Department of Astronomy and Astrophysics, University of Chicago, Chicago, IL 60637, USA}
\affiliation{Kavli Institute for Cosmological Physics, University of Chicago, Chicago, IL 60637, USA}
\author[0000-0001-8251-933X]{A.~Drlica-Wagner}
\affiliation{Fermi National Accelerator Laboratory, P. O. Box 500, Batavia, IL 60510, USA}
\affiliation{Kavli Institute for Cosmological Physics, University of Chicago, Chicago, IL 60637, USA}
\affiliation{Department of Astronomy and Astrophysics, University of Chicago, Chicago, IL 60637, USA}
\author{K.~Wei}
\affiliation{Department of Physics, University of Chicago, Chicago, IL 60637, USA}
\affiliation{Kavli Institute for Cosmological Physics, University of Chicago, Chicago, IL 60637, USA}
\author[0000-0002-9110-6163]{T.~S.~Li}
\affiliation{Department of Astrophysical Sciences, Princeton University, Peyton Hall, Princeton, NJ 08544, USA}
\affiliation{Observatories of the Carnegie Institution for Science, 813 Santa Barbara St., Pasadena, CA 91101, USA}
\author{J.~S{\'a}nchez}
\affiliation{Fermi National Accelerator Laboratory, P. O. Box 500, Batavia, IL 60510, USA}
\author{Y.~Zhang}
\affiliation{Fermi National Accelerator Laboratory, P. O. Box 500, Batavia, IL 60510, USA}
\author[0000-0002-8040-6785]{A.~H.~G.~Peter}
\affiliation{Center for Cosmology and Astro-Particle Physics, The Ohio State University, Columbus, OH 43210, USA}
\author{A.~Feldmeier-Krause}
\affiliation{Kavli Institute for Cosmological Physics, University of Chicago, Chicago, IL 60637, USA}
\author{J.~Prat}
\affiliation{Department of Astronomy and Astrophysics, University of Chicago, Chicago, IL 60637, USA}
\affiliation{Kavli Institute for Cosmological Physics, University of Chicago, Chicago, IL 60637, USA}
\author[0000-0002-2991-9251]{K.~Casey}
\affiliation{Center for Cosmology and Astro-Particle Physics, The Ohio State University, Columbus, OH 43210, USA}
\author[0000-0002-6011-0530]{A.~Palmese}
\affiliation{Fermi National Accelerator Laboratory, P. O. Box 500, Batavia, IL 60510, USA}
\affiliation{Kavli Institute for Cosmological Physics, University of Chicago, Chicago, IL 60637, USA}
\author{C.~S{\'a}nchez}
\affiliation{Department of Physics and Astronomy, University of Pennsylvania, Philadelphia, PA 19104, USA}
\author{J.~DeRose}
\affiliation{Department of Astronomy, University of California, Berkeley,  501 Campbell Hall, Berkeley, CA 94720, USA}
\affiliation{Santa Cruz Institute for Particle Physics, Santa Cruz, CA 95064, USA}
\author[0000-0003-1949-7638]{C.~Conselice}
\affiliation{University of Nottingham, School of Physics and Astronomy, Nottingham NG7 2RD, UK}
\author{L.~Gagnon}
\affiliation{Department of Astronomy and Astrophysics, University of Chicago, Chicago, IL 60637, USA}
\author{T.~M.~C.~Abbott}
\affiliation{Cerro Tololo Inter-American Observatory, National Optical Astronomy Observatory, Casilla 603, La Serena, Chile}
\author{M.~Aguena}
\affiliation{Departamento de F\'isica Matem\'atica, Instituto de F\'isica, Universidade de S\~ao Paulo, CP 66318, S\~ao Paulo, SP, 05314-970, Brazil}
\affiliation{Laborat\'orio Interinstitucional de e-Astronomia - LIneA, Rua Gal. Jos\'e Cristino 77, Rio de Janeiro, RJ - 20921-400, Brazil}
\author[0000-0002-7069-7857]{S.~Allam}
\affiliation{Fermi National Accelerator Laboratory, P. O. Box 500, Batavia, IL 60510, USA}
\author{S.~Avila}
\affiliation{Instituto de Fisica Teorica UAM/CSIC, Universidad Autonoma de Madrid, 28049 Madrid, Spain}
\author{K.~Bechtol}
\affiliation{Physics Department, 2320 Chamberlin Hall, University of Wisconsin-Madison, 1150 University Avenue Madison, WI  53706-1390}
\author{E.~Bertin}
\affiliation{CNRS, UMR 7095, Institut d'Astrophysique de Paris, F-75014, Paris, France}
\affiliation{Sorbonne Universit\'es, UPMC Univ Paris 06, UMR 7095, Institut d'Astrophysique de Paris, F-75014, Paris, France}
\author{S.~Bhargava}
\affiliation{Department of Physics and Astronomy, Pevensey Building, University of Sussex, Brighton, BN1 9QH, UK}
\author[0000-0002-8458-5047]{D.~Brooks}
\affiliation{Department of Physics \& Astronomy, University College London, Gower Street, London, WC1E 6BT, UK}
\author{D.~L.~Burke}
\affiliation{Kavli Institute for Particle Astrophysics \& Cosmology, P. O. Box 2450, Stanford University, Stanford, CA 94305, USA}
\affiliation{SLAC National Accelerator Laboratory, Menlo Park, CA 94025, USA}
\author[0000-0003-3044-5150]{A.~Carnero~Rosell}
\affiliation{Instituto de Astrofisica de Canarias, E-38205 La Laguna, Tenerife, Spain}
\affiliation{Universidad de La Laguna, Dpto. Astrof{\'i}sica, E-38206 La Laguna, Tenerife, Spain}
\author[0000-0002-4802-3194]{M.~Carrasco~Kind}
\affiliation{Department of Astronomy, University of Illinois at Urbana-Champaign, 1002 W. Green Street, Urbana, IL 61801, USA}
\affiliation{National Center for Supercomputing Applications, 1205 West Clark St., Urbana, IL 61801, USA}
\author[0000-0002-3130-0204]{J.~Carretero}
\affiliation{Institut de F\'{\i}sica d'Altes Energies (IFAE), The Barcelona Institute of Science and Technology, Campus UAB, 08193 Bellaterra (Barcelona) Spain}
\author[0000-0002-7887-0896]{C.~Chang}
\affiliation{Department of Astronomy and Astrophysics, University of Chicago, Chicago, IL 60637, USA}
\affiliation{Kavli Institute for Cosmological Physics, University of Chicago, Chicago, IL 60637, USA}
\author{M.~Costanzi}
\affiliation{INAF-Osservatorio Astronomico di Trieste, via G. B. Tiepolo 11, I-34143 Trieste, Italy}
\affiliation{Institute for Fundamental Physics of the Universe, Via Beirut 2, 34014 Trieste, Italy}
\author{L.~N.~da Costa}
\affiliation{Laborat\'orio Interinstitucional de e-Astronomia - LIneA, Rua Gal. Jos\'e Cristino 77, Rio de Janeiro, RJ - 20921-400, Brazil}
\affiliation{Observat\'orio Nacional, Rua Gal. Jos\'e Cristino 77, Rio de Janeiro, RJ - 20921-400, Brazil}
\author[0000-0001-8318-6813]{J.~De~Vicente}
\affiliation{Centro de Investigaciones Energ\'eticas, Medioambientales y Tecnol\'ogicas (CIEMAT), Madrid, Spain}
\author[0000-0002-0466-3288]{S.~Desai}
\affiliation{Department of Physics, IIT Hyderabad, Kandi, Telangana 502285, India}
\author[0000-0002-8357-7467]{H.~T.~Diehl}
\affiliation{Fermi National Accelerator Laboratory, P. O. Box 500, Batavia, IL 60510, USA}
\author{P.~Doel}
\affiliation{Department of Physics \& Astronomy, University College London, Gower Street, London, WC1E 6BT, UK}
\author[0000-0002-1894-3301]{T.~F.~Eifler}
\affiliation{Department of Astronomy/Steward Observatory, University of Arizona, 933 North Cherry Avenue, Tucson, AZ 85721-0065, USA}
\affiliation{Jet Propulsion Laboratory, California Institute of Technology, 4800 Oak Grove Dr., Pasadena, CA 91109, USA}
\author{S.~Everett}
\affiliation{Santa Cruz Institute for Particle Physics, Santa Cruz, CA 95064, USA}
\author[0000-0002-4876-956X]{A.~E.~Evrard}
\affiliation{Department of Astronomy, University of Michigan, Ann Arbor, MI 48109, USA}
\affiliation{Department of Physics, University of Michigan, Ann Arbor, MI 48109, USA}
\author[0000-0002-2367-5049]{B.~Flaugher}
\affiliation{Fermi National Accelerator Laboratory, P. O. Box 500, Batavia, IL 60510, USA}
\author[0000-0003-4079-3263]{J.~Frieman}
\affiliation{Fermi National Accelerator Laboratory, P. O. Box 500, Batavia, IL 60510, USA}
\affiliation{Kavli Institute for Cosmological Physics, University of Chicago, Chicago, IL 60637, USA}
\author[0000-0002-9370-8360]{J.~Garc\'ia-Bellido}
\affiliation{Instituto de Fisica Teorica UAM/CSIC, Universidad Autonoma de Madrid, 28049 Madrid, Spain}
\author[0000-0001-6942-2736]{D.~W.~Gerdes}
\affiliation{Department of Astronomy, University of Michigan, Ann Arbor, MI 48109, USA}
\affiliation{Department of Physics, University of Michigan, Ann Arbor, MI 48109, USA}
\author{R.~A.~Gruendl}
\affiliation{Department of Astronomy, University of Illinois at Urbana-Champaign, 1002 W. Green Street, Urbana, IL 61801, USA}
\affiliation{National Center for Supercomputing Applications, 1205 West Clark St., Urbana, IL 61801, USA}
\author[0000-0003-3023-8362]{J.~Gschwend}
\affiliation{Laborat\'orio Interinstitucional de e-Astronomia - LIneA, Rua Gal. Jos\'e Cristino 77, Rio de Janeiro, RJ - 20921-400, Brazil}
\affiliation{Observat\'orio Nacional, Rua Gal. Jos\'e Cristino 77, Rio de Janeiro, RJ - 20921-400, Brazil}
\author[0000-0003-0825-0517]{G.~Gutierrez}
\affiliation{Fermi National Accelerator Laboratory, P. O. Box 500, Batavia, IL 60510, USA}
\author{W.~G.~Hartley}
\affiliation{D\'{e}partement de Physique Th\'{e}orique and Center for Astroparticle Physics, Universit\'{e} de Gen\`{e}ve, 24 quai Ernest Ansermet, CH-1211 Geneva, Switzerland}
\affiliation{Department of Physics \& Astronomy, University College London, Gower Street, London, WC1E 6BT, UK}
\affiliation{Department of Physics, ETH Zurich, Wolfgang-Pauli-Strasse 16, CH-8093 Zurich, Switzerland}
\author{D.~L.~Hollowood}
\affiliation{Santa Cruz Institute for Particle Physics, Santa Cruz, CA 95064, USA}
\author[0000-0001-6558-0112]{D.~Huterer}
\affiliation{Department of Physics, University of Michigan, Ann Arbor, MI 48109, USA}
\author[0000-0001-5160-4486]{D.~J.~James}
\affiliation{Center for Astrophysics $\vert$ Harvard \& Smithsonian, 60 Garden Street, Cambridge, MA 02138, USA}
\author[0000-0001-8356-2014]{E.~Krause}
\affiliation{Department of Astronomy/Steward Observatory, University of Arizona, 933 North Cherry Avenue, Tucson, AZ 85721-0065, USA}
\author[0000-0003-0120-0808]{K.~Kuehn}
\affiliation{Australian Astronomical Optics, Macquarie University, North Ryde, NSW 2113, Australia}
\affiliation{Lowell Observatory, 1400 Mars Hill Rd, Flagstaff, AZ 86001, USA}
\author[0000-0003-2511-0946]{N.~Kuropatkin}
\affiliation{Fermi National Accelerator Laboratory, P. O. Box 500, Batavia, IL 60510, USA}
\author[0000-0001-9856-9307]{M.~A.~G.~Maia}
\affiliation{Laborat\'orio Interinstitucional de e-Astronomia - LIneA, Rua Gal. Jos\'e Cristino 77, Rio de Janeiro, RJ - 20921-400, Brazil}
\affiliation{Observat\'orio Nacional, Rua Gal. Jos\'e Cristino 77, Rio de Janeiro, RJ - 20921-400, Brazil}
\author{M.~March}
\affiliation{Department of Physics and Astronomy, University of Pennsylvania, Philadelphia, PA 19104, USA}
\author[0000-0003-0710-9474]{J.~L.~Marshall}
\affiliation{George P. and Cynthia Woods Mitchell Institute for Fundamental Physics and Astronomy, and Department of Physics and Astronomy, Texas A\&M University, College Station, TX 77843,  USA}
\author[0000-0002-1372-2534]{F.~Menanteau}
\affiliation{Department of Astronomy, University of Illinois at Urbana-Champaign, 1002 W. Green Street, Urbana, IL 61801, USA}
\affiliation{National Center for Supercomputing Applications, 1205 West Clark St., Urbana, IL 61801, USA}
\author[0000-0002-6610-4836]{R.~Miquel}
\affiliation{Instituci\'o Catalana de Recerca i Estudis Avan\c{c}ats, E-08010 Barcelona, Spain}
\affiliation{Institut de F\'{\i}sica d'Altes Energies (IFAE), The Barcelona Institute of Science and Technology, Campus UAB, 08193 Bellaterra (Barcelona) Spain}
\author[0000-0003-2120-1154]{R.~L.~C.~Ogando}
\affiliation{Laborat\'orio Interinstitucional de e-Astronomia - LIneA, Rua Gal. Jos\'e Cristino 77, Rio de Janeiro, RJ - 20921-400, Brazil}
\affiliation{Observat\'orio Nacional, Rua Gal. Jos\'e Cristino 77, Rio de Janeiro, RJ - 20921-400, Brazil}
\author{F.~Paz-Chinch\'{o}n}
\affiliation{Institute of Astronomy, University of Cambridge, Madingley Road, Cambridge CB3 0HA, UK}
\affiliation{National Center for Supercomputing Applications, 1205 West Clark St., Urbana, IL 61801, USA}
\author[0000-0002-9328-879X]{A.~K.~Romer}
\affiliation{Department of Physics and Astronomy, Pevensey Building, University of Sussex, Brighton, BN1 9QH, UK}
\author[0000-0001-5326-3486]{A.~Roodman}
\affiliation{Kavli Institute for Particle Astrophysics \& Cosmology, P. O. Box 2450, Stanford University, Stanford, CA 94305, USA}
\affiliation{SLAC National Accelerator Laboratory, Menlo Park, CA 94025, USA}
\author[0000-0002-9646-8198]{E.~Sanchez}
\affiliation{Centro de Investigaciones Energ\'eticas, Medioambientales y Tecnol\'ogicas (CIEMAT), Madrid, Spain}
\author{V.~Scarpine}
\affiliation{Fermi National Accelerator Laboratory, P. O. Box 500, Batavia, IL 60510, USA}
\author{S.~Serrano}
\affiliation{Institut d'Estudis Espacials de Catalunya (IEEC), 08034 Barcelona, Spain}
\affiliation{Institute of Space Sciences (ICE, CSIC),  Campus UAB, Carrer de Can Magrans, s/n,  08193 Barcelona, Spain}
\author[0000-0002-1831-1953]{I.~Sevilla-Noarbe}
\affiliation{Centro de Investigaciones Energ\'eticas, Medioambientales y Tecnol\'ogicas (CIEMAT), Madrid, Spain}
\author[0000-0002-3321-1432]{M.~Smith}
\affiliation{School of Physics and Astronomy, University of Southampton,  Southampton, SO17 1BJ, UK}
\author[0000-0002-7047-9358]{E.~Suchyta}
\affiliation{Computer Science and Mathematics Division, Oak Ridge National Laboratory, Oak Ridge, TN 37831}
\author[0000-0003-1704-0781]{G.~Tarle}
\affiliation{Department of Physics, University of Michigan, Ann Arbor, MI 48109, USA}
\author{D.~Thomas}
\affiliation{Institute of Cosmology and Gravitation, University of Portsmouth, Portsmouth, PO1 3FX, UK}
\author[0000-0001-7211-5729]{D.~L.~Tucker}
\affiliation{Fermi National Accelerator Laboratory, P. O. Box 500, Batavia, IL 60510, USA}

\author[0000-0002-7123-8943]{A.~R.~Walker}
\affiliation{Cerro Tololo Inter-American Observatory, National Optical Astronomy Observatory, Casilla 603, La Serena, Chile}

\collaboration{70}{(DES Collaboration)}

\correspondingauthor{\\
D.~Tanoglidis (\href{mailto:dtanoglidis@uchicago.edu}{dtanoglidis@uchicago.edu}) \\
A.~Drlica-Wagner (\href{mailto:kadrlica@fnal.gov}{kadrlica@fnal.gov})}

\begin{abstract}

We present a catalog of \NLSBG extended low-surface-brightness galaxies (LSBGs) identified in $\roughly 5000 \deg^2$ from the first three years of imaging data from the Dark Energy Survey (DES). 
Based on a single-component S\'ersic model fit, we define extended LSBGs as galaxies with $g$-band  effective radii $\Reff(g) > 2.5''$ and mean surface brightness $\mumeaneff(g) > 24.2  \magasecsq$. 
We find that the distribution of LSBGs is strongly bimodal in $(g-r)$ vs.\ $(g-i$) color space. We divide our sample into red ($g-i \geq 0.60$) and blue ($g-i<0.60$) galaxies and study the properties of the two populations.
Redder LSBGs are more clustered than their blue counterparts and are correlated with the distribution of nearby ($z < 0.10$) bright galaxies. 
Red LSBGs constitute $\sim 33\%$ of our LSBG sample, and $\roughly 30\%$ of these are located within 1 deg of low-redshift galaxy groups and clusters (compared to $\sim 8\%$ of the blue LSBGs).
For nine of the most prominent galaxy groups and clusters, we calculate the physical properties of associated LSBGs assuming a redshift derived from the host system.
In these systems, we identify 41 objects that can be classified as ultra-diffuse galaxies, defined as LSBGs with projected physical effective radii $\Reff > 1.5\kpc$ and central surface brighthness $\mu_0(g) > 24.0 \magasecsq$.
The wide-area sample of LSBGs in DES can be used to test the role of environment on models of LSBG formation and evolution.
\end{abstract}

\section{Introduction} \label{sec:intro}

The low-surface-brightness universe is notoriously difficult to characterize due to the significant impact of observational selection effects \citep[e.g.,][]{Disney:1976,McGaugh:1995}.
Low-surface-brightness galaxies (LSBGs) are conventionally defined as galaxies with central surface brightnesses fainter than the night sky \citep{Bothun:1997}.
While these faint galaxies are thought to contribute a minority (a few percent) of the local luminosity and stellar mass density \citep[e.g.,][]{Bernstein:1995, Driver:1999, Hayward:2005,Martin:2019}, they may account for $\roughly 15\%$ of the dynamical mass budget in the present-day universe \citep[e.g.,][]{Driver:1999,ONeil:2000,Minchin:2004}.
However, due to the observational challenges in detecting these faint systems, LSBGs remain difficult to study as an unbiased population.

LSBGs are known to span a wide range of physical sizes and environments, ranging from the ultra-faint satellites of the Milky Way \citep[e.g.,][]{McConnachie:2012,Simon:2019}, to satellites of other nearby galaxies \citep[e.g.,][]{Martin:2013,Merritt:2016,Martin:2016,Danieli:2017,Cohen:2018}, and members of massive galaxy clusters like Virgo \citep[e.g.,][]{Sabatini:2005,Mihos:2015,Mihos:2017}, Perseus \citep[e.g.,][]{Wittmann:2017}, Coma \citep[e.g.,][]{Adami:2006,vanDokkum:2015a,Koda:2015}, Fornax \citep[e.g.,][]{Ferguson:1989,Hilker:1999,Munoz:2015,Venhola:2017}, and other nearby clusters \citep[e.g.,][]{vanderBurg:2016}.
Untargeted searches have also found a large population of LSBGs in the field \citep[e.g.,][]{Zhong:2008, Rosenbaum:2009, Galaz:2011, Greco:2018}.
Understanding how LSBGs come to populate this wide range of environments may inform models of cosmology and galaxy evolution. 
Are LSBGs truly outliers relative to the rest of the galaxy population, or are they merely a natural continuation of the galaxy size--luminosity relation?

The standard model of cosmology (\LCDM) predicts that galaxies form hierarchically, with smaller galaxies forming first and assembling to form larger galaxies, galaxy groups, and galaxy clusters \citep[e.g.,][]{Peebles:1980,Davis:1985,White:1991}.
The formation and growth of galaxies over cosmic time is connected to the growth of the dark matter halos in which they reside (the so-called ``galaxy--halo connection''; e.g., \citealt{Wechsler:2018}).
Many attempts have been made to use the properties of dark matter halos to predict the properties of the galaxies that inhabit them \citep[e.g.,][]{Behroozi:2013,Moster:2013}.
As extremes in the relationship between galaxy size and luminosity, LSBGs provide a litmus test for models that predict galaxy properties from cosmological principles \citep[e.g.,][]{Ferrero:2012, Papastergis:2015}.
It has been suggested that LSBGs form naturally within the \LCDM framework, either primordially in halos with high angular velocity \citep{Dalcanton:1997,Amorisco:2016} or through evolution in dense environments \citep{Tremmel:2019,Martin:2019}.
On the other hand, observations of LSBGs with anomalously low dark matter content \citep{vanDokkum:2018,vanDokkum:2019} may necessitate modified models of galaxy formation \citep[e.g.,][]{Papastergis:2017,Sales:2019} and/or dark matter physics \citep[e.g.,][]{Carleton:2019}.
Disentangling the contributions of various mechanisms for LSBG formation has been historically challenging due to the small volume and highly biased observational samples available.

Over the last few decades, the rapid advance of wide-area, homogeneous, digital imaging has greatly increased our sensitivity to LSBGs.
The Sloan Digital Sky Survey (SDSS) enabled statistical studies of large samples of LSBGs down to central surface brightnesses of $\mu_0(B) \sim 24 \magasecsq$ \citep{Zhong:2008, Rosenbaum:2009, Galaz:2011}.
Smaller telescopes optimized for the low-surface-brightness regime \citep[i.e., the Dragonfly Telephoto Array; ][]{Abraham:2014} have illuminated the populations of LSBGs in nearby groups \citep{Merritt:2016,Danieli:2017,Cohen:2018} and clusters \citep{vanDokkum:2015a,Janssens:2017}, extending down to unprecedented central surface brightnesses of $\mu_0(g) > 27 \magasecsq$.
Recently, the Hyper Suprime-Cam Subaru Strategic Program (HSC SSP) revealed a large population of LSBGs with $\mumeaneff(g) > 24.3 \magasecsq$ in an untargeted search of the first $\roughly 200 \deg^2$ from the Wide layer of the HSC SSP \citep{Greco:2018}.
However, results from these deep photometric surveys are still limited to relatively small areas of sky, limiting our ability to characterize the faintest galaxies in an unbiased manner.

Untargeted searches for LSBGs are essential to understand the role that environment plays in their formation and evolution.
However, such searches are challenging due to the deep imaging and wide area coverage that is required to provide a statistically significant population of LSBGs.
Here we use data from the first three years of the Dark Energy Survey (DES) to detect LSBGs with half-light radii $r_{1/2} > 2.5''$ and mean surface brightness $\mumeaneff(g) >24.2 \magasecsq$ over $\roughly 5000 \deg^2$ of the southern Galactic cap.
Through a combination of classical cut-based selections on measured photometric properties, machine learning (ML) techniques, and visual inspection, we produce a high-purity catalog of \NLSBG LSBGs.
We present the spatial, morphological, and photometric properties of this sample based on detailed multi-band S\'ersic model fits.

This paper is organized as follows.
In \secref{data} we describe the DES data set and object catalog used for our search. 
In \secref{pipeline} we describe our multi-step selection and measurement pipeline, resulting in our catalog of LSBGs.
In \secref{efficiency} we estimate the efficiency of our catalog selection method by comparing against deeper data around the Fornax galaxy cluster.
In \secref{properties}, we describe the observed properties of this sample, and in \secref{clustering} we examine the statistical clustering of LSBGs.
In \secref{clusters}, we examine the properties of LSBGs that are close in projection to nearby galaxy groups and clusters.
We summarize the results of this work in \secref{summary}.

\section{DES Data} \label{sec:data}

DES is an optical--near infrared imaging survey covering $\roughly 5000 \deg^2$ of the southern Galactic cap using the Dark Energy Camera \citep[DECam;][]{Flaugher:2015} on the 4-m Blanco Telescope at the Cerro Tololo Inter-American Observatory (CTIO).
The DECam focal plane comprises 62 2k$\times$4k CCDs dedicated to science imaging and 12 2k$\times$2k CCDs for guiding, focus, and alignment.
The DECam field of view covers $3\deg^2$ with a central pixel scale of $0.263''$.
DES observes with a dithered exposure pattern to account for gaps between CCDs \citep{Neilsen:2019} and combines the individual exposures into coadded images that are $0.73 \times 0.73 \deg$ in size \citep{Morganson:2018}.
The median sky brightness levels in the DES exposures are $g = 22.01, r = 21.15$, and $i = 19.89 \magasecsq$ \citep{DES:2018}.

We use data collected from the first three years of DES observing (DES Y3).
This data set shares the same single-image processing, image coaddition, and object detection as the first DES data release \citep[DR1;][]{DES:2018}.
In particular, object detection was performed on $r+i+z$ coadded detection images using \SExtractor \citep{Bertin:2006}. 
Photometric measurements were performed in each band using \SExtractor in ``dual image'' mode using the band of interest in combination with the detection image.
The depth of the DES Y3 object catalog at signal-to-noise ratio $({\rm S/N}) = 10$ based on the \SExtractor adaptive aperture fit (\var{MAG\_AUTO}) is $g=23.52$, $r=23.10$, and $i=22.51$ \citepalias{DES:2018}.
The DES pipeline was optimized for the detection and measurement of galaxies at cosmological distances, which are generally faint and relatively small in projected size.

Sky background estimation is an important component in the detection of extended LSBGs.
In DES Y3, sky background estimation and subtraction were performed in two phases \citep{Morganson:2018}. 
First, the background was fit using a principal components analysis (PCA) algorithm applied to the full focal plane binned into $128\times128$ superpixels that are $\roughly 1\arcmin$ in size \citep{Bernstein:2018}.
Next, \SExtractor was used to fit the residual local background on each CCD using a bicubic spline fit to $256\times256$ pixel blocks, which are again $\roughly 1\arcmin$ in size \citep{Bertin:2006,Morganson:2018}. 
For comparison, the half-light radii of the LSBGs in this study range from $2.5\arcsec$ to $\roughly 20\arcsec$ in radius.
Background modeling may reduce the efficiency for detecting larger and lower surface-brightness sources, and we leave further background modeling optimization to future work.

We estimated the surface-brightness contrast on $10'' \times 10''$ scales for each DES coadd tile using the \code{sbcontrast} module from Multi-Resolution Filtering packaged developed for the Dragonfly Telephoto Array \citep{vanDokkum:2020}.\footnote{\url{https://github.com/AstroJacobLi/mrf}} This procedure bins each coadd image on the desired scale, subtracts a local background from each binned pixel based on the surrounding 8 pixels, and calculates the variation among the binned and background-subtracted pixels \citep[e.g.,][]{Gilhuly:2020}.  We applied this procedure to each DES coadd tile after masking bad pixels and sources detected by \SExtractor. We find that on $10'' \times 10''$ scales, the median surface brightness limit at $3\sigma$ is $g = 28.26^{+0.09}_{-0.13}, r = 27.86^{+0.10}_{-0.15}, i = 27.37^{+0.10}_{-0.13} \magasecsq$, where the upper and lower bounds represent the 16th and 84th percentiles of the distribution over DES tiles (\appref{sbcontrast}).\footnote{The uncertainty within individual tiles is sharply peaked at a median value of $0.004 \magasecsq$.} These values can be directly compared to the $3\sigma$ surface-brightness contrast of $g = 28.616, r = 28.936 \magasecsq$ reported for Dragonfly observations of NGC 4565 \citep{Gilhuly:2020}. However, we note that the DES source detection pipeline has not been optimized for the detection of large, low surface-brightness sources, and so the source detection threshold cannot be directly compared to other catalogs optimized to this purpose.

\section{LSBG Catalog} 
\label{sec:pipeline}

Here we describe the pipeline used to identify and measure LSBGs in the DES Y3 data.
Briefly, we start with a generic catalog of \SExtractor detections and use the morphological and photometric properties to identify a subset of LSBG candidates.
We train a machine learning algorithm to remove artifacts and visually inspect the resulting candidate list to assemble a high-purity catalog of LSBGs.
We then fit a S\'ersic profile to each identified LSBG in order to  determine photometric properties in a manner that is consistent with previous work \citep[e.g.][]{Greco:2018}.
Our full catalog of DES LSBGs is available as supplemental material.\footnote{\url{https://des.ncsa.illinois.edu/releases/other/y3-lsbg}}

\subsection{Initial sample selection}
\label{sec:sample}

We began with the DES Y3 Gold coadd object catalog (v2.2) assembled from \SExtractor detections \citep{Sevilla:2020}.
We first removed objects classified as point-like based on the $i$-band \SExtractor \code{SPREAD\_MODEL} parameter (see \appref{select} and \citealt{Sevilla:2020} for more details).
Following \citet{Greco:2018}, we defined our initial sample of candidate LSBGs based on angular size and surface brightness.
Because these cuts were primarily intended to reject imaging artifacts, no correction for interstellar extinction was applied at this stage.
We required that sources have half-light radii in the $g$ band (as estimated by \SExtractor \code{FLUX\_RADIUS}) to be in the range $2.5'' < r_{1/2}(g) < 20\arcsec$\footnote{After assembling our catalog, we inspected all the candidates ($\roughly 1{,}500$) satisfying our color and surface brightness cuts and having $r_{1/2}(g) > 20''$. We found 6 LSBGs that were subsequently included in our catalog.} and mean surface brightness $24.2 < \mumeaneff(g) <  28.8 \magasecsq$.\footnote{Note that there is a difference in the mean surface brightness selection, compared to \citet{Greco:2018} that uses $24.3 < \mumeaneff(g) <  28.8 \magasecsq$. 
Our definition is slightly more inclusive, and the reader should keep this in mind when comparing to the HSC catalog from \citet{Greco:2018}.}
We also restricted our selection to objects with colors (based on the \SExtractor \code{MAG\_AUTO} magnitudes) in the range:
\begin{eqnarray}
- 0.1 < g - i < 1.4 \\
(g-r) > 0.7 \times (g-i) - 0.4\\
(g-r) < 0.7 \times (g-i) + 0.4.
\end{eqnarray}
These color cuts were guided by the HSC SSP analysis of \citet{Greco:2018}, and were found to produce similar results in DES. Furthermore, we required the objects in our catalog to have ellipticity $<0.7$, to eliminate some high-ellipticity spurious artifacts (i.e., diffraction spikes).
Our complete selection criteria are presented in \appref{select}.
After performing the cuts described above, our sample consisted of 419,895 objects from an initial catalog of $\roughly 400$ million objects.

\subsection{Machine Learning Classification} \label{sec:ML}

Visual inspection of a few thousand candidates passing the cuts described in the previous section revealed that $\lesssim 8\%$ of the objects passing these selections were LSBGs. 
The most common sources of contamination were:
\begin{enumerate}[wide, labelwidth=!, labelindent=0pt, itemsep=0pt]
  \item Faint, compact objects blended in the diffuse light from nearby bright stars or giant elliptical galaxies.
  \item Bright regions of Galactic cirrus.
  \item Knots and star-forming regions in the arms of large spiral galaxies.
  \item Tidal ejecta connected to high-surface-brightness host galaxies.
\end{enumerate}
The large size and low purity of our initial candidate list was well suited to the application of conventional ML classification algorithms. 
Our goal with ML classification was to reject a large fraction of false positives while retaining high completeness for true LSBGs.

\subsubsection{Training Set}\label{sec:training}

\begin{figure*}[t]
\centering
\includegraphics[width=0.75\textwidth]{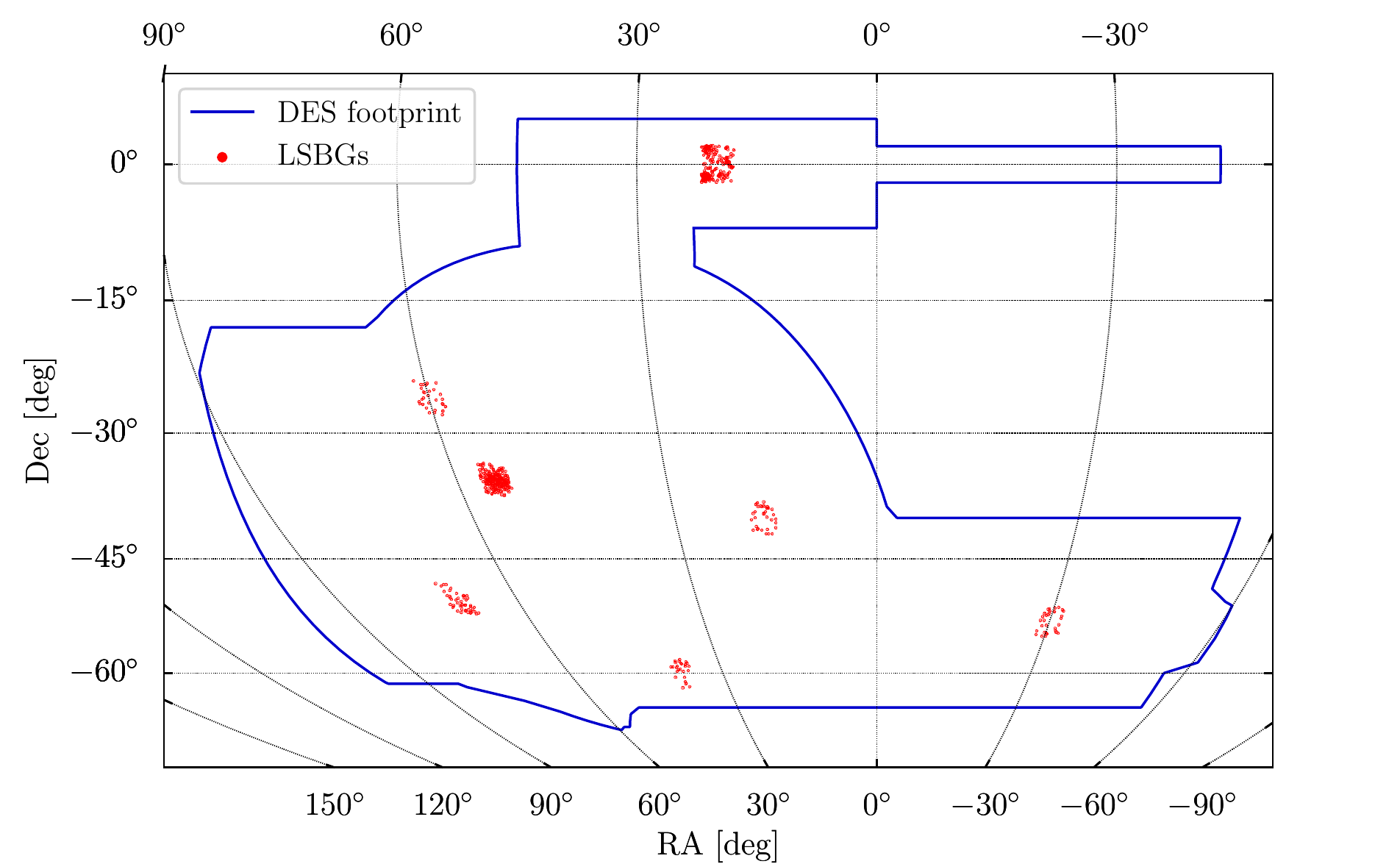}
\vspace{-0.2cm}
\caption{The distribution of the objects visually classified as LSBGs in the seven $4^\circ \times 4^\circ$ regions used to create the labeled set for classification and validation. The Fornax galaxy cluster is located at (RA, DEC) $\sim (55^\circ, -35^\circ)$.}
\label{fig:training_set}
\end{figure*}

In order to train a supervised ML classification algorithm, we required a sample of objects where the true classification was known.
To avoid biases when training the classifier, we seek to assemble a labeled training sample that is representative of the full LSBG candidate sample. 
We created a labeled sample by visually inspecting all objects that pass the cuts defined in \secref{sample} in seven patches spread over the DES footprint, comprising $\roughly 100 \deg^2$ (\figref{training_set}). One of these regions was centered on the Fornax galaxy cluster, which is known to contain a high concentration of LSBGs \citep[e.g.,][]{Munoz:2015}, while the locations of the other regions were selected at random.
Our training set consists of 7760 visually inspected objects, of which 640 were classified as LSBGs. 

\subsubsection{Features and Classifiers} \label{sec:class}

We split the labeled objects into two sets: $75\%$ of the labeled objects were used as a training set, while the remaining $25\%$ were used as a validation set. We used the validation set to evaluate the performance of different classifiers and tune their hyperparameters. 
Since the ML classifier was used solely as a precursor to visual inspection, we were not concerned with precisely characterizing its performance. 
Thus, rather than allocating an independent testing sample, we used our entire labeled data set for training and validation.

In the classification, we used 18 features derived from the \SExtractor measured properties without correcting for interstellar extinction.  
Specifically, we used:
\begin{enumerate}[wide, labelwidth=!, labelindent=0pt, itemsep=0pt]
  \item The adaptive aperture magnitudes in the $g,r,i$ bands, \code{MAG\_AUTO}.
  \item The colors $(g-r)$, $(g-i)$, and $(i-r)$ derived from the adaptive aperture magnitudes.
  \item The size of a circular isophote containing half the flux in the $g,r,i$ bands, \code{FLUX\_RADIUS}.
  \item The effective surface brightness in the $g,r,i$ bands, \code{MU\_EFF\_MODEL}.
  \item The maximum surface brightness measured by \SExtractor in the $g,r,i$ bands, \code{MU\_MAX}.
  \item The semi-major and semi-minor axes of the isophotal ellipse containing half the light, \code{A\_IMAGE} and \code{B\_IMAGE}.
  \item The isophotal ellipticity, $1 - \code{B\_IMAGE}/\code{A\_IMAGE}$.
\end{enumerate}

\noindent We tested a number of popular classification algorithms, as implemented in the Python library \code{scikit-learn} \citep{Pedregosa:2011}.\footnote{\url{https://scikit-learn.org/stable/}}.
Specifically, we tested naive Bayes, AdaBoost, nearest neighbor, random forest, linear support vector machines (SVM), and SVM with radial basis function (RBF) kernel classifiers. 
Due to the relatively small size of our training set (and specifically the small number of positive instances), we did not attempt classification using deep learning techniques.

Our goal was to find a classifier that minimized the false-negative rate (FNR)---i.e., true LSBGs classified as false detections---while keeping the true-positive rate (TPR) reasonably high. 
In other words, we favored completeness over purity in the sample classified as LSBGs. 
This choice was motivated by our goal to reduce the candidate sample to a tractable size for visual inspection (which would reject the remaining false positives), without losing many real LSBGs in the process. 

Note that the samples in our training data were heavily imbalanced: from the 5820 objects ($7760 \times 0.75$) only $480$ ($640\times 0.75$) were true LSBGs. 
Class imbalance can lead to low accuracy in predicting the label of objects belonging to the less frequent class. 
We dealt with this by weighting the classes using the \code{class\_weight} parameter. 
Setting this parameter equal to \code{"balanced"} assigns each class a weight that is inversely proportional to its frequency, $w_j = n/2n_j$, where $w_j$ is the weight of the $j-$th class and  $n, n_j$ are the total number of observations and observations of the $j-$th class, respectively.

\begin{figure}[ht!]
\centering
\epsscale{1.2}
\plotone{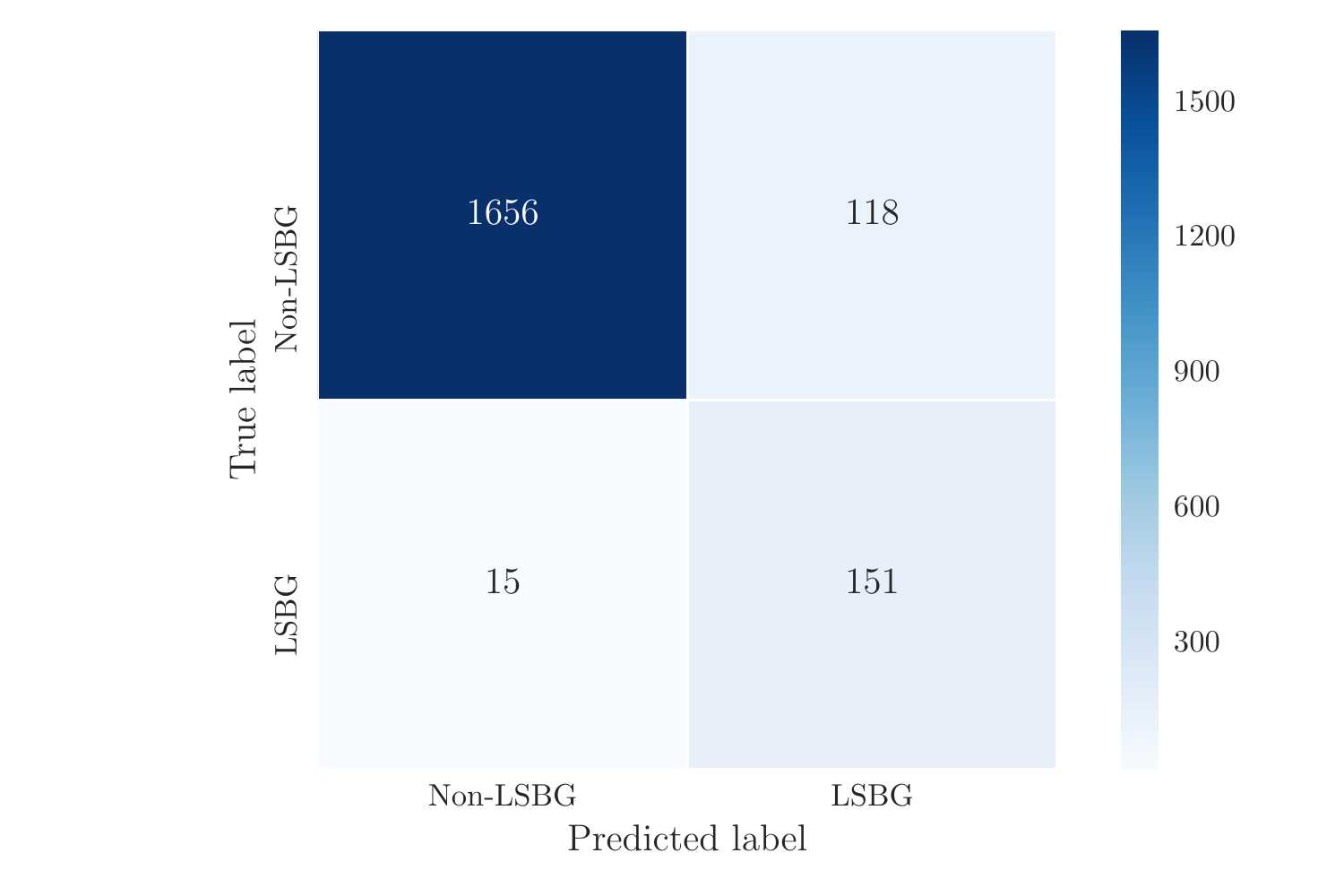}
\caption{The confusion matrix of our final SVM classifier evaluated on the validation set. The quoted numbers correspond to the number of the validation instances (objects) based on their true and predicted label. The false-negative rate is  $\sim 9\%$.}
\label{fig:conf_mat}
\end{figure}

We found that the optimal classifier for our specified goal was an SVM classifier with an RBF kernel and parameters $C=10^4$ and $\gamma = 0.012$ (These parameters are related to the sensitivity to the missclassification rate of training examples vs simplicity of the decision boundary, and the influence of a single training example, respectively. For more details on SVMs see, e.g., \citet{Elements}). In \figref{conf_mat}, we present the confusion matrix for this classifier, evaluated on the validation set. 
We see that the FNR, defined as the fraction of true LSBGs classified as non-LSBGs (${\rm FNR = FN/(FN + TP)}$), is $\roughly 9\%$. We visually inspected the 15 LSBGs rejected by the SVM classifier, as well as examples of LSBGs that were correctly classified. Comparing the two cases, we find that the rejected objects are systematically fainter (about one magnitude in mean surface brightness) than the LSBGs that passed the classification step.

From the same plot, we expect that $\roughly 44\%$ of the objects classified as LSBGs are false positives. 
Subsequent visual inspection (\secref{visual}) showed that the number of false positives was consistent with the estimate presented here.

Using the optimized classifier, as described in the above section, we classified the 419,895 LSBG candidates that were selected by the cuts defined in \secref{sample}. The classification returned 44,979 objects classified as LSBGs, thus reducing the sample by about an order of magnitude.

\subsection{Visual Inspection} \label{sec:visual}

\begin{figure*}[t]
\centering
\epsscale{1.0}
\plotone{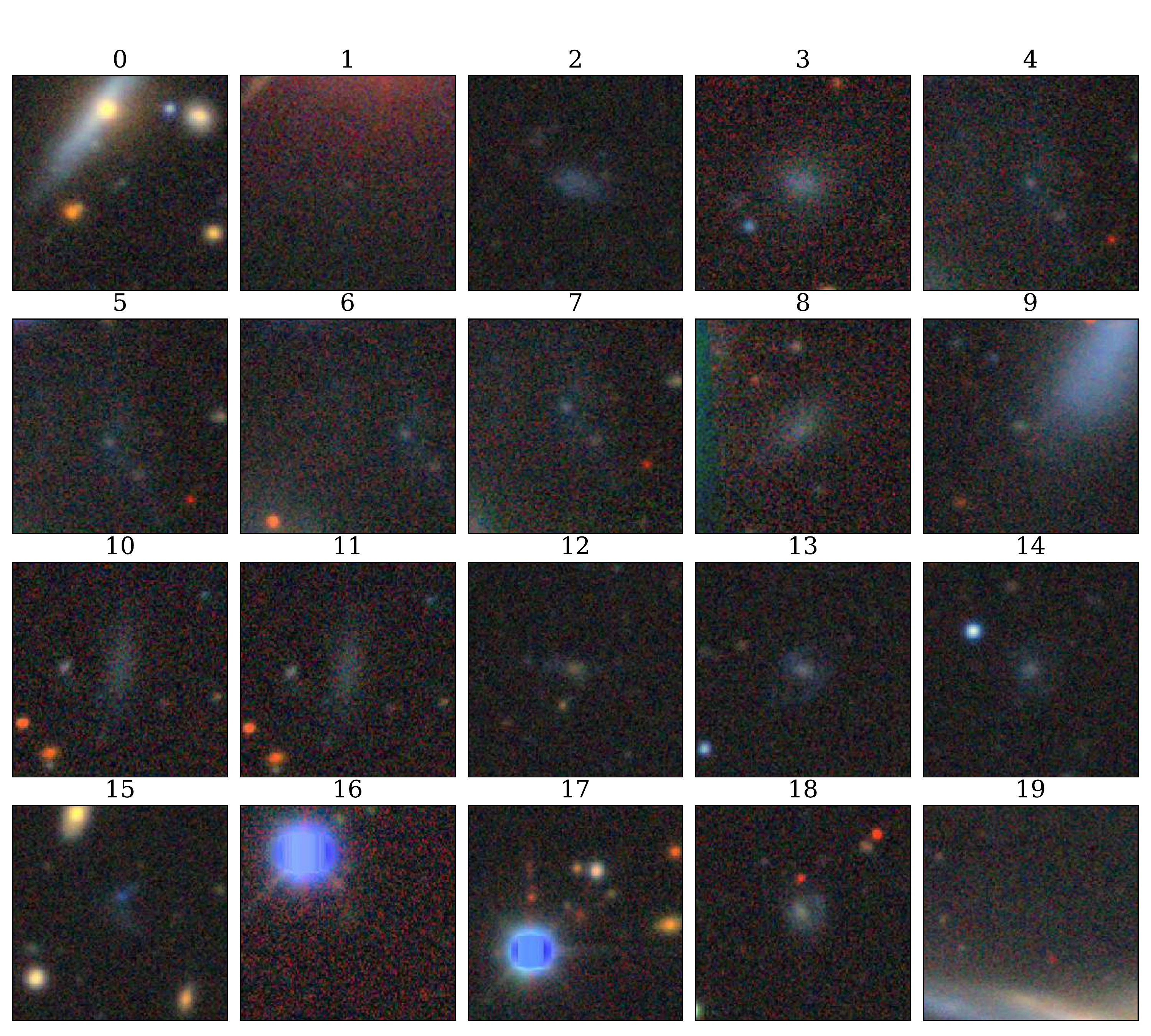}
\caption{$30\arcsec \times 30\arcsec$ cutouts of 20 candidates, positively classified by our machine learning algorithm (\secref{ML}). Candidates 2, 3, 8, 11, 12, 13, 14, 15, and 18 are visually classified as LSBGs, while the other candidates are rejected as false positives and/or duplicates.}
\label{fig:cutouts}
\end{figure*}

The next step in the generation of our LSBG sample was visual inspection of objects that were classified as LSBGs by our ML classifier.
We generate $30\arcsec \times 30\arcsec$ cutouts centered at the coordinates of each of the candidates, and we inspect candidates in batches of 500. 
For cutout generation, we use the DESI Legacy Imaging Surveys sky viewer to access the DES DR1 images.\footnote{\url{http://legacysurvey.org/}}

\figref{cutouts} shows cutouts around 20 candidates passing our ML classifier. 
Our visual inspection procedure classified candidates 2, 3, 8, 11, 12, 13, 14, 15, and 18 as LSBGs. 
Some of these objects are elliptical galaxies while others are spirals. 
We see that candidates 10 and 11 represent the same object, as do 4, 5, 6, and 7. These duplicates come from \SExtractor shredding larger galaxies into smaller constituents. When we find sources that have been shredded in this way, we make an effort to ``stitch'' the segmentation maps back together for the \galfitm (Section \ref{sec:galfit}).
In these cases, we picked the candidate that was best centered on the galaxy; in the example presented here, these are candidates 11 and 4. 
To avoid further contamination from duplicates in our sample, we also ran an automated spatial cross-match on our final catalog to remove duplicate objects separated by $<4\arcsec$.
Candidates 0, 1, 9, 16, 17, and 19 were rejected by visual inspection as false positives. 
For some candidates (i.e., number 4), it is not immediately clear whether they are isolated LSBGs or tidal debris from larger nearby galaxies.
In these cases,  we used the DES Sky Viewer\footnote{\url{https://desportal2.cosmology.illinois.edu/sky/}} to inspect the region surrounding the candidate. The DES Sky Viewer provides flexible zooming and scaling, and we ended up rejecting candidate 4, because it is a point-like object blended with the diffuse light of a large galaxy centered outside of the cutout.
We note that we make no attempt to distinguish between small, low-luminosity, nearby LSBGs and large, luminous, distant LSBGs.

After visual inspection, our sample contains 21,292 objects. Although we tried to minimize false positives, this sample may still contain a small fraction of low-surface-brightness contaminants such as:

\begin{enumerate}[wide, labelwidth=!, labelindent=0pt, itemsep=0pt]
  \item Ejecta from large galaxies that reside outside the small angular size of the cutouts. 
  \item Small background galaxies in the halos of bright stars.
  \item Recent mergers with extended halos of stellar debris.
\end{enumerate}

\subsection{S\'ersic Model Fitting}\label{sec:galfit}

To compare the properties of our LSBG catalog against similar catalogs in the literature \citep[e.g.,][]{Greco:2018}, we fit each galaxy with a single-component S\'ersic light profile.
We use \galfitm, a multi-band implementation of \galfit developed in the context of the \code{MegaMorph} project \citep[][]{Peng:2002,Barden:2012,Haussler:2013}, to perform a multi-band fit for each galaxy using the DES coadd images from the $g$, $r$, and $i$ bands. 
We started by creating square cutout images centered on each galaxy. 
The cutout size was set to be $10 \times the \var{FLUX\_RADIUS}$ of each galaxy (rounded up to the nearest 50 pixel step). 
A minimum cutout size of $201 \times 201 \pix$ ($\roughly 50\arcsec$ on a side) was used for small galaxies. 
We assembled a mask in each band by combining the segmentation map from the DES detection coadd (a combination of the $r,i,z$ images) with the bad pixel mask from each individual band.
The \galfitm ``sigma image'' was derived from the inverse variance weights plane produced by \code{SCAMP} \citep{Bertin:2006} for each of the DES coadded images.

Large LSBGs are sometimes segmented into several catalog objects by \SExtractor.
Since we are using the segmentation map as a mask, regions of the image associated with other \SExtractor sources are excluded from the \galfitm analysis by default.
These ``siblings'' of the LSBG often consist of foreground stars, background galaxies, and various stellar overdensities associated with the LSBG itself (e.g., globular clusters, star forming regions, nuclei of recently merged satellites, etc.), as well as spurious shredding of the (mostly) smooth emission of the LSBG.
To avoid unnecessary masking, we visually inspect the segmentation maps of each LSBG in our sample.
We remove mask regions associated with spurious shredding, while retaining masks associated with compact, high-surface brightness objects.
Approximately 5\% of our LSBG sample had segmentation maps modified in this way.

The parameters of the S\'ersic model fit were initialized based on the values of the \SExtractor catalog.
The centroid was initialized at the position derived by \SExtractor, and was constrained within 10\% of the  \code{FLUX\_RADIUS}.
The S\'ersic effective radius was similarly initialized based on the \code{FLUX\_RADIUS} and was constrained to be within a factor of 2 from this initial value.
The S\'ersic index was initialized at a value of $n = 1.0$ and was constrained to lie within the range $0.2 < n < 5.0$.
The \galfitm package uses a series of Chebyshev polynomials to parameterize the morphological parameters as a function of wavelength \citep{Haussler:2013}. 

When performing the fit with \galfitm, we tied the centroid position, S\'ersic index, ellipticity, and position angle across the three bands.
In contrast, the flux normalization of the model was allowed to vary independently in each band according to a quadratic function of wavelength, and the effective radius was fit in each band as a linear function of wavelength.
This has the effect of constraining color gradients to vary monotonically with wavelength.
We visually inspect the residuals of each fit to identify and correct catastrophic errors.
The resulting best-fit S\'ersic model parameters are provided as supplemental material.

While the S\'ersic model fit provides consistent properties across all objects in our sample and allows comparison to similar catalogs in the literature, it is not a sufficiently complex model to provide a good fit for all LSBGs.
In particular, we note that a subset of our objects would be fit better through the inclusion of a nuclear point source, while others show clear indications of irregular, peculiar, or spiral structure.
We provide a local estimate of the reduced $\chi^2$ ($\chi^2$ per degree of freedom) of our model in each band calculated within the central region of each LSBG.
This information can be used to identify objects that were poorly fit by the simple S\'ersic model, and can be followed up with more detailed modeling.
The most common modeling issue comes from the existence of compact nuclear sources, which often lead to local $\chi^2 > 3$.

\subsection{Extinction Correction and Final Cuts}

We corrected for the effects of Galactic interstellar extinction on the magnitudes and other derived quantities (color and surface brightness) of our sample. 
We used the fiducial DES interstellar extinction coefficients \citepalias[see Section 4.2 of][]{DES:2018}.
Briefly, these were derived from the $E(B-V)$ maps of \citet{Schlegel1998} with the normalization adjustment of \citet{Schlafly:2011} using the reddening law of \citet{Fitzpatrick:1999} with $R_V = 3.1$.
For the remainder of this paper, we refer only to the extinction-corrected properties of our sample.

As a final step in defining our LSBG sample, we require that galaxies have $\Reff(g)>2.5''$ and $\mumeaneff(g)>24.2 \magasecsq$ \footnote{Note that there is no consensus in the literature about the definition of the effective radius of the LSBGs. Some authors use the semi-major axis $\Reff=a$ of the ellipse used in the Sérsic model fit, while others use the circularized effective radius, defined as $\overline{\Reff} = \Reff\sqrt{b/a}$. We use the first option, and then we estimate the mean surface brightness as the total flux contained within the ellipse over its area.} based on the extinction-corrected S\'ersic profile fit. 
After performing these cuts, our final sample contains \NLSBG LSBGs distributed over the $\roughly 5000 \deg^2$ DES Y3 footprint. Interestingly, the average angular number density of LSBGs in DES Y3 ($\roughly 4.5 \deg^{-2}$) is similar to that found in the first $\roughly 200 \deg^2$ of HSC SSP \citep[$\roughly 3.9 \deg^{-2}$,][]{Greco:2018}.

\section{Detection Efficiency around the Fornax Cluster} \label{sec:efficiency}

\begin{figure}[t!]
\centering
\epsscale{1.1}
\plotone{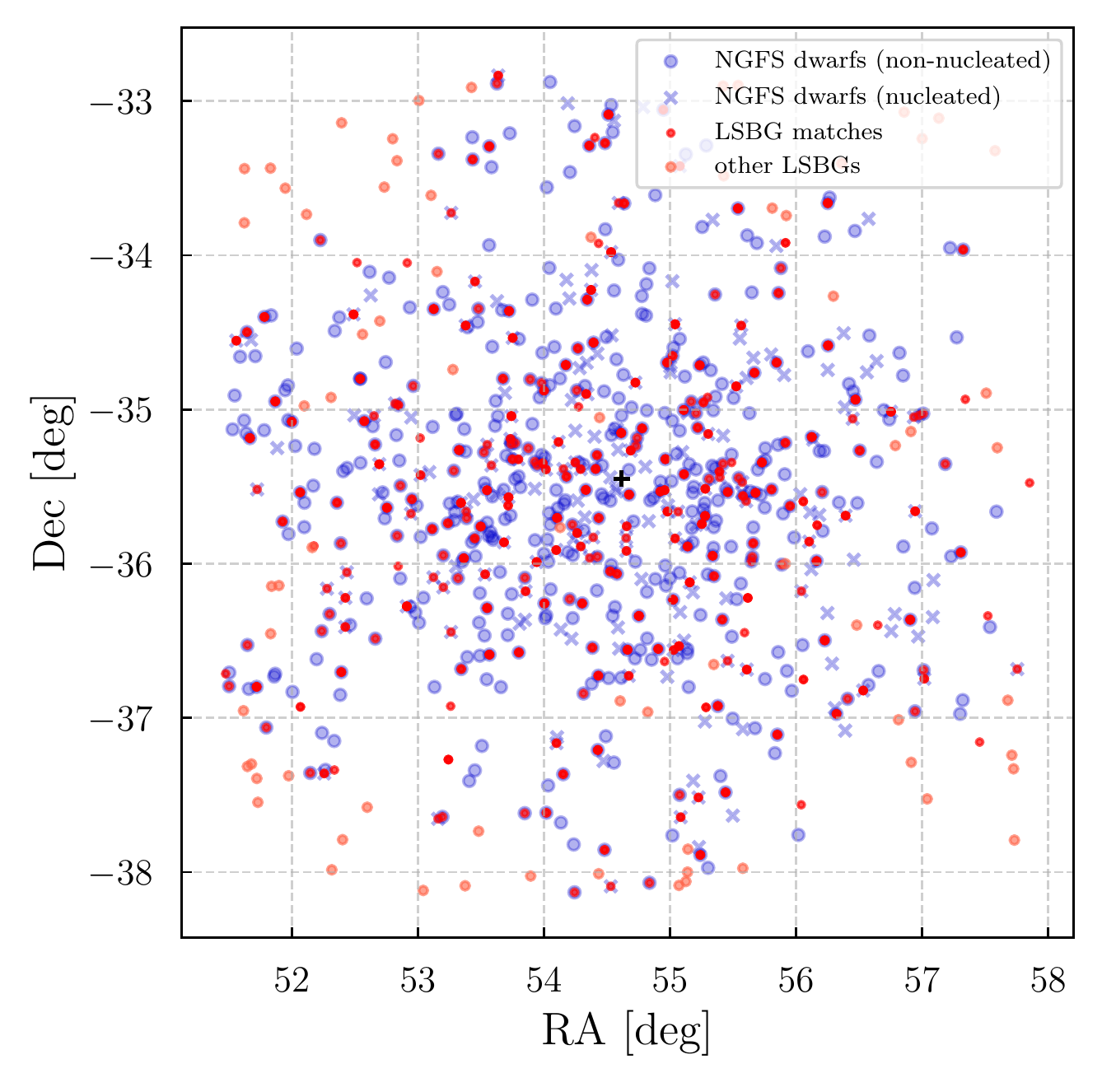}
\caption{The dwarf galaxies present in the NGFS catalog (in blue) and the matches from our DES LSBG catalog (red). The NGFS catalog is separated into nucleated (denoted by an `X') and non-nucleated (circles) galaxies. We plot DES LSBGs that were not matched to NGFS objects in light red (these are generally located outside the NGFS area). The black cross denotes the nominal center of the Fornax cluster.
}
\label{fig:Efficiency}
\end{figure}

\begin{deluxetable*}{lccc}
\tablenum{1}
\tablecaption{Detection efficiency around the Fornax Cluster}
\tablewidth{10pt}
\tablehead{
\colhead{Cuts applied}&\colhead{All galaxies}&\colhead{Nucleated}&\colhead{Non-nucleated} 
}
\startdata
No cuts & 76.6\% & 89.5\%  & 71.6\%\\
Surface-brightness cut only & 63.1\%  & 58.6\%  & 64.9\%\\
Angular size cut only & 56.4\% & 81.8\%  & 46.4\%\\
Both cuts & 43.4\% & 52.5\%& 40.3\%\\
Final result (after ML/Vis. inspection) & 37.7\% & 46.9\%& 34.1\%\\ 
\enddata
\tablecomments{
\label{tab:Efficiency_Summary}
Efficiency of our LSBG selection procedure estimated by comparing to the NGFS catalog \citep{Eigenthaler:2018,Ordenes:2018}.
We calculate the fraction of NGFS objects included in the DES LSBG sample after performing each step in sample selection. 
We also present the efficiency for nucleated and non-nucleated subsamples separately.}
\end{deluxetable*}






To estimate the efficiency of our multi-step LSBG selection procedure, we compare our LSBG catalog to similar catalogs produced with deeper data (note that here by deeper we refer to the point-source depth, not the surface brightness).
The Fornax galaxy cluster (Abell S373) resides within the DES footprint  and is known to host a large population of faint galaxies \citep[e.g.,][]{Ferguson:1989,Hilker:1999,Munoz:2015,Venhola:2017}. 
In particular, the Next Generation Fornax Survey \citep[NGFS;][]{Munoz:2015} has used DECam to image the region around Fornax to an ${\rm S/N} = 5$ point-source depth of $g=26.1$ and $i=25.3$, which is approximately 2 magnitudes deeper than the DES Y3 imaging in this region of the sky.
The NGFS has assembled catalogs of dwarf galaxies covering $\roughly 30 \deg^2$ around the Fornax cluster.
The NGFS has reported a total dwarf galaxy population of 643 galaxies, which is split into nucleated (181) and non-nucleated (462) galaxies \citep{Eigenthaler:2018,Ordenes:2018}.

The NGFS dwarf galaxy catalogs were assembled through visual inspection of the DECam data surrounding Fornax. 
The NGFS catalog creation process was specifically focused on identifying dwarf galaxies/LSBGs, and it did not apply any cuts similar to those that we imposed on the photometric DES catalog.
This makes the NFGS an interesting independent data set to quantitatively evaluate the efficiency of our catalog creation and LSBG sample selection procedures. 

We match the NGFS catalogs from \citet{Eigenthaler:2018} and \citet{Ordenes:2018} with the DES Y3 Gold catalog using a matching radius of $3\arcsec$ (we find that using a larger matching radius does not significantly increase the number of matches).
In \tabref{Efficiency_Summary}, we report the fraction of objects from the NGFS catalog that are matched to objects in the DES Y3 Gold catalog before any cuts, and the resulting change in the matched fraction of galaxies as we apply each of the LSBG selection criteria defined in \secref{pipeline}. 
This allows us to estimate the efficiency of each cut and the completeness of our final LSBG sample relative to the NGFS sample.
We also examine the efficiency of our selection to nucleated and non-nucleated galaxies separately, since the non-nucleated galaxies in the NGFS were found to be fainter and smaller than their nucleated counterparts.

\tabref{Efficiency_Summary} shows that $\roughly77\%$ of the NGFS galaxies were matched to objects in the DES Y3 Gold catalog generated with \SExtractor.
As expected, the recovery fraction is higher for the nucleated LSBGs where the DES detection efficiency reaches $\sim 90\%$.
Our surface-brightness cut significantly reduces the number of detected objects, affecting nucleated galaxies more strongly due to their higher central surface brightnesses. 
The angular size cut, $r_{1/2} > 2.5''$, results in a more significant reduction in the efficiency for recovering non-nucleated galaxies. 
We expect that this angular size cut will result in an even more severe reduction in the number of distant LSBGs that pass our cuts, since more distant galaxies will be required to have larger physical sizes.

After applying both surface-brightness and size criteria, the detection efficiency drops to $43.4\%$ overall, with a detection efficiency of $52.2\%$ and $40.3\%$ for the nucleated and non-nucleated subsamples, respectively. We further examine the decrease in efficiency from applying our machine learning classification and visual inspection. We find that the drop in efficiency (difference between the last two rows of \tabref{Efficiency_Summary}) corresponds to an absolute drop of $\roughly 13\%$ in the number of LSBGs in the field that were not detected. That number is consistent with our expectation that the ML classification has ${\rm FNR} \sim 10\%$ (\figref{conf_mat}). Furthermore, visual inspection of misclassified galaxies showed that most were either extremely faint/hard to distinguish from random background fluctuations or too compact to be included in our LSBG catalog.

\figref{Efficiency} shows a scatter plot of the NGFS dwarfs, matched LSBGs from our catalog, and unmatched LSGBs in the region around the Fornax cluster. 
Some of them ($\roughly 5$) are close to an NGFS object and would have been matched with a slightly larger matching radius.
This figure also shows the presence of LSBGs detected in our catalog but not present in the NGFS catalog. Most of these galaxies reside outside of the NGFS footprint. 
Within half the projected virial radius of the Fornax cluster \citep[$\sim 700\kpc$,][]{Drinkwater:2001}, we find 11 LSBGs not present in the NGFS catalog. 

Overall, our analysis here shows that our pipeline is able to retrieve most of NGFS LSBGs, as we defined them based on the surface-brightness and radius cuts.

NGFS has the benefit of having been conducted with the same instrument as DES, thus optimal for comparison with our catalog. However, completeness estimates are not provided. The Fornax Deep Survey (FDS) provides a catalog of 564 dwarf galaxies around Fornax, together with completeness estimates from simulations \citep{Venhola:2017,Venhola2018}. This catalog is $\ge 50\%$ complete at a mean surface brightness (in the $r$ band) of $\mumeaneff(r) = 26.0 \magasecsq$.

We match our sample with the FDS catalog using a matching radius of $3''$. Before applying any cuts, we find that $\sim 92\%$ of the galaxies in FDS are also present in the DES data. We repeat this matching after applying cuts of  $\mumeaneff(r)>24.2 \magasecsq$ and $\Reff(r)>2.5''$ (only $r$-band data were provided for FDS) to both the DES catalog and the FDS catalog. We find that $\sim 66\%$ of the galaxies in the FDS catalog are contained in the DES catalog. A more detailed analysis of efficiency as a function of surface brightness and radius is not very informative given the small number of galaxies that pass the LSBG selection. However, we find that the DES LSBG catalog is $80-90\%$ complete for the lowest- and highest-surface-brightness galaxies.

\section{LSBG Properties} \label{sec:properties}

\begin{figure*}[t!]
\centering
\subfigure[]{\includegraphics[width=1.05\columnwidth]{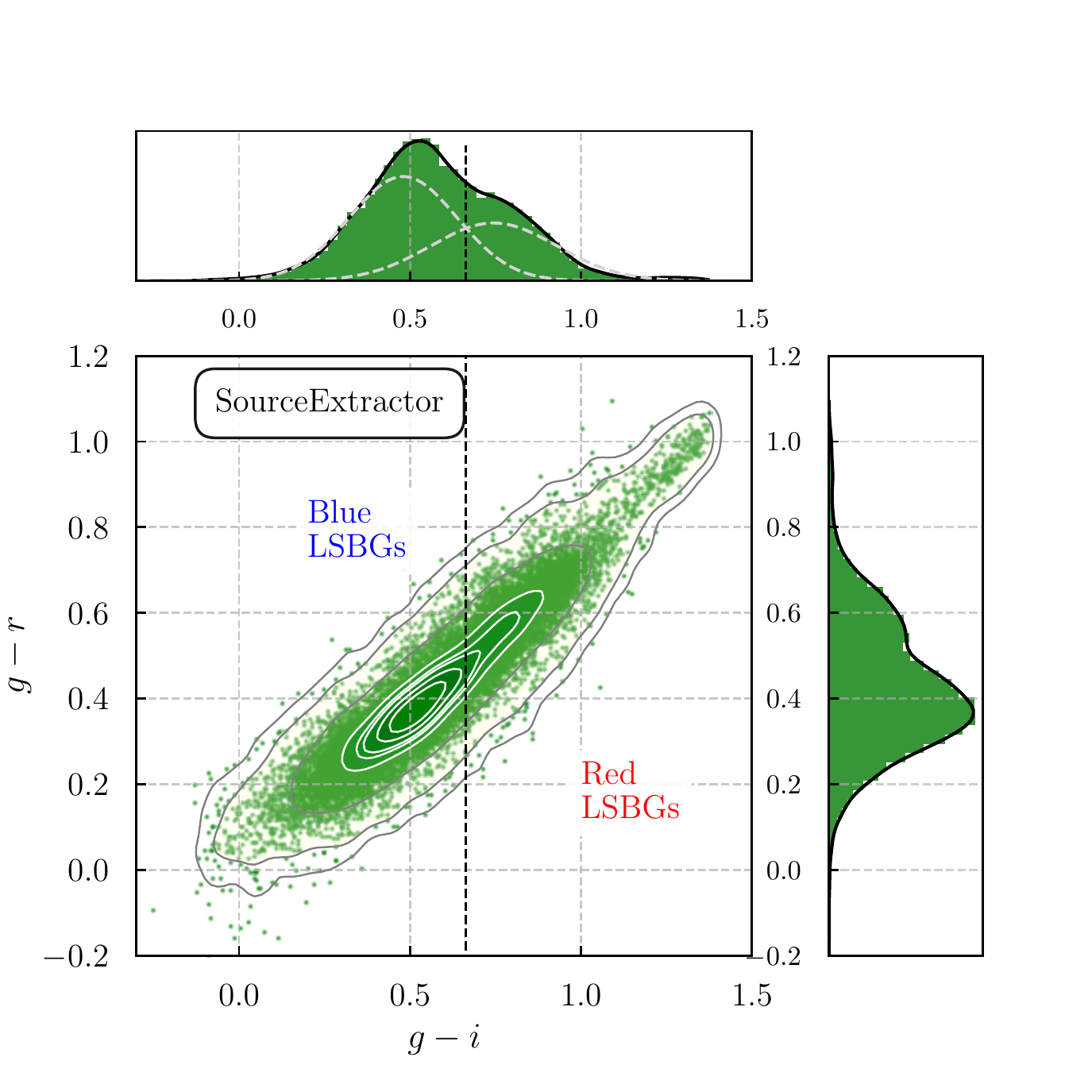}}
\subfigure[]{\includegraphics[width=1.05\columnwidth]{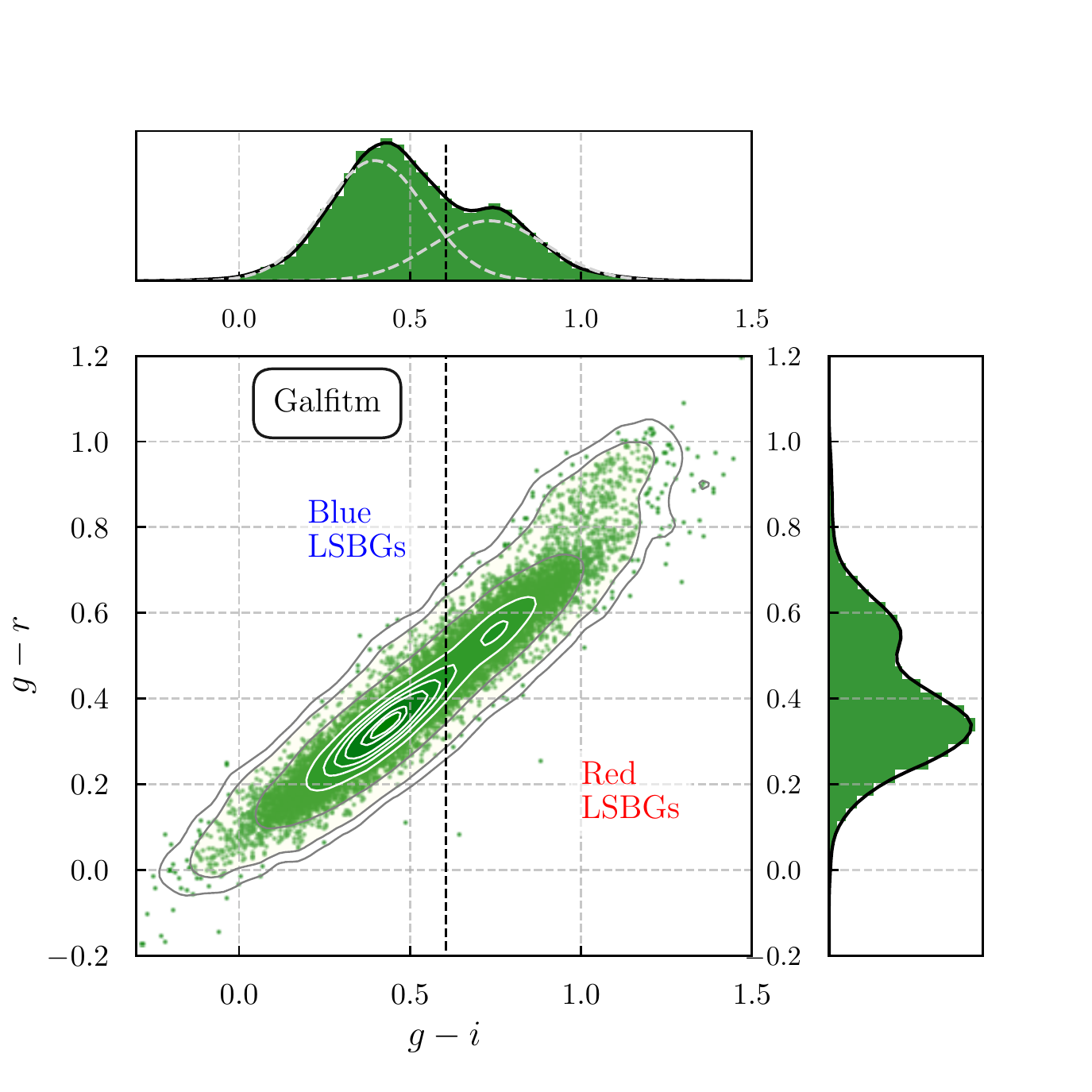}}
\caption{Color--color diagram of our LSBG sample, using (a) \SExtractor \code{MAG\_AUTO} parameters and (b) magnitudes derived by fitting with \code{galfitm}. In both cases, we observe a bimodality in the $g-i$ and $g-r$ color distributions. We separate the total sample into red and blue galaxies, based on their $g-i$ color value: we fit the $g-i$ distribution with a Gaussian mixture model with two Gaussians (gray dashed lines in the top panels) and find the intersection point. This is at $g-i = 0.66$ and $g-i = 0.60$ for the \SExtractor and \code{galfitm} cases, respectively (black vertical dashed lines). We use the intersection point derived from the \code{galfitm} distribution to define red and blue LSBG samples.}
\label{fig:col_col}
\end{figure*}

The large sky area covered by DES ($\sim 5000 \deg^2$) gives us a unique opportunity to study the statistical properties of the LSBG population.
Our search results in a sample of \NLSBG LSBGs with effective radii $\Reff(g) > 2.5\arcsec$ and extinction-corrected mean effective surface brightnesses $\mumeaneff(g) > 24.2 \magasec^{-2}$.
This is the largest such catalog of LSBGs to date. 
In this section, we divide our catalog of LSBGs into red and blue subsamples and compare the properties of these samples to each other and to previous results \citep[i.e.,][]{Greco:2018}.

The optical colors of galaxies are indicative of their stellar populations.
Colors are known to correlate strongly with galaxy morphology and environment. 
Galaxies are conventionally divided based on color into two well-known sequences of red and blue galaxies \citep[e.g.,][]{Strateva:2001, Blanton:2009}.
Less is known about how the colors of LSBGs correlate with  morphology, star formation history, and environment. 
For example, \citet{ONeil:1997} found that classical disk LSBGs span a range of blue and red colors.
Similar to high-surface-brightness galaxies (HSBGs), blue colors are generally associated with actively star forming spiral or irregular systems, while red colors tend to be indicative of spheroidal or elliptical morphology \citep[e.g.,][]{Larson:1980,Strateva:2001,Baldry:2004,Lintott:2011}.
Red galaxies are found preferentially in denser environments, where quenching from massive hosts prevents ongoing star formation \citep{Bamford:2009,Geha:2017,Roman:2017b}.
\citet{Greco:2018} found that LSBGs detected in HSC showed a clear bimodality in color, with two apparently distinct populations separated at $g' - i' = 0.64$ (where $g'$ and $i'$ are used to indicate extinction-corrected magnitudes in the HSC filters). 
They found that blue LSBGs had a brighter mean surface brightness, while galaxies that are large ($\Reff > 6\arcsec$) and faint ($\mumeaneff(g) > 26 \magasec^{-2}$) are almost exclusively red.

In \figref{col_col}, we present the distribution of our LSBG sample in the $g-i$ vs. $g-r$ color space. We show the color-color diagrams derived from the \SExtractor \code{MAG\_AUTO} quantities (left panel), and  the magnitudes derived from the \code{galfitm}  S\'ersic model fit (right panel). The color distributions are similar and present signs of bimodality that are slightly more prominent using colors from the S\'ersic model fit.  
Having established the similarity of the color distributions derived from these two fits, in the remainder of this paper, we quote photometric parameters (magnitudes, colors, surface brightness)  derived from the \code{galfitm} model.
Thus, photometric and structural parameters (S\'ersic index, effective radius) come from the same model fit and can be consistently compared to results in the literature.

We separate the total LSBG sample into red and blue subsamples, according to their $g-i$ color. To do so, we use the following procedure: we fit a two-component Gaussian mixture model (GMM) to the 1D $g-i$ color distribution. The components can be seen in the top panels of \figref{col_col} (dashed gray lines).  
We find that the two Gaussians intersect at $g-i = 0.60$ (\code{galfit} case; for comparison using the distribution coming from the \SExtractor quantities the same point is at $g-i = 0.66$). 
We define a red galaxy sample as galaxies with $g-i \geq 0.60$ (7,671 galaxies), and a blue galaxy sample as galaxies with $g-i < 0.60$ (16,119 galaxies).
Note that in the upper-right corner in both panels a ``tail" of objects is clearly visible. Inspecting them visually and checking the $\chi^2$ of their \galfit model fit,  we found that most of these are poorly fitted spiral LSBGs.

Our  $g-i$ separation threshold is bluer than that of \citeauthor{Greco:2018} ($g'-i' = 0.64$ in the HSC bandpass).\footnote{From a comparison of matched point sources in the HSC SSP Wide and DES Y3 Gold catalogs, we find that the difference between HSC and DES colors is $\Delta (g-i) = 0.013$ for sources with $0.3 < (g' - r') < 0.6$.}
Note that \citeauthor{Greco:2018} used the median of the distribution to separate the two populations, which was effective as the two populations had similar size. 
However, the DES LSBG sample is dominated by blue galaxies, which shifts the median to $(g-i) = 0.60$.
The median colors of our red and blue LSBG subsamples are $g-i = 0.76$ and $g-i = 0.40$, respectively.

\begin{figure*}[t]
\centering
\subfigure[]{\includegraphics[width=0.49\textwidth]{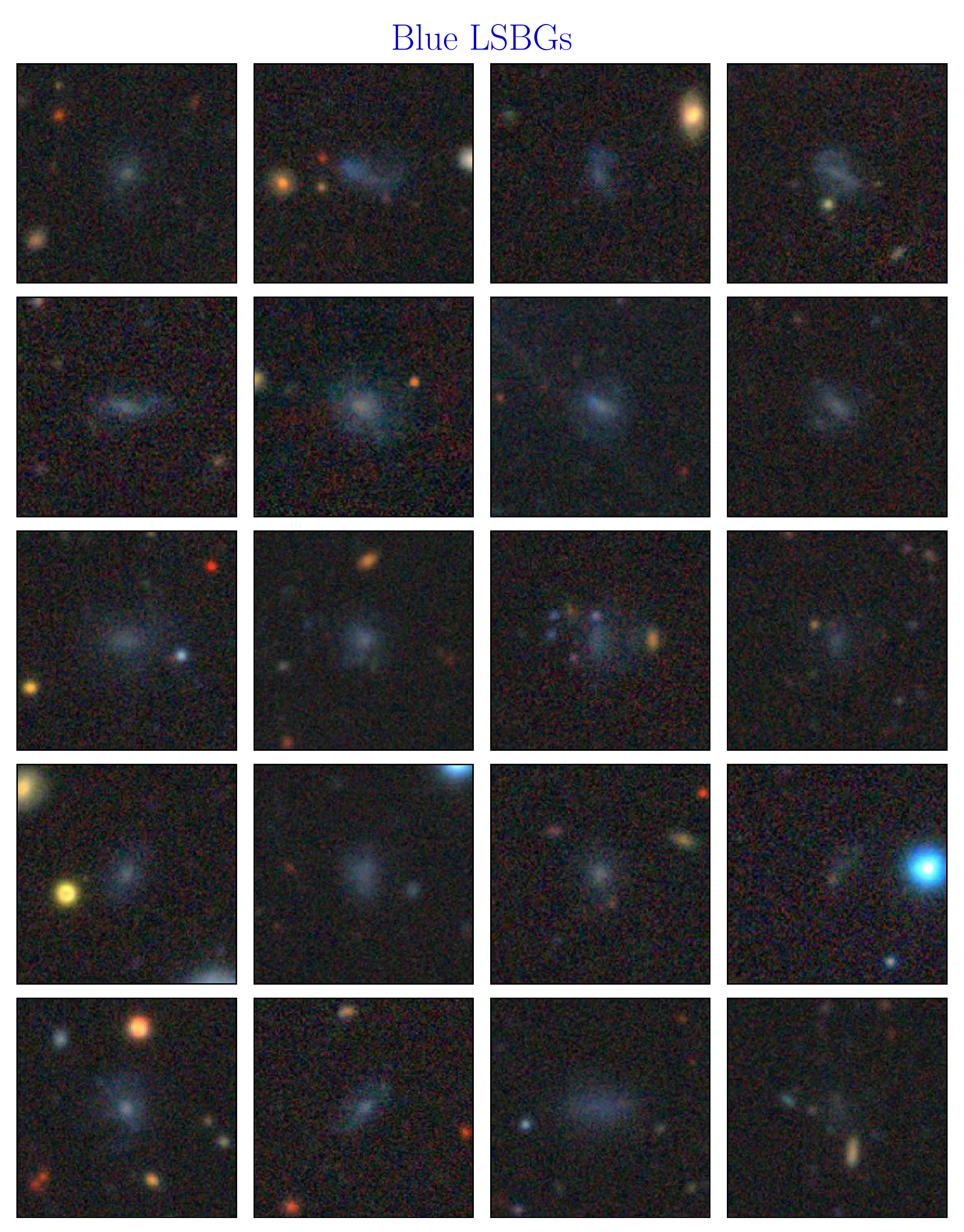}}
\subfigure[]{\includegraphics[width=0.49\textwidth]{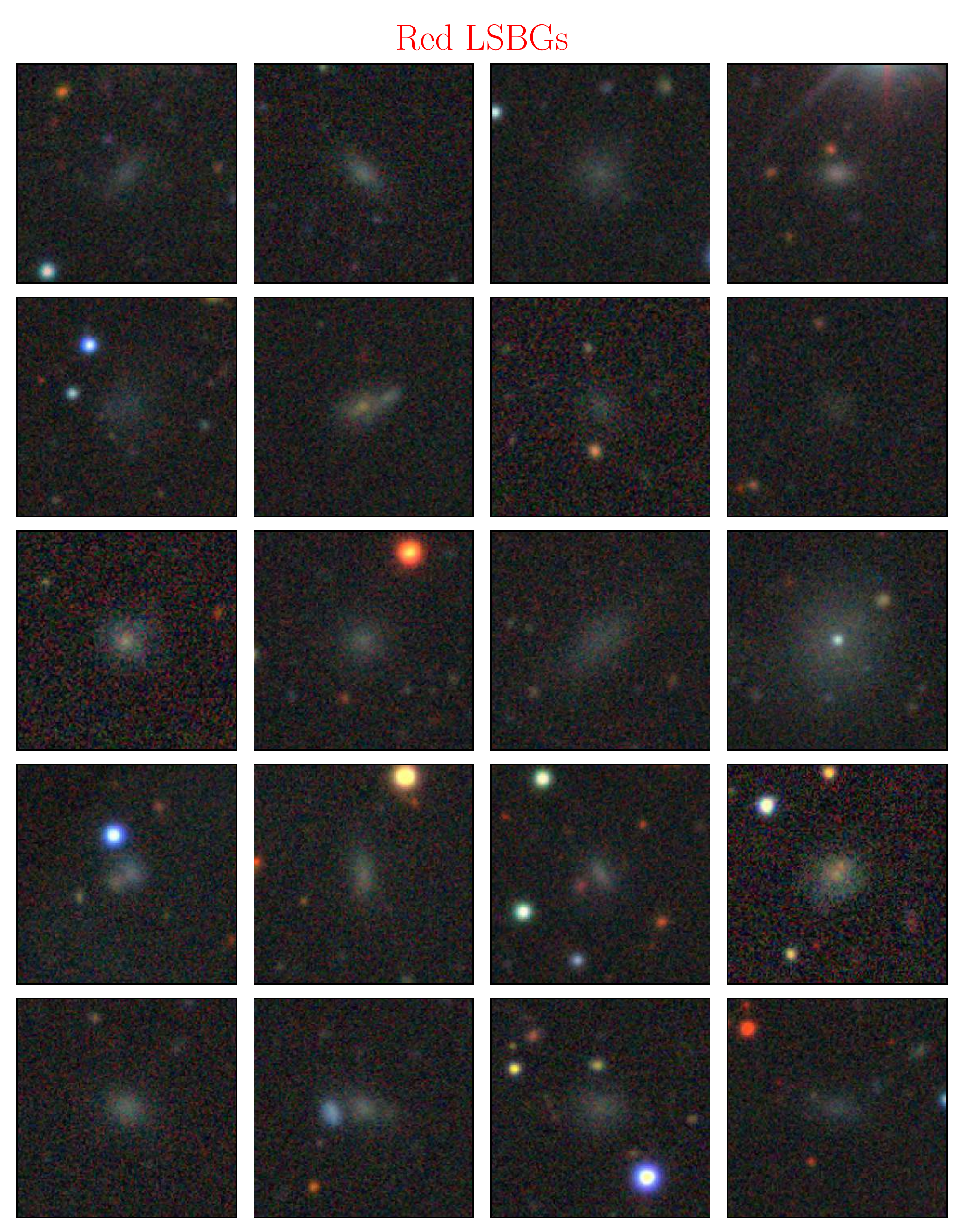}}
\caption{Examples of (a) blue  and (b) red LSBGs in our sample. We randomly selected red galaxies with $g-i$ above the median for the red population ($g-i>0.76$) and blue galaxies below the median of the blue population ($g-i < 0.40$) to make the color difference more prominent. Each cutout is $30\arcsec \times 30\arcsec$ in size.}
\label{fig:blue_red_examples}
\end{figure*}

In \figref{blue_red_examples} we show examples of randomly selected blue galaxies with $g-i < 0.40$ (below the median of the blue population) and red galaxies with $g-i > 0.76$ (above the median of the red population). As we can see, the two subsamples show morphological differences. The blue sample is composed primarily of irregular galaxies and galaxies with signs of spiral structure. The red sample consists predominantly of nucleated and non-nucleated spherical and elliptical galaxies. 

\begin{figure*}[t]
\centering
\subfigure[]{\includegraphics[width=1.05\columnwidth]{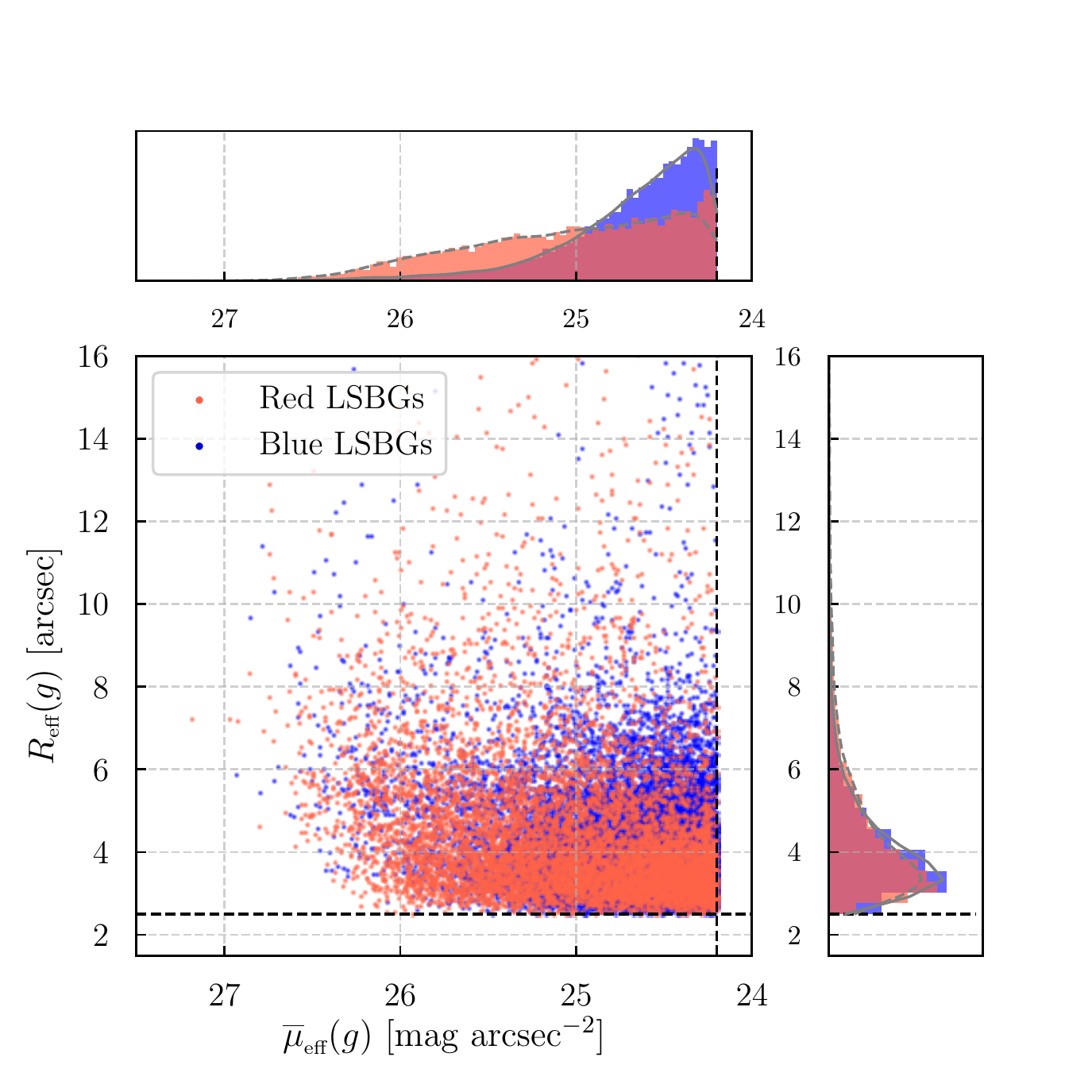}}
\subfigure[]{\includegraphics[width=1.05\columnwidth]{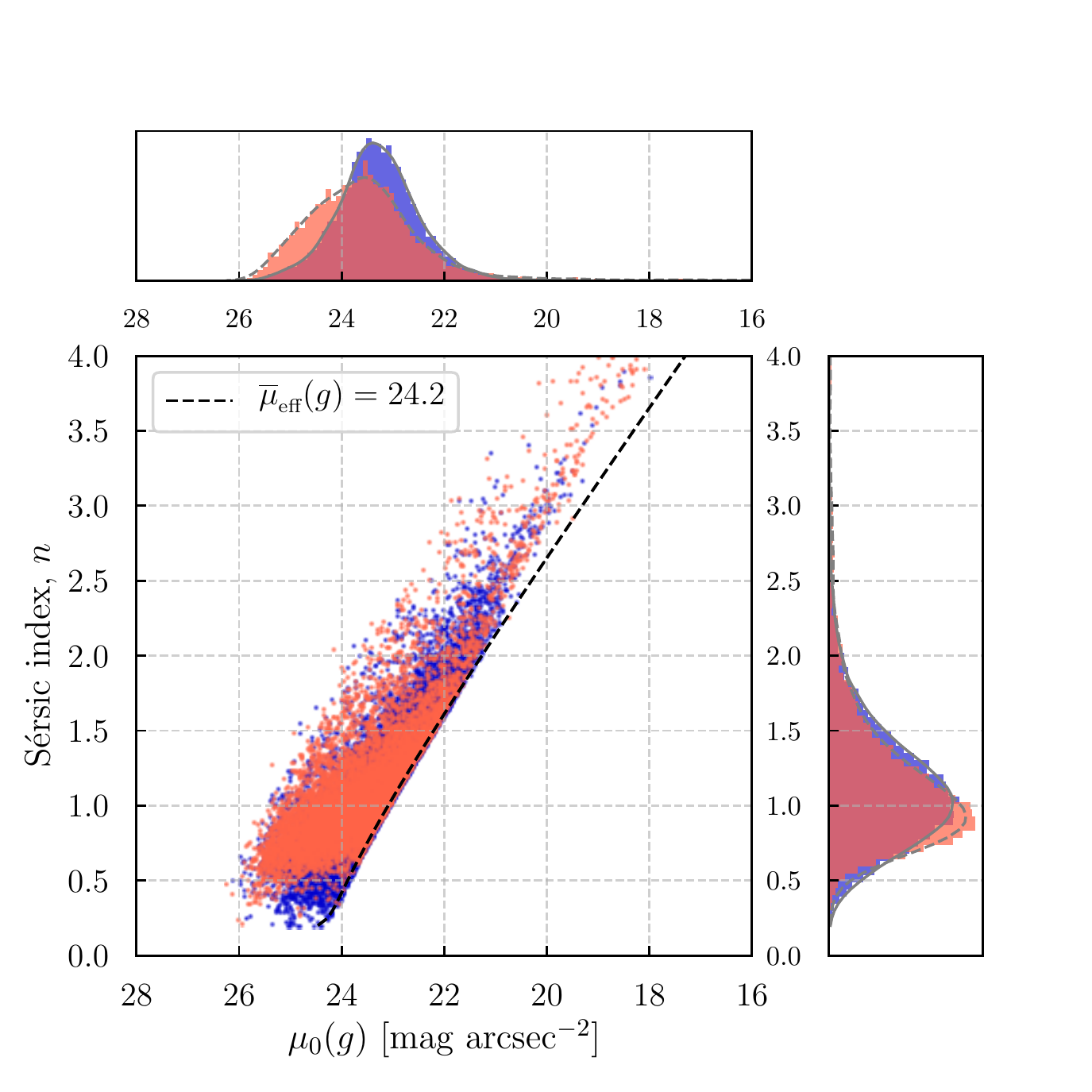}}
\caption{(a) Joint distribution of the red and blue LSBGs in the space of effective radius, $\Reff$, and mean surface brightness (within the effective radius),  $\mumeaneff$, both in the $g$ band. The two populations are defined according to the $g-i$ color criterion described in \secref{properties}. The dashed horizontal and vertical lines correspond to the limits of the selection criteria $r_{1/2} > 2.5''$ and $\mumeaneff(g) > 24.2 \magasec^{-2}$, respectively. Note that although surface brightness is independent of distance, and thus the scatter shown here reflects the intrinsic properties of our sample, much of the scatter in the angular effective radius comes from the fact that the LSBGs lie at different distances. (b) S\'ersic index, $n$, versus central surface brightness, $\mu_0(g)$ \citep[e.g.,][]{Graham:2005}, for the galaxies in our red and blue subsamples. The black dashed line corresponds to our selection criterion, $\mumeaneff(g) = 24.2 \magasec^{-2}$.
}
\label{fig:eff_rad}
\end{figure*}

In the left panel of \figref{eff_rad}, we present the joint distribution of our red and blue LSBG samples in the space of effective radius, $\Reff(g)$, and mean surface brightness (within the effective radius),  $\mumeaneff(g)$. Both populations have sizes ranging from $2.5'' - 16''$. Despite the wide range in angular sizes, most LSBGs in our sample ($90\%$) have radii less than $6''$, with a median of $\roughly 4''$. Note that the scatter in angular sizes does not necessarily mean that our galaxies occupy a wide range in physical sizes; much of the scatter comes from the fact that our sample contains galaxies at different distances. For example, in \secref{clusters}, we show that overdensities in the distribution of LSBGs are associated with galaxy clusters that lie in a range of distances between $\roughly 20 \Mpc$ and $\roughly 100 \Mpc$. For a typical galaxy size of $\roughly 1\kpc$, that translates into a range of angular sizes between $2\arcsec$--$10\arcsec$.

We find that the red galaxy population has a larger tail toward lower surface brightness (larger values of $\mumeaneff(g)$), while the blue galaxies tend to have higher mean surface brightness.
The $50^{\rm th}$, $80^{\rm th}$, and $90^{\rm th}$ percentiles in surface brightness are $\mumeaneff(g) = 24.6, 24.9, 25.2  \magasec^{-2}$ for the red sample and $\mumeaneff(g) = 24.9, 25.6, 25.9  \magasec^{-2}$ for the blue sample.
This result is interesting in the context of early studies that showed no pronounced relationship between color and surface brightness \citep[e.g.,][]{Bothun:1997}.
However, extrapolating the size--luminosity relationship for red and blue galaxies in SDSS \citep{Shen:2003} suggests that at lower luminosities, red galaxies should be larger than their blue counterparts. 
A similar result has been shown for the LSBG sample from HSC SSP \citep[][]{Greco:2018}.

In the right panel of \figref{eff_rad} we plot the S\'ersic index, $n$, versus the central surface brightness, $\mu_0(g)$, for our red and blue LSBG samples \citep[e.g.,][]{Graham:2005}.
The distribution in the S\'ersic index is similar for two samples, with $0.2 \lesssim n \lesssim 4.0$ and median of $n \sim 1.0$. 
We do note that the red LSBGs tend to be underrepresented in the regime of small S\'ersic index, $n < 0.7$.
Unsurprisingly, we find that blue galaxies tend to have higher central surface brightness; however, the difference in central surface brightness between red and blue galaxies is not as striking as the difference in mean surface brightness.
The median of the red population is at $\mu_0(g) = 23.6 \magasec^{-2}$, while that of the blue population at $\mu_0(g) = 23.3 \magasec^{-2}$.

\section{Clustering of LSBGs}
\label{sec:clustering}

\subsection{Clustering of Red and Blue LSBGs} \label{sec:red_blue_clustering}

\begin{figure*}[t]
\centering
\subfigure[]{\includegraphics[width=2.05\columnwidth]{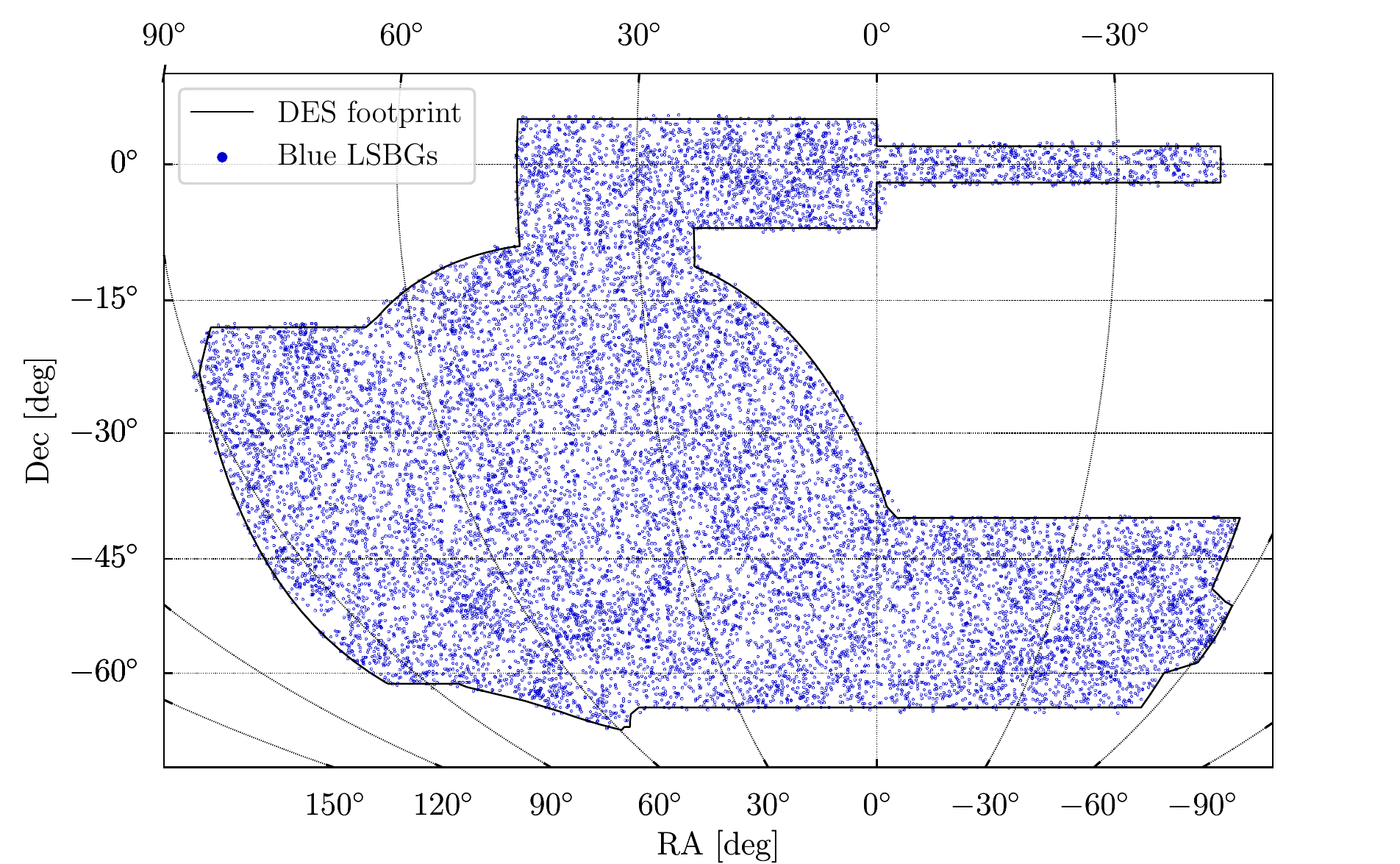}}
\subfigure[]{\includegraphics[width=2.05\columnwidth]{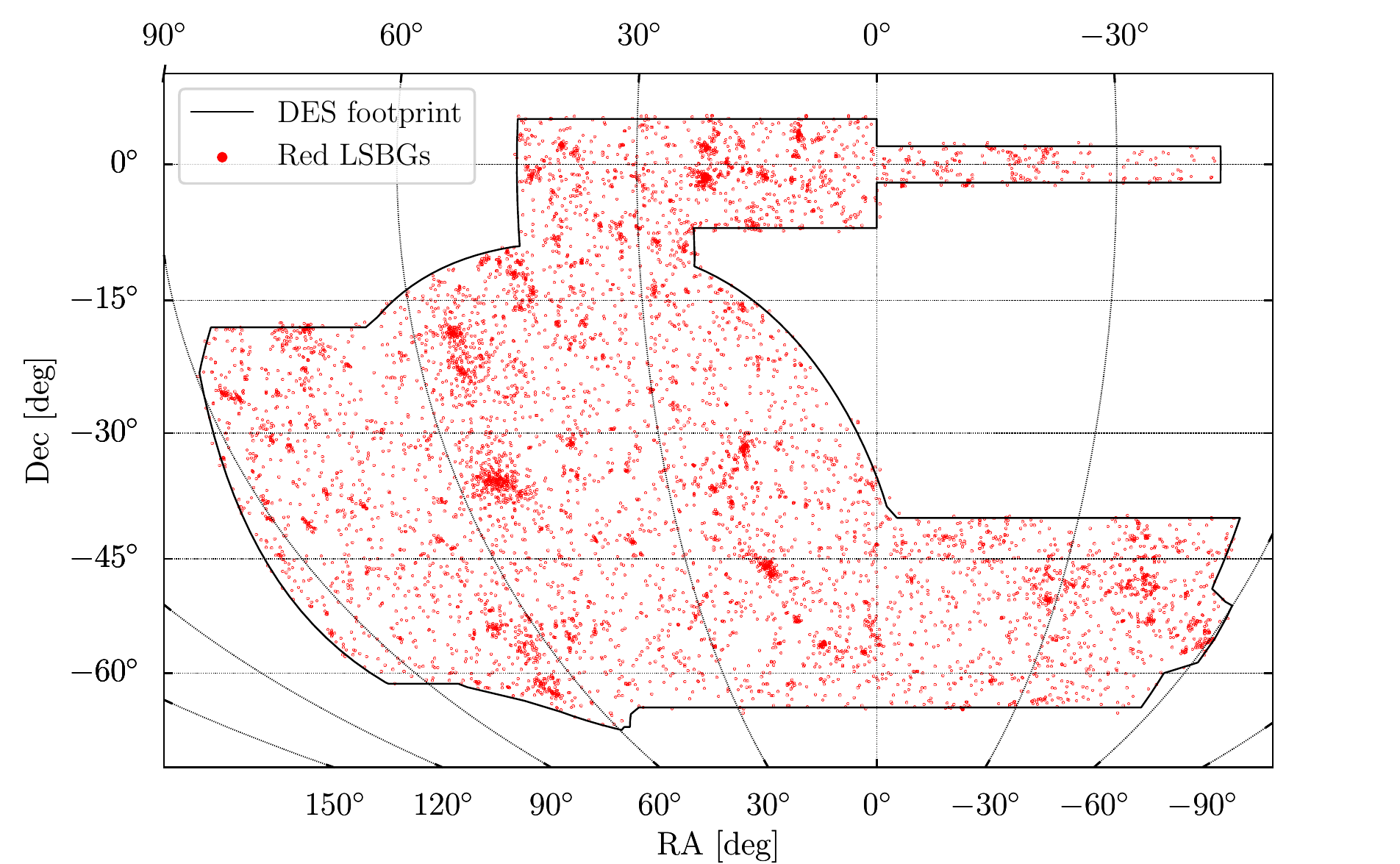}}
\caption{Sky positions of (a) blue LSBGs ($g-i < 0.60$; 7,671 galaxies) and (b) red LSBGs ($g-i \geq 0.60$; 16,119 galaxies) within the DES footprint. The distribution of the red LSBGs is more strongly clustered than that of the blue LSBGs.}
\label{fig:red_and_blues}
\end{figure*}

\citeauthor{Greco:2018} tentatively suggested that the spatial distribution of LSBGs in the HSC SSP may be correlated with low-redshift galaxies from the NASA-Sloan Atlas\footnote{\url{http://nsatlas.org/}}. 
However, due to the relatively small area covered by their HSC SSP data set ($\roughly 200\deg^2$), they were unable to make any firm statistical statement about possible correlations.
Our DES Y3 LSBG catalog covers a contiguous region $\roughly 25$ times larger than that of \citeauthor{Greco:2018}, allowing us to perform a detailed exploration of the spatial distribution of LSBGs.
In particular, we are able to \emph{separately} explore the clustering of our red and blue LSBG subsamples (as defined in \secref{properties}).
In \figref{red_and_blues}, we present the spatial distribution of blue and red LSBGs over the DES footprint. 
We find a  stark contrast in the spatial distribution of these two LSBG subpopulations: red LSBGs are highly clustered, while blue galaxies are more uniformly distributed.

\begin{figure}[t!]
\centering
\epsscale{1.2}
\plotone{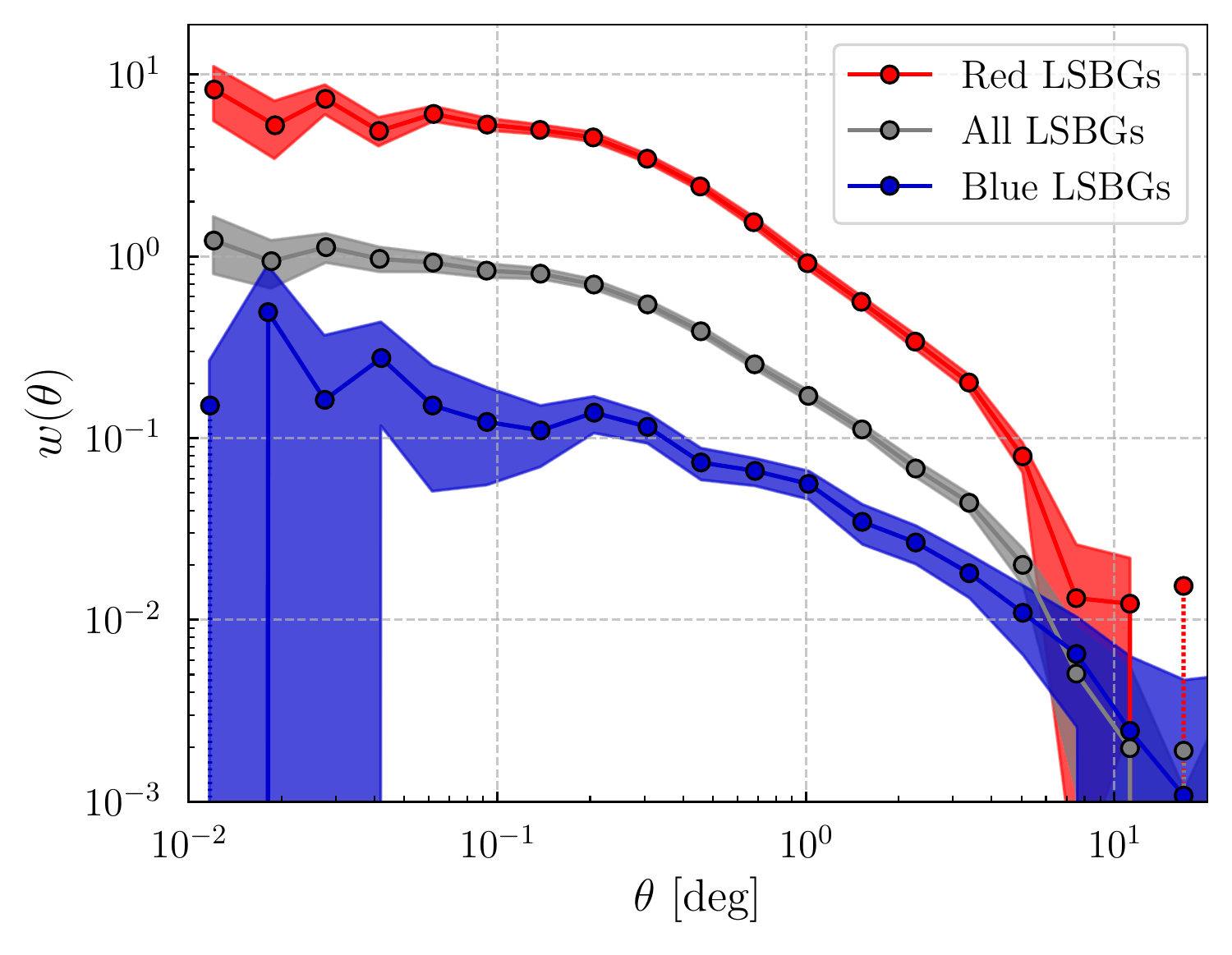}
\vspace{0.4cm}
\caption{The angular autocorrelation function of the total LSBG sample (dark gray line), and the red and blue LSBG subsamples (red and blue lines, accordingly). The errors were calculated using the jackknife method. The correlation function of the red LSBGs has a higher amplitude than that of the blue LSBGs across all angular scales.}
\label{fig:2pt_blue_red}
\end{figure}

To quantify the clustering of our LSBG sample and the red/blue subsamples, we calculate the angular two-point autocorrelation function of LSBGs, $w(\theta)$ \citep[e.g.,][]{Peebles:1980,Connolly:2002}. 
We use \code{treecorr} \citep{Jarvis:2015}\footnote{\url{https://github.com/rmjarvis/TreeCorr}} to calculate $w(\theta)$ using the estimator of \citet{Landy:1993} with a random sample of points drawn from the DES Y3 Gold footprint mask derived from the DES imaging data using \code{mangle} \citep[e.g.,][]{Swanson:2008}.
In \figref{2pt_blue_red} we plot $w(\theta)$ for the full LSBG sample, as well as the red and blue subsamples (gray, red, and blue curves, respectively). 
We estimate the errors on $w(\theta)$ using jackknife resampling \citep[e.g.,][]{Efron:1983}.
As expected from \figref{red_and_blues}, we find that the amplitude of the autocorrelation function of red LSBGs is more than an order of magnitude larger than that of blue LSBGs at angular scales $\theta \lesssim 3^\circ$.

The differences in clustering amplitude between red and blue galaxies has been studied extensively in spectroscopic surveys \citep[e.g.,][]{Zehavi:2002,Zehavi:2005,Zehavi:2011,Law-Smith:2017}.
In particular, it has been noted that there is a strong difference in the amplitude and shape of the autocorrelation function of intrinsically faint red galaxies relative to brighter and/or bluer galaxies \citep[e.g.,][]{Norberg:2002, Hogg:2003, Zehavi:2005, Swanson:2008, Cresswell:2009, Zehavi:2011}.
We find the same pronounced difference in the amplitude and shape of $w(\theta)$ for red LSBGs relative to the blue LSBG subsample and the power-law behavior observed in higher-surface brightness galaxies, $w(\theta) \propto \theta^{-0.7}$ \citep[e.g.,][]{Connolly:2002, Maller:2005, Zehavi:2011, Wang:2013}.
The observed shape of the angular autocorrelation function of red LSBGs (which is also manifested in the total LSBG population)  can be produced if the LSBG sample has a preferred scale for clustering.
We find that we can reproduce the shape of the LSBG $w(\theta)$ by selectively enhancing overdense regions at scales of a few degrees.

Previous theoretical modeling has suggested that the strong clustering of faint red galaxies is the result of these galaxies being dominantly satellites of massive dark matter halos \citep{Berlind:2005, Wang:2009, Zehavi:2011}.
\citet{Zehavi:2011} note a strong inflection in the clustering of faint red galaxies ($M_r < -19$) at a scale of $\roughly 3 h^{-1} \Mpc$.
By mapping this physical scale to the enhanced clustering observed in the red LSBG sample at angular scales of $\theta \lesssim 3^\circ$, we derive an estimated distance of $\roughly 40 \Mpc$ for the clustered red LSBG sample.

To assess whether the difference in clustering observed between red and blue LSBGs could be attributed solely to a difference in stellar mass, we subdivide our red and blue LSBG samples into samples of faint red galaxies ($21 < g < 22$) and bright blue galaxies ($19.5 < g < 20.5$).
Blue galaxies generally have a higher luminosity at a given stellar mass than red galaxies \citep[e.g.,][]{Conroy:2013}.
Following \citet{Greco:2018}, we find that the $(g-i)$ colors of our blue and red LSBGs are well represented by a simple stellar population from \citet{Marigo:2017} with ${\rm [Fe/H]} = -0.4$ and an age of $1 \Gyr$ and $4 \Gyr$, respectively.
We find that these populations differ in total absolute $g$-band magnitude by $\Delta(M_g) \sim 1.5$.
We also find that the angular autocorrelation functions of the bright red and faint blue samples do not differ significantly from the total red and blue LSBG samples, respectively.
This suggests that the difference in clustering shape and amplitude cannot be attributed to a difference in stellar mass alone. 

Some authors have argued that observations support a decrease in the number of LSBGs close to the cores of galaxy clusters \citep[e.g.,][]{vanderBurg:2016,Wittmann:2017}.
Such a suppression could reduce the clustering power on small scales, leading to a flattening in the autocorrelation function.
However, rigorously testing for a suppression in the abundance of LSBGs in dense regions would require end-to-end simulations with injected LSBGs to characterize the DES detection efficiency as a function of local galaxy density.  \citep[e.g., using a tool like \code{Balrog};][]{Suchyta:2016,Everett:2020}.
We leave a detailed characterization of the DES selection function for LSBGs to future work.

\begin{figure}[t!]
\centering
\epsscale{1.2}
\plotone{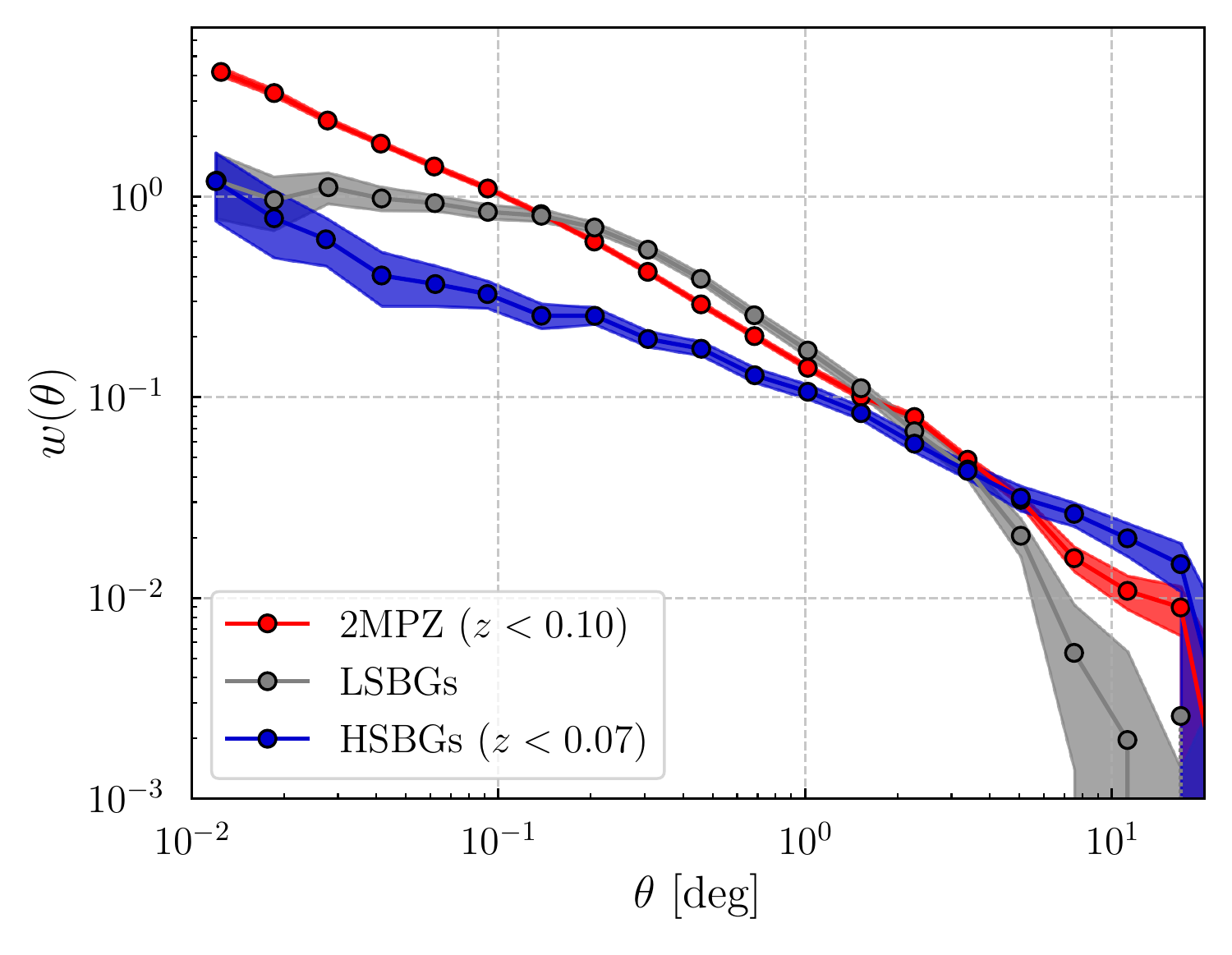}
\vspace{0.4cm}
\caption{The angular autocorrelation function of all LSBGs (gray line), the HSBG sample extracted from the DES data (blue line) and the 2MPZ sample (red line). We see that the LSBG exhibits a turnover at lower angular scales that is not observed either at the HSBG or 2MPZ samples.
}
\label{fig:2pt_HSB_MPZ}
\end{figure}

\subsection{Comparison to other galaxy samples} \label{sec: comparison_to_HSBGs}

We compare the clustering properties of our LSBG sample to two other galaxy samples: a catalog of HSBGs extracted from the DES Y3 Gold catalog, and an external sample of low-redshift galaxies from the 2MASS Photometric Redshift (2MPZ) catalog.
Our goals here are twofold: (1) to compare the clustering of DES galaxies as a function of surface brightness and (2) to use the superior redshifts of the 2MPZ sample to approximately determine the redshift distribution of our LSBGs.

We construct an HSBG sample from the DES Y3 Gold catalog by applying the same star--galaxy separation, color, and ellipticity cuts described in \secref{sample} and summarized in \appref{select}. 
We do not apply any angular size restriction on the HSBG sample, but rather we require that the HSBGs have mean surface brightness $20.0 < \mumeaneff(g) <  22.0 \magasecsq$.
Ideally, we would be able to compare the clustering of LSBGs and HSBGs with the same stellar mass and redshift distributions.
Since the redshift distribution of the LSBGs is unknown, we 
scanned over a range of redshifts for the HSBGs using redshifts estimated trough the Directional Neighbourhood Fitting algorithm  \citep[DNF;][]{DeVicente2016} derived from the DES multi-object fitting (MOF) photometry.

For each redshift-selected sample of HSBGs, we select a random subset of galaxies that produces the same distribution in $g$-band apparent magnitude as our LSBG sample in the range $18 < g < 22$ (see \appref{magnitude_dists}). 
We compare the clustering amplitude of the LSBG and HSBG samples, and find that the best match is achieved for a photometric redshift cut of $z<0.07$. 
However, even for this optimal selection, we find  less clustering in the HSBG sample than the LSBG sample in the intermediate angular range $\theta \sim 0.1^\circ -- 4^\circ$  (\figref{2pt_HSB_MPZ}).
We note that it is likely that the HSBG sample is contaminated by distant galaxies due to the large photometric redshift uncertainty of DES, which is $\sigma_{68}(z) \sim 0.1$ overall and is known to have a large outlier fraction at low redshift \citep[e.g.,][]{Hoyle:2018}.

We perform a similar analysis for the 2MPZ catalog \citep{Bilicki:2014}, an optical-IR all-sky photometric redshift catalog based on SuperCOSMOS, 2MASS, and WISE extending to $z \sim 0.3$ (peaking at $z \sim 0.07$).
We select this catalog due to its uniform sky coverage and accurate photometric redshifts ($\sigma_z = 0.015$). 
We note that 2MPZ has a very different selection function than DES, as it requires detection in the IR bands. 
By matching 2MPZ galaxies with galaxies in  the DES Y3 Gold catalog, we retrieved information about DES-measured magnitude and surface-brightness distribution of 2MPZ galaxies. 
We find that the DES-measured mean surface brightness for matched 2MPZ galaxies is  significantly brighter ($19.0 < \mumeaneff(g) <  23.0 \magasecsq$) than  the LSBG sample. 
The $g$-band magnitude (\code{MAG\_AUTO}) of the 2MPZ sample lies in the range $14.0 < g < 18.5$, while the LSBG sample range is $18 < g < 22$ (see \appref{magnitude_dists}, \figref{g_band_mags}). 
We thus expect the 2MPZ sample to consist of brighter, higher stellar mass galaxies compared to the LSBG sample.
As before, we identified a redshift cut that resulted in an angular autocorrelation function that is best-matched to that of the LSBGs. 
In the case of 2MPZ galaxies, we find that this is achieved with a redshift cut of $z<0.10$. 

In \figref{2pt_HSB_MPZ} we plot the angular autocorrelation function, $w(\theta)$, of the LSBGs (gray line), the DES HSBGs with $z < 0.07$ (blue line), and the 2MPZ catalog with $z<0.10$ (red line).
We find that both the DES HSBG and 2MPZ samples have lower clustering amplitude than the LSBG sample at intermediate angular scales ($ 0.1^\circ \lesssim \theta \lesssim 4^\circ$).
Overall, we find that the amplitude of the angular correlation function of LSBGs is better matched by the 2MPZ catalog than the DES HSBG catalog.

\begin{figure*}[t]
\centering
\subfigure[]{\includegraphics[width=1.\columnwidth]{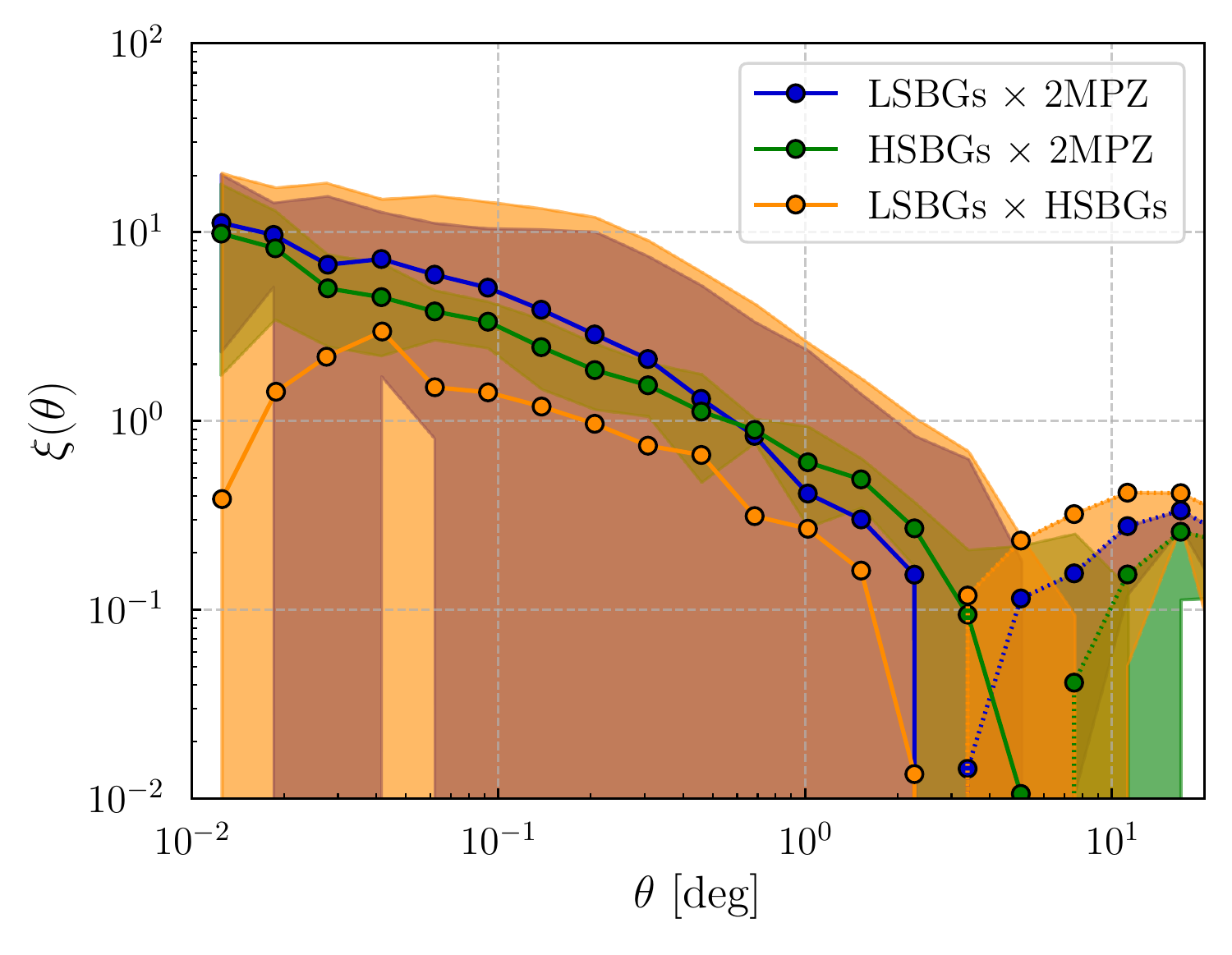}}
\subfigure[]{\includegraphics[width=1.027\columnwidth]{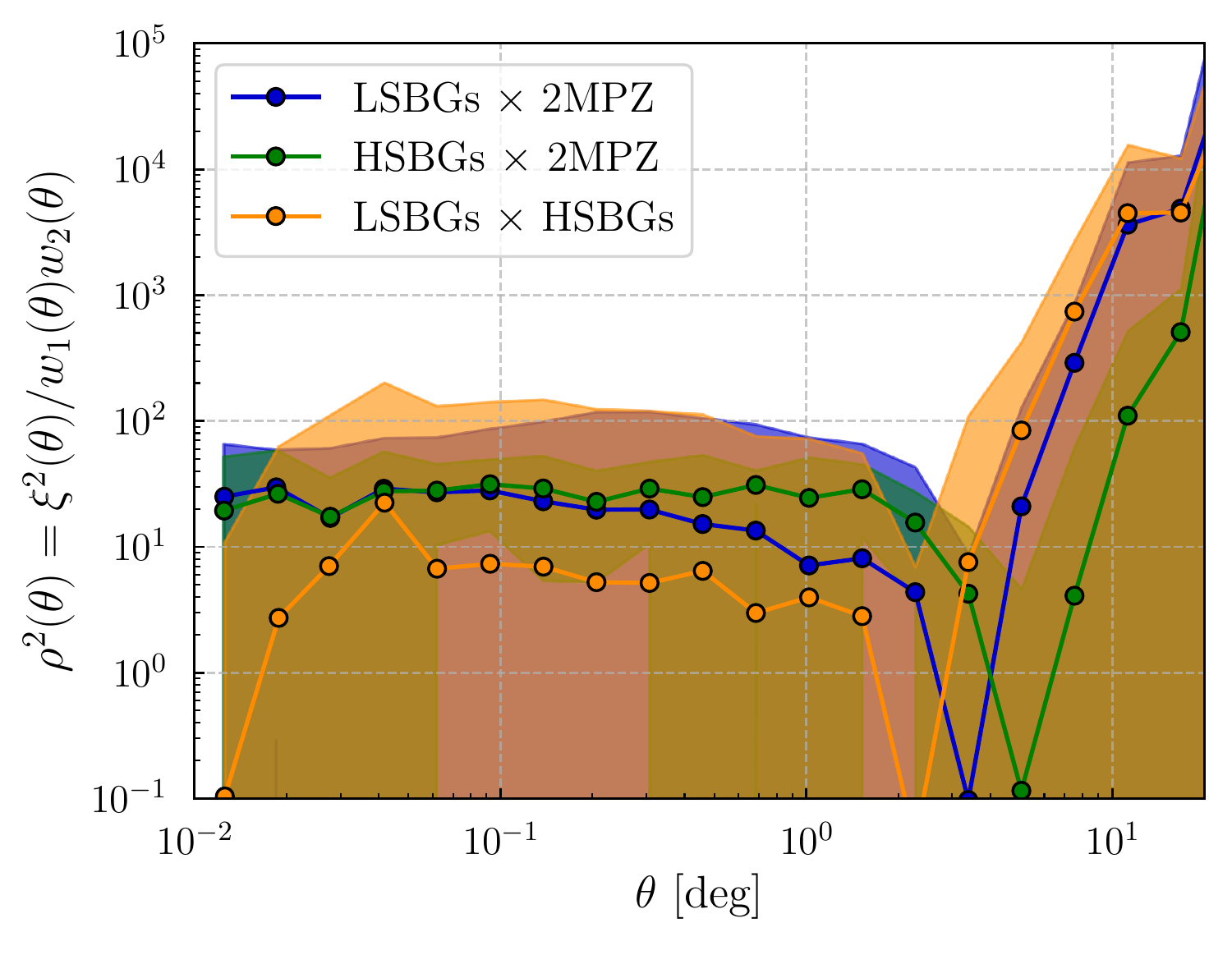}}
\caption{(a) The cross-correlation function, $\xi(\theta)$, between (i) the DES LSBG and HSBG samples (orange line), (ii) the LSBG and 2MPZ samples (blue line), and (iii) the  DES HSBG and 2MPZ samples (green line). 
(b) The square of the cross-correlation coefficient between the same samples as in panel (a),  in order to cancel out the contribution of the different galaxy biases and compare the different cross-correlation levels. In both panels, the shaded regions correspond to the errors in the estimated cross-correlations.}
\label{fig:cross_correlations}
\end{figure*}

\subsection{Cross-correlation between galaxy samples}

The previous autocorrelation analysis compares the clustering properties of the LSBG, HSBG and 2MPZ catalogs individually.
However, it does not indicate whether these galaxy samples probe the underlying matter density field in a similar way, i.e., whether the peaks and troughs in their distributions coincide on a statistical basis. 
Galaxies are known to be biased traces of the underlying matter density field. 
For large angular scales, the two fields are connected by a (linear) galaxy bias factor, $b_g$, defined as $\delta_g(z) \equiv b_g(z) \delta_m(z)$, where $\delta$ refers to the overdensity field and the subscripts $g$ and $m$ refer to galaxies and matter, respectively. 
In general, these are functions of redshift, while the bias factor is different for different galaxy samples. 
The galaxy angular autocorrelation function can be defined as $w(\theta) = \langle \delta_g(\hat{\mathbf{n}})\delta_g(\hat{\mathbf{n}}+\theta)\rangle= b^2_g \langle \delta_m(\hat{\mathbf{n}})\delta_m(\hat{\mathbf{n}}+\theta)\rangle$, where $\hat{\mathbf{n}}$ is the direction in the sky. 

To address whether the galaxy samples studied in the previous section trace the matter density field in a similar way, we calculate the cross-correlation function, $\xi(\theta)$, between the LSBG and HSBG samples, the LSBG and the 2MPZ samples, and the HSBG and 2MPZ samples (left panel of \figref{cross_correlations}).
The cross-correlation between two galaxy samples (labeled 1 and 2) is given by $\xi_{12}(\theta) = \langle \delta_{g,1}(\hat{\mathbf{n}})\delta_{g,2}(\hat{\mathbf{n}}+\theta)\rangle = b_{g,1}b_{g,2}\langle \delta_m(\hat{\mathbf{n}})\delta_m(\hat{\mathbf{n}}+\theta)\rangle$. 
We define the cross-correlation coefficient between the two samples as
\begin{equation}
\rho_{12}(\theta) = \frac{\xi_{12}(\theta)}{\sqrt{w_1(\theta)w_2(\theta)}},
\end{equation}
where $w_{1,2}(\theta)$ are the autocorrelation functions of the individual samples. 
In this case, we can cancel the corresponding bias factors present in the different samples, and we can compare the correlations between the matter fields probed by the two samples. 
We plot the (square of the) cross-correlation coefficient between the same samples as those described above in the right panel of \figref{cross_correlations}.

Although the uncertainties are large, we find that the 2MPZ$\times$LSBG sample exhibits a larger cross-correlation signal than  the LSBG$\times$HSBG. 
This likely reflects the better agreement between the redshift distributions of the LSBG and 2MPZ samples, which is expected due to the superior redshift information provided by the 2MPZ.
The stronger cross-correlation signal motivates our use of the 2MPZ sample when constructing radial profiles of HSBGs associated with the prominent peaks in the LSBG distribution.

\section{Associations with Galaxy Clusters and Groups}
\label{sec:clusters}

In the previous section, we described a statistical study of the clustering of LSBGs, which can also be demonstrated visually when plotting the positions of LSBGs (\figref{red_and_blues}).
In this section, we instead focus on identifying the most prominent spatial overdensities of LSBGs and associating them with known galaxy clusters, galaxy groups, and individual bright galaxies.
Associating peaks in the LSBG distribution to external catalogs provides useful information, such as:
\begin{enumerate}[wide, labelwidth=!, labelindent=0pt, itemsep=0pt]
\item Associating a peak in the LSBG distribution with a galaxy system at a known distance allows us to estimate the distances to the LSBGs (assuming a physical association between the LSBGs and reference object). Distances allow us to estimate the intrinsic properties of the LSBGs, such as physical size and luminosity.
\item Defining a sample of likely LSBG cluster members allows us to compare the properties of the LSBGs in cluster environments to those in the field. Such comparisons can be useful for testing models of LSBG formation and evolution.
For example, we can compare the radial distributions of LSBG and HSBG cluster members to test for observable signatures of environmental effects that may be responsible for the formation of LSBGs.
\item Peaks in the LSBG density that are not associated to known clusters or groups can be potentially interesting, indicating different clustering patterns for LSBGs and HSBGs.
\end{enumerate}

\begin{figure*}[t]
\centering
\includegraphics[width=1.0\textwidth]{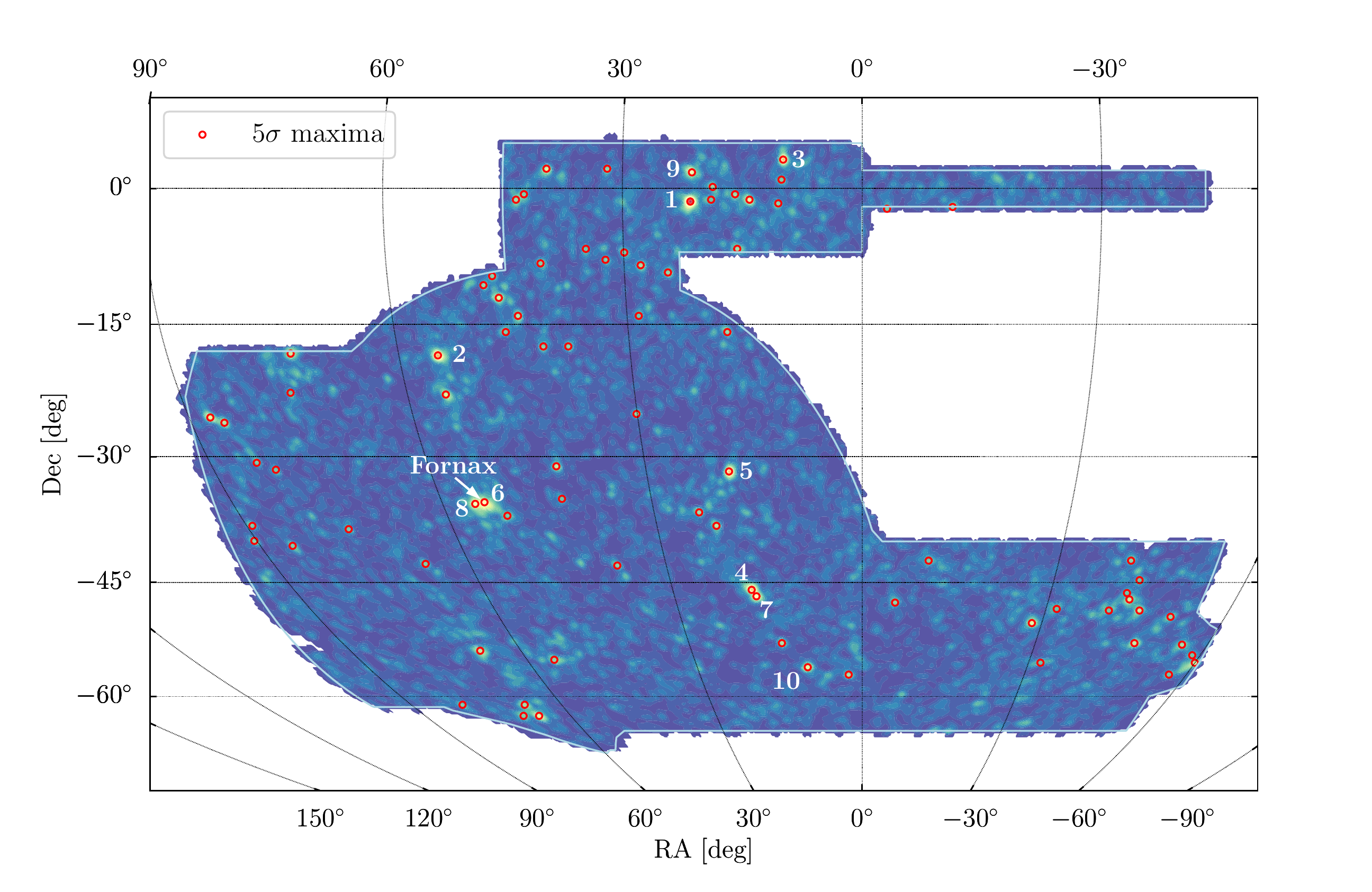}
\vspace{-0.2cm}
\caption{KDE map of the distribution of our LSBG sample. Blue regions denote areas of low density, while regions of high density are indicated in yellow/red. Open red circles indicate the positions of the 82 prominent density peaks identified as described in \secref{clusters}. We have labeled the 10 most prominent peaks, which are summarized in \tabref{Peaks_table}.}
\label{fig:KDE_peaks}
\end{figure*}

\begin{deluxetable*}{ccccccc}
\tablenum{2}
\tablecaption{Characteristics of the 10 most prominent density peaks and their associations}
\tablewidth{10pt}
\tablehead{
\colhead{Peak}&\colhead{(RA,Dec)$_{\rm peak}$}&\colhead{Best}&\colhead{(RA,Dec)$_{\rm assoc}$}&
\colhead{Redshift}&\colhead{Distance}&\colhead{$N(<0.5^\circ)$}\\
\colhead{Number}&\colhead{(deg,deg)}&\colhead{Association}&\colhead{(deg,deg)}&\colhead{$z$}&
\colhead{(Mpc)}& 
}
\startdata
1 & (21.5012, -1.4286) & Abell 194  & (21.4200, -1.4072)& 0.018 & 75.07 $\pm$ 5.26 & 68\\
2 & (54.9388, -18.4712)  & RXC J0340.1-1835  & (55.0475, -18.5875) & 0.0057& 23.41 $\pm$ 1.64 & 48\\
3 & (9.8887, 3.1829) & NGC 199 & (9.8882, 3.1385) & 0.0153 & 62.81 $\pm$ 4.41 &46\\
4 & (17.4972, -45.9398) & Abell 2877 & (17.6017, -45.9228) & 0.0247 & 106.61 $\pm$ 7.45 &41\\ 
5 & (18.4983, -31.7043) & Abell S141  & (18.4758, -31.7519) & 0.020 &84.80 $\pm$ 5.94 & 42\\
6 & (53.9377, -35.3133) & Fornax (Abell S373) & (54.6162, -35.4483)  & 0.0046 & 18.97 $\pm$ 1.33 &32\\
7 & (16.8965, -46.7418) &  Abell 2870 & (16.9299, -46.9165) & 0.0237 & 102.03 $\pm$ 3.89 & 36\\
8 & (55.3393, -35.5138) & Fornax (Abell S373)  & (54.6162, -35.4483) & 0.0046 & 18.97 $\pm$ 1.33 & 28\\ 
7 & (21.3014, 1.7794) & RXC J0125.5+0145  & (21.3746, 1.7627)  &  0.01739 & 72.32 $\pm$ 5.10 & 28\\ 
10 & (9.8888, -55.9649) & Abell 2806 & (10.0270, -56.1167) & 0.0277 & 120.23 $\pm$ 8.42 & 32
\enddata
\tablecomments{
\label{tab:Peaks_table}
Characteristics of the 10 most prominent overdensities in the spatial distribution of LSBGs: (1) peak label, (2) centroid of the density peaks, (3) best association  (see \secref{clusters}), (4) coordinates of best associations, (5)-(6) redshift and the distance to the associations, retrieved from the NASA Extragalactic Database, and (7) number of LSBGs that lie within $0.5^\circ$ from the center of each peak.}
\end{deluxetable*}

\begin{figure*}[t]
\centering
\includegraphics[width=0.95\textwidth]{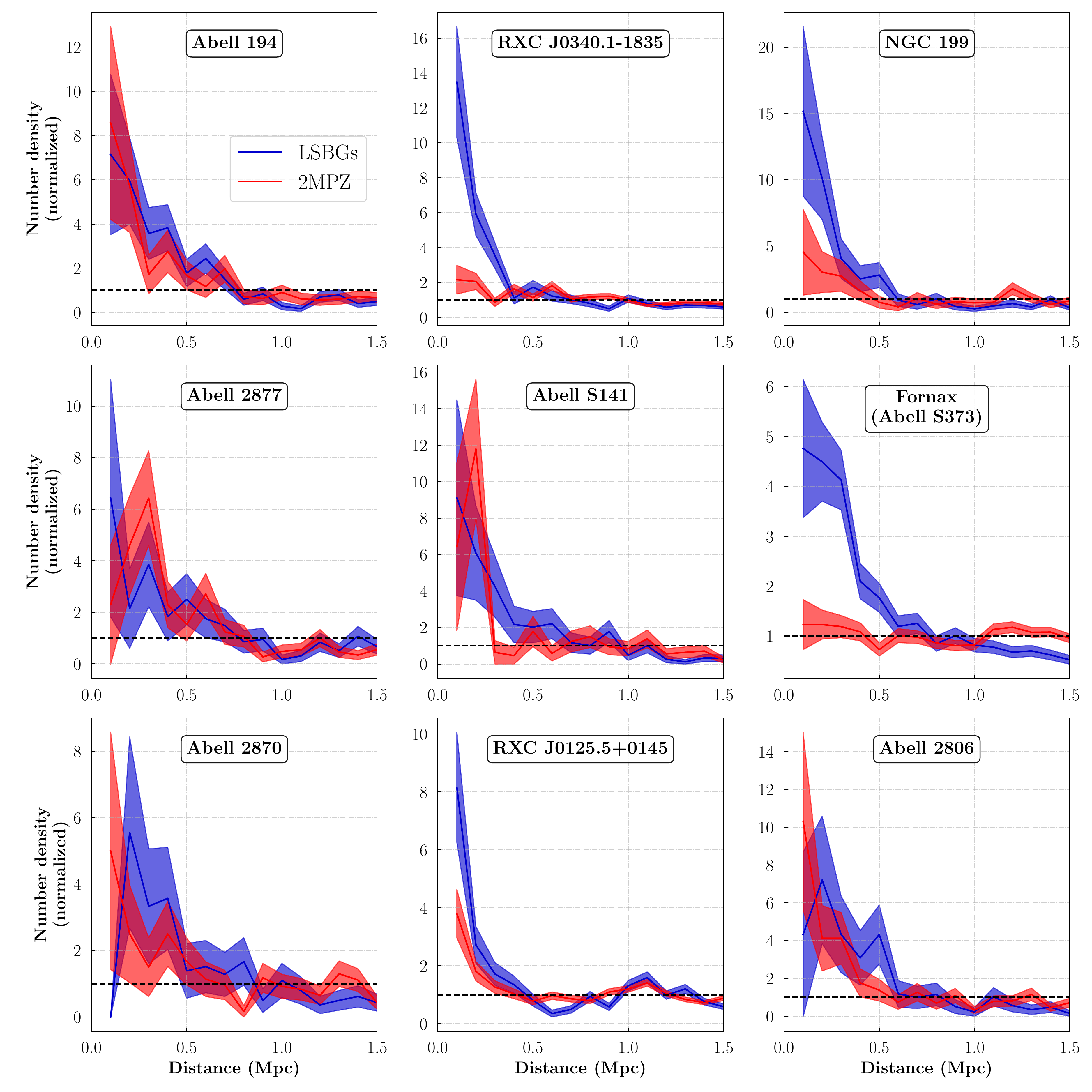}\vspace{+0.1cm}
\caption{Normalized radial profiles of the distribution of LSB galaxies (blue) and galaxies from the 2MPZ catalog (red) around the associations of the most prominent LSBG overdensity peaks, presented in \tabref{Peaks_table}. We have assumed that all galaxies that are within a radius that corresponds to a physical scale of $1.5$ Mpc at the distance of the association belong to that association. The normalization constant corresponds to the mean number density of galaxies within the 1.5 Mpc radius.}
\label{fig:Rad_profiles}
\end{figure*}

We use kernel density estimation (KDE) to estimate the projected density of our full LSBGs sample.
We apply a Gaussian smoothing kernel with a bandwidth of $0.3 \degree$, using the haversine distance metric to account for the cosine dependence on declination \citep{Pedregosa:2011}. 
The kernel bandwidth was selected to be similar to the characteristic angular scale of the overdensities present in \figref{red_and_blues}. 
This kernel size is further motivated by the radial profiles of LSBGs around peaks (see \figref{Rad_profiles}), where it is seen that the typical scale of cluster cores is of the order of $\roughly 0.5\Mpc$. 
The median distance of clusters associated to our sample is $\roughly 80 \Mpc$, which results into a typical angular size of $\roughly 0.35 \degree$. 
For more distant clusters, that typical angular size is smaller ($\roughly 0.28 \degree$ at a distance of 100\Mpc), while for the closest clusters,  the typical angular size is significantly larger (e.g., for Fornax at a distance of $\roughly 19\Mpc$ this scale is $1.5 \degree$).
In fact, a bandwidth of $0.3\degree$ resolves the Fornax cluster into two peaks.

The resulting KDE map is presented in \figref{KDE_peaks}, with  blue regions representing areas of lower density and yellow/red regions representing areas of higher density. 
To detect outliers in this map, we perform an iterative sigma-clipping procedure where at each step, values that exceed the median by $5\sigma$ or more are rejected. 
We find the local maxima in the regions of the KDE map that are above the $5\sigma$ threshold value returned from sigma-clipping. 
We locate 82 peaks passing our criteria, which are indicated with  red open circles in \figref{KDE_peaks}. 
We furthermore number the 10 most prominent of them (as defined by their KDE value) and present their coordinates in \tabref{Peaks_table}. 
In the seventh column of that table, we also present the number of LSBGs within $0.5$ degrees from the center of each peak. 
The complete catalog can be found in a machine-readable form in the supplemental material.

Next, we cross-match our list of high-density LSBG peaks with known overdensities in the low-redshift universe. 
Specifically, we cross-match against:
\begin{enumerate}[wide, labelwidth=!, labelindent=0pt, itemsep=0pt]
\item The Abell catalog of rich clusters \citep[southern survey,][]{Abell:1989}.
\item The ROSAT-ESO Flux Limited X-ray (REFLEX) Galaxy cluster survey \citep{Reflex:2004}.
\item A catalog of galaxy groups built from the sample of the 2MASS Redshift Survey \citep{Tully:2015}. We keep only those groups that have more than five members.
\item Bright galaxies from the revised New General Catalogue \citep{NGC:1999}.
\end{enumerate}

For each peak in the LSBG distribution, we overplotted the distribution of LSBGs and external catalog objects in a region $\pm 0.5^\circ$  from the nominal center of the peak. 
To identify associations (if any), we selected the object from the external catalogs that is closest to the center of the LSBG peak, giving priority to objects according to ordering listed above. 
For example, if an LSBG peak is matched to both an NGC galaxy and an Abell cluster, we select the Abell cluster as the association.
From the 82 peaks, we find that 32 are associated with an Abell cluster, 11 with a REFLEX cluster, 10 with a 2MASS group, 16 with an NGC galaxy, while 13 peaks have no association assigned by our criteria. 
We used the DES Sky Viewer tool to visually inspect the regions around the 13 LSBG peaks that were not associated with objects in our external catalogs. 
In seven cases, we identified nearby bright galaxies/galaxy clusters that were not included in the external catalogs we used for the matching. 
Interestingly, in six cases we did not find an obvious nearby galaxy cluster, galaxy group, or bright nearby galaxy. As an interesting case, we mention a peak at (RA,DEC)$\sim (-50.978^\circ,-49.348^\circ)$ with 18 LSBGs in a $0.5^\circ$ area around it. We leave the more detailed study of these systems for future work.

In \tabref{Peaks_table} we present the coordinates of the ten most prominent LSBG overdensities and their best associations, along with the coordinates, redshifts, and distances of these associations (retrieved from the NASA Extragalactic Database)\footnote{\url{https://ned.ipac.caltech.edu/}}. 
We also report the number of LSBGs within $0.5^\circ$ from the center of each peak. 
Note that two peaks are both associated with the Fornax cluster (Abell S373).
The full table of associations can be found in the supplemental material, where we provide an additional column characterizing the quality of association: I (very good), II (good), to III (not so good). The quality of the association was determined based on the projected, angular distance of the association from the peak and the presence (or absence) of other potential associations in the vicinity of the peak. Our classification is qualitative, though, and is just a guide for follow-up research. For the cases where we did not find an association using any of the catalogs mentioned above, we visually inspected the region around the peak using the DES Sky Viewer. If there was not any visible high-surface-brightness counterpart around, we indicated quality = I, otherwise (visible clusters of bright galaxies) we indicated quality = III.

By assuming a physical association between these LSBG overdensities and the matched external systems, we can use the known distances of the external systems to estimate the distance to the associated LSBGs.
This information is otherwise absent due to our inability to accurately estimate the photometric redshift for these galaxies from the DES data alone.
In the remainder of this section, we will use distance information from the nine most prominent associations to (i) study the radial distribution of LSBGs around clusters and (ii) derive the size--luminosity relation for associated LSBGs.

\subsection{Radial Profiles} \label{sec:Rad_prof}

Comparing the distribution of LSBGs and HSBGs in dense environments may help illuminate the processes governing the formation and evolution of LSBGs. 
In \figref{Rad_profiles} we plot the number density of LSBGs and 2MPZ galaxies with redshift $z<0.10$ around the nine most prominent associated systems (clusters and NGC galaxies; \tabref{Peaks_table}). 
For each of these nine associations, we select all LSBGs and 2MPZ galaxies that reside within an angle corresponding to $1.5 \Mpc$ at the distance of each associated object.
We calculate the radial profiles of LSBGs and 2MPZ galaxies in fifteen annuli of width $0.1\Mpc$.
In order to compare the LSBGs and 2MPZ galaxies on the same scale, we normalize the number densities to the mean number density of galaxies in each sample within the 1.5\Mpc region---i.e., a flat line with unit amplitude indicates a homogeneous distribution of galaxies within the 1.5\Mpc region. 
We estimate the uncertainty on our radial profile by combining the Poisson uncertainties on the measured number of galaxies per annulus and the total number of galaxies in the 1\Mpc region.

In all cases, we find that the LSBG distribution is peaked within $0.5\Mpc$ and flattens  at distances $\mbox{$\gtrsim 1 \Mpc$}$. 
We find that the normalized number density of LSBGs peaks at similar amplitudes for most systems, with the most peaked overdensity found around the lenticular galaxy NGC 199.
This may be expected given that this association represents the dwarf satellite population of a single central bright galaxy. 
We find three cases where the normalized radial distributions of the LSBG and 2MPZ samples appear quite different. 
RXC J0340.1$-$1835 and Fornax are at significantly lower redshift than the other systems, $z = 0.0057$ and $z = 0.0046$, respectively (the next closest associated system is NGC~1200 at $z = 0.013$.)
The 2MPZ catalog includes just a few objects with such low redshifts; there are only 24 objects with $z<0.005$ and 42 objects with $z<0.006$. 
Thus, in these two cases it is likely that the 2MPZ sample consists of background galaxies. 
The third case where the distribution of 2MPZ and LSBG galaxies differ is around NGC~199. 
Again, the LSBGs are much more peaked than the 2MPZ sample, suggesting that the observed LSBG overdensity is caused by dwarf galaxies surrounding a single central host.
Despite the small sample size, we can say qualitatively that the radial distribution of LSBGs and 2MPZ galaxies appear to largely agree.
We use the Kolmogorov--Smirnov test to quantitatively evaluate the similarity of the radial distributions of LSBGs and 2MPZ galaxies surrounding these systems.
We calculate the $p$-values for the null hypothesis that the two galaxy samples are drawn from the same underlying distribution. 
We find that for RXC J0340.1$-$1835 and Fornax, $p \ll 0.01$ (thus strongly rejecting the null hypothesis), $p=0.015$ for NGC 199 (making the null hypothesis unlikely), while for all the other systems $p>0.1$.


\subsection{Size--Luminosity Relation} \label{sec: Size_Lum}

\begin{figure*}[t]
\centering
\includegraphics[width=0.9\textwidth]{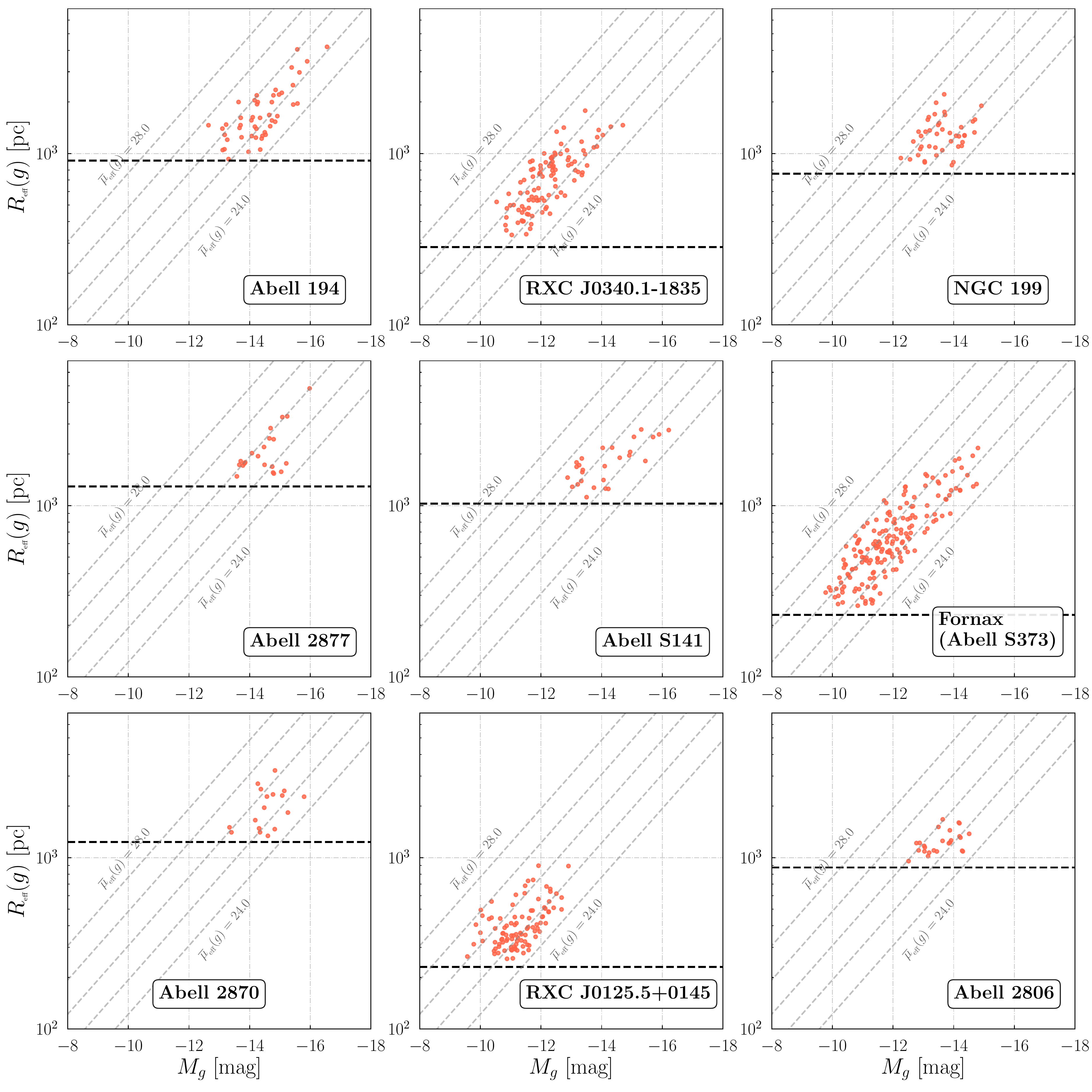}\vspace{-0.2cm}
\caption{Size--luminosity relation for LSBGs around the associations of the most prominent overdensity peaks, presented in \tabref{Peaks_table}. We have assumed that all LSBGs within an angle corresponding to a physical radius of 0.5 Mpc at the distance of the association belong to it. With the dashed horizontal lines, we show the physical scale corresponding to the radius cut $r_{1/2}(g) >2.5''$ at the distance of the cluster. We also show (dashed, diagonal gray lines) the lines of constant mean surface-brightness.}
\label{fig:Size_Lumin}
\end{figure*}

Distance information from our external catalog systems allows us to calculate the physical properties of associated LSBGs. 
For the nine most prominent peaks in the LSGB distribution, we assume that all LSBGs that reside within a projected distance of \CHECK{$0.5\Mpc$} are associated to these systems and reside at the same distance. 
Using this distance, we can estimate the physical effective radii (in pc) and absolute magnitudes of these LSBGs. 

\begin{figure}[ht!]
\centering
\epsscale{1.1}
\plotone{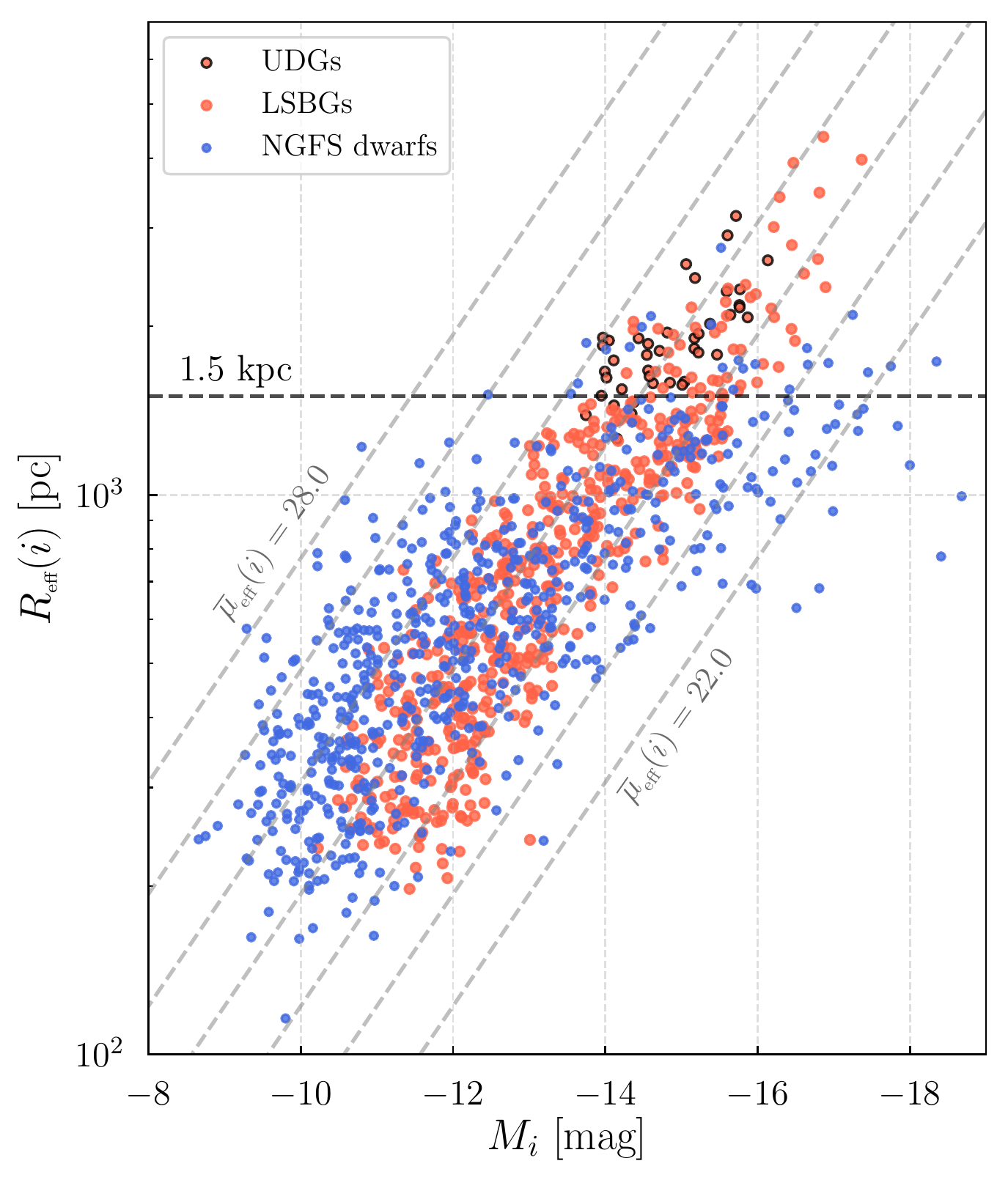}
\vspace{0.1cm}
\caption{Size--luminosity relation of LSBGs around the nine most prominent overdensities (red points) in the $i$ band. The sample consists of 555 galaxies. For comparison, we overplot the dwarf galaxies found around Fornax in the NGFS survey \citep{Eigenthaler:2018,Ordenes:2018}. 41 galaxies in our sample have effective radii exceeding 1.5 kpc in the $g$ band (black circles) and central surface brightness $\mu_0(g) > 24.0 \magasecsq$, which is a conventional definition for ultra-diffuse galaxies \citep[UDG;][]{vanDokkum:2015a}.}
\label{fig:Size_Lumin_all}
\end{figure}

In \figref{Size_Lumin}, we present the size--luminosity relationship for the LSBGs around these nine peaks, based on the physical effective radius, $\Reff(g)$, and the absolute magnitude in the $g$ band, $M_g$. 
We see that the number of LSBGs associated with each system varies significantly; the smallest number of LSBGs (17) is associated with Abell 2870, while the largest number of LSBGs (175) are associated to Fornax. 
In \figref{Size_Lumin} we also indicate the physical scale corresponding to the angular selection criterion, $\Reff(g) > 2.5\arcsec$, at the distance of the associated system (dashed black line). 
Since Fornax is the closest cluster, this angular selection criterion corresponds to the smallest physical size ($\roughly 230 \pc$), resulting in more faint galaxies passing the selection. 
Similarly, RXC J0340.1$-$1835 is also a nearby cluster and has a large number of LSBGs (102). We also show lines of constant mean surface brightness.  
The bright-end limit is largely set by the requirement $\mumeaneff(g) > 24.2 \magasecsq$ used to produce our catalog.   Only two associated galaxies have surface brightness $\mumeaneff(g) > 27.0 \magasecsq$.

In \figref{Size_Lumin_all}, we combine the observations of LSBGs from the nine clusters in a single size--luminosity plot. We compare the distribution of our sample to that of the dwarf galaxies discovered in the NGFS survey, described in  \secref{efficiency}. Since the NGFS only provides magnitudes and effective radii in the $i$ band \citep{Eigenthaler:2018,Ordenes:2018}, we choose to plot against the $i$-band quantities of our sample. 
We see that the two samples occupy a similar region in the size--luminosity parameter space, with the NGFS sample spanning a larger range of absolute magnitudes.
The NGFS extends to fainter absolute magnitudes due to their deeper imaging data, while the lack of an explicit surface-brightness cut extends their sample to brighter magnitudes.

Recently, much attention has been paid to the class of ultra-diffuse galaxies (UDGs), which have been conventionally defined as galaxies with  central surface brightness  $\mu_0(g) > 24.0$ and effective radius $\Reff(g) > 1.5\kpc$ \citep[e.g.,][]{vanDokkum:2015a}.
The LSBGs in our associated sample span a wide range of physical sizes, from \CHECK{$0.26 \kpc \lesssim \Reff(g) \lesssim 4.83 \kpc$}, with a median of \CHECK{$\Reff(g) = 0.8 \kpc$} (the $i$-band values presented in \figref{Size_Lumin_all} are $0.20 \kpc \lesssim \Reff(i) \lesssim 4.36 \kpc$ with a median of $\Reff(i) = 0.75 \kpc$). 
The lower limit is largely set by our angular size selection criterion, translated to a physical size for the nearest cluster (Fornax). 
We find 41 galaxies have size $\Reff(g) > 1.5\kpc$ and surface brightness $\mu_0(g) > 24.0 \magasecsq$, thus satisfying the conventional UDG definition.
We note again that our angular size selection requires distant galaxies to have larger physical sizes.
  
The sample covers a wide range of absolute $g$-band magnitude, $-9.8 \gtrsim M_g \gtrsim -16.5$, with a median of $M_g \sim -12.4$. We see that the galaxies in the sample discussed here span the same range in mean surface brightness ($24.2 \lesssim \mumeaneff(g) \lesssim 27.0 \magasecsq$), regardless of their sizes: both small and large galaxies populate the range of surface brightnesses. 
Thus, UDGs seem to be a natural continuation of the LSBG population in the regime of large size and low surface brightness, and not a distinct population that is well separated in the size--luminosity space from other LSBGs \citep[a similar conclusion was drawn by][]{Conselice2018}.

\section{Summary and Conclusions} \label{sec:summary}

In this paper, we have selected and analyzed \NLSBG extended,  LSBGs from the first three years of DES imaging data. 
Our sample selection pipeline consists of the following steps:

\begin{enumerate}[wide, labelwidth=!, labelindent=0pt, itemsep=0pt]
\item We selected objects from the DES Y3 Gold catalog based on \SExtractor parameters. The most important selections were based on the half-light radius, $r_{1/2} > 2.5''$ and mean surface brightness, 
$\mumeaneff(g) >24.2 \magasecsq$. The selection criteria are summarized in Appendix \ref{app:select}.
\item We applied an SVM classifier tuned to reduce the incidents of false negatives (LSBGs classified as non-LSBGs). This reduced the number of false-positive candidates by an order of magnitude.
\item A visual inspection that eliminated the remaining false positives to produce a high-purity sample of LSBGs.
\item We fit each galaxy with a single-component S\'ersic profile, and we made a final selection based on the derived size and surface brightness. 
\end{enumerate}
We divided the total LSBG sample into two subsamples according to their $g-i$ color. 
We study the photometric, structural and spatial clustering properties of the red ($g-i \geq 0.60$) and blue ($g-i < 0.60$) subsamples. 
Our main findings are the following:

\begin{enumerate}[wide, labelwidth=!, labelindent=0pt, itemsep=0pt]
\item The distributions in angular size (effective radius) are similar for the two subsamples with the red population having slightly higher median value ($\roughly 3.90 \arcsec$) compared to the blue population ($\roughly 3.76 \arcsec$).
\item Both samples have a similar median S\'ersic index of $n \sim 1.0$.
\item The mean surface-brightness distributions differ noticeably between the two populations: blue galaxies tend to be brighter. We note this behavior is not as prominent as previously observed by \citet{Greco:2018}. The distribution in the central surface brightness, $\mu_0(g)$, does not present as large a difference between the two subsamples.
\item The spatial distribution of red LSBGs is much more clustered than that of blue LSBGs, which have an almost homogeneous distribution. This is quantified in the two-point angular correlation function, which is an order of magnitude higher for the red subsample than the blue subsample. 
\end{enumerate}

Furthermore, we compared the clustering of the full LSBG sample with a sample of HSBGs selected from the DES and with an external catalog of low-redshift galaxies from the 2MPZ.
We find a similar autocorrelation amplitude (and also a high cross-correlation signal) between the LSBG sample and the 2MPZ catalog with a redshift cut of $z < 0.1$ (which is indicative of the low redshift of our LSBG sample). An interesting feature is the lower amplitude of clustering for LSBGs at angular scales less than $\sim 0.1$ deg.

The spatial distribution of LSBGs contains prominent overdensities.  
We cross-match the 82 most prominent overdensities with external catalogs of galaxy clusters, galaxy groups, and individual bright galaxies. 
The association of peaks with objects (clusters, groups, and galaxies) of known distance provides us with distance information for a subset of LSBGs. 
The distances of associated systems range from $\sim 19\Mpc$ ( Fornax cluster) to $\sim 354\Mpc$ (Abell 2911), with a median distance of $82\Mpc$. 
The mean distance is $106\Mpc$ with a standard deviation of $\sim 66\Mpc$.

By associating LSBGs with other systems at known distances, we are able to further explore the physical properties of some LSBGs and their host systems. 
In particular, we present:
\begin{enumerate}[wide, labelwidth=!, labelindent=0pt, itemsep=0pt]
\item Projected radial profiles of the distribution of the LSBG and 2MPZ galaxies around the nine most prominent associations.
We find that in galaxy clusters, the radial distributions of these two galaxy samples are similar.
\item A physical size--absolute magnitude relationship for LSBGs belonging to the nine most prominent associations.  We find that LSBGs in our sample, span a range in physical size (effective radius) from $\roughly 0.26 \kpc$ up to $\roughly 4.83 \kpc$, with a median size of $0.8\kpc$. Out of the 555 LSBGs studied, 41 can be classified as UDGs--i.e., have effective radii $\Reff(g) > 1.5\kpc$ and central surface brightness $\mu_0(g) > 24.0 \magasecsq$. UDGs appear to be a continuation of the LSBG population.
\end{enumerate}

Our catalog is the largest catalog of LSBGs ($\Reff(g) > 2.5\arcsec$ and $\mumeaneff(g) > 24.2  \magasecsq$) assembled to date.
We have presented a general statistical analysis of our catalog, with the hope of enabling more detailed analyses of individual systems and the ensemble population.
Future quantitative comparisons can test galaxy formation models in the low-surface-brightness regime, including studies of properties of LSBGs in different environments (clusters/field) and constraints on the mean mass of LSBGs using weak lensing \citep[e.g.,][]{Sifon:2018}. 
Our sample can also be used to better prepare for the next generation galaxy surveys (e.g., with the Vera C.\ Rubin Observatory). Automated selection procedures result in a large false-positives fraction, necessitating the visual inspection of LSBG candidates. 
However, visual inspection will become infeasible for the large data sets collected by future surveys. 
Our LSBG sample can serve as training set for machine and deep learning algorithms, in the hope of fully automating the selection process. 
The potential of such algorithms will be further explored in upcoming projects. Furthermore, we plan to build upon the know-how we developed constructing the catalog presented in this paper to study LSBGs using the upcoming, deeper data from the total six years of DES observations.

\section{Acknowledgments}\label{sec:ack}
\
We thank Erin Kado-Fong for feedback related to the LSBG catalog and an anonymous reviewer for useful suggestions that helped improve this paper.
This work was supported by the University of Chicago and the Department of Energy under section H.44 of Department of Energy contract No.\ DE-AC02-07CH11359 awarded to Fermi Research Alliance, LLC. 
This work was partially funded by Fermilab LDRD 2018-052. 
\NEW{This material is based upon work supported by the National Science Foundation under grant Nos.\ AST-1615838, AST-2006340, and AST-2008110.}

Funding for the DES Projects has been provided by the US Department of Energy, the US National Science Foundation, the Ministry of Science and Education of Spain, 
the Science and Technology Facilities Council of the United Kingdom, the Higher Education Funding Council for England, the National Center for Supercomputing 
Applications at the University of Illinois at Urbana-Champaign, the Kavli Institute of Cosmological Physics at the University of Chicago, 
the Center for Cosmology and Astro-Particle Physics at the Ohio State University,
the Mitchell Institute for Fundamental Physics and Astronomy at Texas A\&M University, Financiadora de Estudos e Projetos, 
Funda{\c c}{\~a}o Carlos Chagas Filho de Amparo {\`a} Pesquisa do Estado do Rio de Janeiro, Conselho Nacional de Desenvolvimento Cient{\'i}fico e Tecnol{\'o}gico and 
the Minist{\'e}rio da Ci{\^e}ncia, Tecnologia e Inova{\c c}{\~a}o, the Deutsche Forschungsgemeinschaft, and the Collaborating Institutions in the Dark Energy Survey. 

The Collaborating Institutions are Argonne National Laboratory, the University of California at Santa Cruz, the University of Cambridge, Centro de Investigaciones Energ{\'e}ticas, 
Medioambientales y Tecnol{\'o}gicas-Madrid, the University of Chicago, University College London, the DES-Brazil Consortium, the University of Edinburgh, 
the Eidgen{\"o}ssische Technische Hochschule (ETH) Z{\"u}rich, 
Fermi National Accelerator Laboratory, the University of Illinois at Urbana-Champaign, the Institut de Ci{\`e}ncies de l'Espai (IEEC/CSIC), 
the Institut de F{\'i}sica d'Altes Energies, Lawrence Berkeley National Laboratory, the Ludwig-Maximilians Universit{\"a}t M{\"u}nchen and the associated Excellence Cluster Universe, 
the University of Michigan, NFS's NOIRLab, the University of Nottingham, The Ohio State University, the University of Pennsylvania, the University of Portsmouth, 
SLAC National Accelerator Laboratory, Stanford University, the University of Sussex, Texas A\&M University, and the OzDES Membership Consortium.

Based in part on observations at Cerro Tololo Inter-American Observatory at NSF's NOIRLab (NOIRLab Prop. ID 2012B-0001; PI: J. Frieman), which is managed by the Association of Universities for Research in Astronomy (AURA) under a cooperative agreement with the National Science Foundation.

The DES data management system is supported by the National Science Foundation under grant Nos. AST-1138766 and AST-1536171.
The DES participants from Spanish institutions are partially supported by MICINN under grants ESP2017-89838, PGC2018-094773, PGC2018-102021, SEV-2016-0588, SEV-2016-0597, and MDM-2015-0509, some of which include ERDF funds from the European Union. IFAE is partially funded by the CERCA program of the Generalitat de Catalunya.
Research leading to these results has received funding from the European Research
Council under the European Union's Seventh Framework Program (FP7/2007-2013) including ERC grant agreements 240672, 291329, and 306478.
We  acknowledge support from the Brazilian Instituto Nacional de Ci\^encia
e Tecnologia (INCT) do e-Universo (CNPq grant 465376/2014-2).

This manuscript has been authored by Fermi Research Alliance, LLC, under contract No. DE-AC02-07CH11359 with the US Department of Energy, Office of Science, Office of High Energy Physics.

\vspace{5mm}
\facilities{Blanco, DECam}

\software{\code{astropy} \citepalias{2013A&A...558A..33A},  
          \code{Cloudy} \citep{2013RMxAA..49..137F}, 
          \code{healpix} \citep{Gorski:2005},
          \code{matplotlib} \citep{Hunter:2007},
          \code{numpy} \citep{numpy:2011},
          \code{SExtractor} \citep{1996A&AS..117..393B},
          \code{scipy} \citep{scipy:2001},
          \code{scikit-learn} \citep{Pedregosa:2011}.
          }

\bibliographystyle{apj}
\bibliography{main}

\begin{thebibliography}{}
\expandafter\ifx\csname natexlab\endcsname\relax\def\natexlab#1{#1}\fi

\bibitem[{{Abell} {et~al.}(1989){Abell}, {Corwin}, \& {Olowin}}]{Abell:1989}
{Abell}, G.~O., {Corwin}, Jr., H.~G., \& {Olowin}, R.~P. 1989, \apjs, 70, 1

\bibitem[{{Abraham} \& {van Dokkum}(2014)}]{Abraham:2014}
{Abraham}, R.~G., \& {van Dokkum}, P.~G. 2014, \pasp, 126, 55

\bibitem[{{Adami} {et~al.}(2006){Adami}, {Scheidegger}, {Ulmer}, {Durret},
  {Mazure}, {West}, {Conselice}, {Gregg}, {Kasun}, {Pell{\'o}}, \&
  {Picat}}]{Adami:2006}
{Adami}, C., {Scheidegger}, R., {Ulmer}, M., {et~al.} 2006, \aap, 459, 679

\bibitem[{{Amorisco} \& {Loeb}(2016)}]{Amorisco:2016}
{Amorisco}, N.~C., \& {Loeb}, A. 2016, \mnras, 459, L51

\bibitem[{{Astropy Collaboration} {et~al.}(2013){Astropy Collaboration},
  {Robitaille}, {Tollerud}, {Greenfield}, {Droettboom}, {Bray}, {Aldcroft},
  {Davis}, {Ginsburg}, {Price-Whelan}, {Kerzendorf}, {Conley}, {Crighton},
  {Barbary}, {Muna}, {Ferguson}, {Grollier}, {Parikh}, {Nair}, {Unther},
  {Deil}, {Woillez}, {Conseil}, {Kramer}, {Turner}, {Singer}, {Fox}, {Weaver},
  {Zabalza}, {Edwards}, {Azalee Bostroem}, {Burke}, {Casey}, {Crawford},
  {Dencheva}, {Ely}, {Jenness}, {Labrie}, {Lim}, {Pierfederici}, {Pontzen},
  {Ptak}, {Refsdal}, {Servillat}, \& {Streicher}}]{2013A&A...558A..33A}
{Astropy Collaboration}, {Robitaille}, T.~P., {Tollerud}, E.~J., {et~al.} 2013,
  \aap, 558, A33

\bibitem[{{Baldry} {et~al.}(2004){Baldry}, {Glazebrook}, {Brinkmann},
  {Ivezi{\'c}}, {Lupton}, {Nichol}, \& {Szalay}}]{Baldry:2004}
{Baldry}, I.~K., {Glazebrook}, K., {Brinkmann}, J., {et~al.} 2004, \apj, 600,
  681

\bibitem[{{Bamford} {et~al.}(2009){Bamford}, {Nichol}, {Baldry}, {Land},
  {Lintott}, {Schawinski}, {Slosar}, {Szalay}, {Thomas}, {Torki}, {Andreescu},
  {Edmondson}, {Miller}, {Murray}, {Raddick}, \& {Vandenberg}}]{Bamford:2009}
{Bamford}, S.~P., {Nichol}, R.~C., {Baldry}, I.~K., {et~al.} 2009, \mnras, 393,
  1324

\bibitem[{{Barden} {et~al.}(2012){Barden}, {H{\"a}u{\ss}ler}, {Peng},
  {McIntosh}, \& {Guo}}]{Barden:2012}
{Barden}, M., {H{\"a}u{\ss}ler}, B., {Peng}, C.~Y., {McIntosh}, D.~H., \&
  {Guo}, Y. 2012, \mnras, 422, 449

\bibitem[{{Behroozi} {et~al.}(2013){Behroozi}, {Wechsler}, \&
  {Conroy}}]{Behroozi:2013}
{Behroozi}, P.~S., {Wechsler}, R.~H., \& {Conroy}, C. 2013, \apj, 770, 57

\bibitem[{{Berlind} {et~al.}(2005){Berlind}, {Blanton}, {Hogg}, {Weinberg},
  {Dav{\'e}}, {Eisenstein}, \& {Katz}}]{Berlind:2005}
{Berlind}, A.~A., {Blanton}, M.~R., {Hogg}, D.~W., {et~al.} 2005, \apj, 629,
  625

\bibitem[{{Bernstein} {et~al.}(1995){Bernstein}, {Nichol}, {Tyson}, {Ulmer}, \&
  {Wittman}}]{Bernstein:1995}
{Bernstein}, G.~M., {Nichol}, R.~C., {Tyson}, J.~A., {Ulmer}, M.~P., \&
  {Wittman}, D. 1995, \aj, 110, 1507

\bibitem[{{Bernstein} {et~al.}(2018){Bernstein}, {Abbott}, {Armstrong},
  {Burke}, {Diehl}, {Gruendl}, {Johnson}, {Li}, {Rykoff}, {Walker}, {Wester},
  \& {Yanny}}]{Bernstein:2018}
{Bernstein}, G.~M., {Abbott}, T.~M.~C., {Armstrong}, R., {et~al.} 2018, \pasp,
  130, 054501

\bibitem[{{Bertin}(2006)}]{Bertin:2006}
{Bertin}, E. 2006, in Astronomical Society of the Pacific Conference Series,
  Vol. 351, Astronomical Data Analysis Software and Systems XV, ed.
  C.~{Gabriel}, C.~{Arviset}, D.~{Ponz}, \& S.~{Enrique}, 112

\bibitem[{{Bertin} \& {Arnouts}(1996)}]{1996A&AS..117..393B}
{Bertin}, E., \& {Arnouts}, S. 1996, \aaps, 117, 393

\bibitem[{{Bilicki} {et~al.}(2014){Bilicki}, {Jarrett}, {Peacock}, {Cluver}, \&
  {Steward}}]{Bilicki:2014}
{Bilicki}, M., {Jarrett}, T.~H., {Peacock}, J.~A., {Cluver}, M.~E., \&
  {Steward}, L. 2014, \apjs, 210, 9

\bibitem[{{Blanton} \& {Moustakas}(2009)}]{Blanton:2009}
{Blanton}, M.~R., \& {Moustakas}, J. 2009, \araa, 47, 159

\bibitem[{{B{\"o}hringer} {et~al.}(2004){B{\"o}hringer}, {Schuecker}, {Guzzo},
  {Collins}, {Voges}, {Cruddace}, {Ortiz-Gil}, {Chincarini}, {De Grandi},
  {Edge}, {MacGillivray}, {Neumann}, {Schindler}, \& {Shaver}}]{Reflex:2004}
{B{\"o}hringer}, H., {Schuecker}, P., {Guzzo}, L., {et~al.} 2004, \aap, 425,
  367

\bibitem[{{Bothun} {et~al.}(1997){Bothun}, {Impey}, \& {McGaugh}}]{Bothun:1997}
{Bothun}, G., {Impey}, C., \& {McGaugh}, S. 1997, \pasp, 109, 745

\bibitem[{{Carleton} {et~al.}(2019){Carleton}, {Errani}, {Cooper},
  {Kaplinghat}, {Pe{\~n}arrubia}, \& {Guo}}]{Carleton:2019}
{Carleton}, T., {Errani}, R., {Cooper}, M., {et~al.} 2019, \mnras, 485, 382

\bibitem[{{Cohen} {et~al.}(2018){Cohen}, {van Dokkum}, {Danieli}, {Romanowsky},
  {Abraham}, {Merritt}, {Zhang}, {Mowla}, {Kruijssen}, {Conroy}, \&
  {Wasserman}}]{Cohen:2018}
{Cohen}, Y., {van Dokkum}, P., {Danieli}, S., {et~al.} 2018, \apj, 868, 96

\bibitem[{{Connolly} {et~al.}(2002){Connolly}, {Scranton}, {Johnston},
  {Dodelson}, {Eisenstein}, {Frieman}, {Gunn}, {Hui}, {Jain}, {Kent},
  {Loveday}, {Nichol}, {O'Connell}, {Postman}, {Scoccimarro}, {Sheth},
  {Stebbins}, {Strauss}, {Szalay}, {Szapudi}, {Tegmark}, {Vogeley}, {Zehavi},
  {Annis}, {Bahcall}, {Brinkmann}, {Csabai}, {Doi}, {Fukugita}, {Hennessy},
  {Hindsley}, {Ichikawa}, {Ivezi{\'c}}, {Kim}, {Knapp}, {Kunszt}, {Lamb},
  {Lee}, {Lupton}, {McKay}, {Munn}, {Peoples}, {Pier}, {Rockosi}, {Schlegel},
  {Stoughton}, {Tucker}, {Yanny}, \& {York}}]{Connolly:2002}
{Connolly}, A.~J., {Scranton}, R., {Johnston}, D., {et~al.} 2002, \apj, 579, 42

\bibitem[{{Conroy}(2013)}]{Conroy:2013}
{Conroy}, C. 2013, \araa, 51, 393

\bibitem[{{Conselice}(2018)}]{Conselice2018}
{Conselice}, C.~J. 2018, Research Notes of the American Astronomical Society,
  2, 43

\bibitem[{{Cresswell} \& {Percival}(2009)}]{Cresswell:2009}
{Cresswell}, J.~G., \& {Percival}, W.~J. 2009, \mnras, 392, 682

\bibitem[{{Dalcanton} {et~al.}(1997){Dalcanton}, {Spergel}, {Gunn}, {Schmidt},
  \& {Schneider}}]{Dalcanton:1997}
{Dalcanton}, J.~J., {Spergel}, D.~N., {Gunn}, J.~E., {Schmidt}, M., \&
  {Schneider}, D.~P. 1997, \aj, 114, 635

\bibitem[{{Danieli} {et~al.}(2017){Danieli}, {van Dokkum}, {Merritt},
  {Abraham}, {Zhang}, {Karachentsev}, \& {Makarova}}]{Danieli:2017}
{Danieli}, S., {van Dokkum}, P., {Merritt}, A., {et~al.} 2017, \apj, 837, 136

\bibitem[{{Davis} {et~al.}(1985){Davis}, {Efstathiou}, {Frenk}, \&
  {White}}]{Davis:1985}
{Davis}, M., {Efstathiou}, G., {Frenk}, C.~S., \& {White}, S.~D.~M. 1985, \apj,
  292, 371

\bibitem[{{De Vicente} {et~al.}(2016){De Vicente}, {S{\'a}nchez}, \&
  {Sevilla-Noarbe}}]{DeVicente2016}
{De Vicente}, J., {S{\'a}nchez}, E., \& {Sevilla-Noarbe}, I. 2016, \mnras, 459,
  3078

\bibitem[{{DES Collaboration} {et~al.}(2018){DES Collaboration}, {Abbott},
  {Abdalla}, {Allam}, {Amara}, {Annis}, {Asorey}, {Avila}, {Ballester},
  {Banerji}, {Barkhouse}, {Baruah}, {Baumer}, {Bechtol}, {Becker},
  {Benoit-L{\'e}vy}, {Bernstein}, {Bertin}, {Blazek}, {Bocquet}, {Brooks},
  {Brout}, {Buckley-Geer}, {Burke}, {Busti}, {Campisano}, {Cardiel-Sas},
  {Carnero Rosell}, {Carrasco Kind}, {Carretero}, {Castander}, {Cawthon},
  {Chang}, {Chen}, {Conselice}, {Costa}, {Crocce}, {Cunha}, {D'Andrea}, {da
  Costa}, {Das}, {Daues}, {Davis}, {Davis}, {De Vicente}, {DePoy}, {DeRose},
  {Desai}, {Diehl}, {Dietrich}, {Dodelson}, {Doel}, {Drlica-Wagner}, {Eifler},
  {Elliott}, {Evrard}, {Farahi}, {Fausti Neto}, {Fernandez}, {Finley},
  {Flaugher}, {Foley}, {Fosalba}, {Friedel}, {Frieman}, {Garc{\'\i}a-Bellido},
  {Gaztanaga}, {Gerdes}, {Giannantonio}, {Gill}, {Glazebrook}, {Goldstein},
  {Gower}, {Gruen}, {Gruendl}, {Gschwend}, {Gupta}, {Gutierrez}, {Hamilton},
  {Hartley}, {Hinton}, {Hislop}, {Hollowood}, {Honscheid}, {Hoyle}, {Huterer},
  {Jain}, {James}, {Jeltema}, {Johnson}, {Johnson}, {Kacprzak}, {Kent},
  {Khullar}, {Klein}, {Kovacs}, {Koziol}, {Krause}, {Kremin}, {Kron}, {Kuehn},
  {Kuhlmann}, {Kuropatkin}, {Lahav}, {Lasker}, {Li}, {Li}, {Liddle}, {Lima},
  {Lin}, {L{\'o}pez-Reyes}, {MacCrann}, {Maia}, {Maloney}, {Manera}, {March},
  {Marriner}, {Marshall}, {Martini}, {McClintock}, {McKay}, {McMahon},
  {Melchior}, {Menanteau}, {Miller}, {Miquel}, {Mohr}, {Morganson}, {Mould},
  {Neilsen}, {Nichol}, {Nogueira}, {Nord}, {Nugent}, {Nunes}, {Ogand o}, {Old},
  {Pace}, {Palmese}, {Paz-Chinch{\'o}n}, {Peiris}, {Percival}, {Petravick},
  {Plazas}, {Poh}, {Pond}, {Porredon}, {Pujol}, {Refregier}, {Reil}, {Ricker},
  {Rollins}, {Romer}, {Roodman}, {Rooney}, {Ross}, {Rykoff}, {Sako}, {Sanchez},
  {Sanchez}, {Santiago}, {Saro}, {Scarpine}, {Scolnic}, {Serrano},
  {Sevilla-Noarbe}, {Sheldon}, {Shipp}, {Silveira}, {Smith}, {Smith}, {Smith},
  {Soares-Santos}, {Sobreira}, {Song}, {Stebbins}, {Suchyta}, {Sullivan},
  {Swanson}, {Tarle}, {Thaler}, {Thomas}, {Thomas}, {Troxel}, {Tucker},
  {Vikram}, {Vivas}, {Walker}, {Wechsler}, {Weller}, {Wester}, {Wolf}, {Wu},
  {Yanny}, {Zenteno}, {Zhang}, {Zuntz}, {Juneau}, {Fitzpatrick}, {Nikutta},
  {Nidever}, {Olsen}, \& {Scott}}]{DES:2018}
{DES Collaboration}, {Abbott}, T.~M.~C., {Abdalla}, F.~B., {et~al.} 2018,
  \apjs, 239, 18

\bibitem[{{Disney}(1976)}]{Disney:1976}
{Disney}, M.~J. 1976, \nat, 263, 573

\bibitem[{{Drinkwater} {et~al.}(2001){Drinkwater}, {Gregg}, \&
  {Colless}}]{Drinkwater:2001}
{Drinkwater}, M.~J., {Gregg}, M.~D., \& {Colless}, M. 2001, \apjl, 548, L139

\bibitem[{{Driver}(1999)}]{Driver:1999}
{Driver}, S.~P. 1999, \apjl, 526, L69

\bibitem[{Efron \& Gong(1983)}]{Efron:1983}
Efron, B., \& Gong, G. 1983, The American Statistician, 37, 36

\bibitem[{{Eigenthaler} {et~al.}(2018){Eigenthaler}, {Puzia}, {Taylor},
  {Ordenes-Brice{\~n}o}, {Mu{\~n}oz}, {Ribbeck}, {Alamo-Mart{\'\i}nez},
  {Zhang}, {{\'A}ngel}, {Capaccioli}, {C{\^o}t{\'e}}, {Ferrarese}, {Galaz},
  {Grebel}, {Hempel}, {Hilker}, {Lan{\c{c}}on}, {Mieske}, {Miller}, {Paolillo},
  {Powalka}, {Richtler}, {Roediger}, {Rong}, {S{\'a}nchez-Janssen}, \&
  {Spengler}}]{Eigenthaler:2018}
{Eigenthaler}, P., {Puzia}, T.~H., {Taylor}, M.~A., {et~al.} 2018, \apj, 855,
  142

\bibitem[{{Everett} {et~al.}(in prep.){Everett}, {Yanny},
  {et~al.}}]{Everett:2020}
{Everett}, S., {Yanny}, B., {et~al.} in prep., in prep

\bibitem[{{Ferguson}(1989)}]{Ferguson:1989}
{Ferguson}, H.~C. 1989, \aj, 98, 367

\bibitem[{{Ferland} {et~al.}(2013){Ferland}, {Porter}, {van Hoof}, {Williams},
  {Abel}, {Lykins}, {Shaw}, {Henney}, \& {Stancil}}]{2013RMxAA..49..137F}
{Ferland}, G.~J., {Porter}, R.~L., {van Hoof}, P.~A.~M., {et~al.} 2013, \rmxaa,
  49, 137

\bibitem[{{Ferrero} {et~al.}(2012){Ferrero}, {Abadi}, {Navarro}, {Sales}, \&
  {Gurovich}}]{Ferrero:2012}
{Ferrero}, I., {Abadi}, M.~G., {Navarro}, J.~F., {Sales}, L.~V., \& {Gurovich},
  S. 2012, \mnras, 425, 2817

\bibitem[{{Fitzpatrick}(1999)}]{Fitzpatrick:1999}
{Fitzpatrick}, E.~L. 1999, \pasp, 111, 63

\bibitem[{{Flaugher} {et~al.}(2015){Flaugher}, {Diehl}, {Honscheid}, {Abbott},
  {Alvarez}, {Angstadt}, {Annis}, {Antonik}, {Ballester}, {Beaufore},
  {Bernstein}, {Bernstein}, {Bigelow}, {Bonati}, {Boprie}, {Brooks},
  {Buckley-Geer}, {Campa}, {Cardiel-Sas}, {Castand er}, {Castilla}, {Cease},
  {Cela-Ruiz}, {Chappa}, {Chi}, {Cooper}, {da Costa}, {Dede}, {Derylo},
  {DePoy}, {de Vicente}, {Doel}, {Drlica-Wagner}, {Eiting}, {Elliott}, {Emes},
  {Estrada}, {Fausti Neto}, {Finley}, {Flores}, {Frieman}, {Gerdes},
  {Gladders}, {Gregory}, {Gutierrez}, {Hao}, {Holland}, {Holm}, {Huffman},
  {Jackson}, {James}, {Jonas}, {Karcher}, {Karliner}, {Kent}, {Kessler},
  {Kozlovsky}, {Kron}, {Kubik}, {Kuehn}, {Kuhlmann}, {Kuk}, {Lahav}, {Lathrop},
  {Lee}, {Levi}, {Lewis}, {Li}, {Mand richenko}, {Marshall}, {Martinez},
  {Merritt}, {Miquel}, {Mu{\~n}oz}, {Neilsen}, {Nichol}, {Nord}, {Ogando},
  {Olsen}, {Palaio}, {Patton}, {Peoples}, {Plazas}, {Rauch}, {Reil}, {Rheault},
  {Roe}, {Rogers}, {Roodman}, {Sanchez}, {Scarpine}, {Schindler}, {Schmidt},
  {Schmitt}, {Schubnell}, {Schultz}, {Schurter}, {Scott}, {Serrano}, {Shaw},
  {Smith}, {Soares-Santos}, {Stefanik}, {Stuermer}, {Suchyta}, {Sypniewski},
  {Tarle}, {Thaler}, {Tighe}, {Tran}, {Tucker}, {Walker}, {Wang}, {Watson},
  {Weaverdyck}, {Wester}, {Woods}, {Yanny}, \& {DES
  Collaboration}}]{Flaugher:2015}
{Flaugher}, B., {Diehl}, H.~T., {Honscheid}, K., {et~al.} 2015, \aj, 150, 150

\bibitem[{{Galaz} {et~al.}(2011){Galaz}, {Herrera-Camus}, {Garcia-Lambas}, \&
  {Padilla}}]{Galaz:2011}
{Galaz}, G., {Herrera-Camus}, R., {Garcia-Lambas}, D., \& {Padilla}, N. 2011,
  \apj, 728, 74

\bibitem[{{Geha} {et~al.}(2017){Geha}, {Wechsler}, {Mao}, {Tollerud}, {Weiner},
  {Bernstein}, {Hoyle}, {Marchi}, {Marshall}, {Mu{\~n}oz}, \& {Lu}}]{Geha:2017}
{Geha}, M., {Wechsler}, R.~H., {Mao}, Y.-Y., {et~al.} 2017, \apj, 847, 4

\bibitem[{{Gilhuly} {et~al.}(2020){Gilhuly}, {Hendel}, {Merritt}, {Abraham},
  {Danieli}, {Lokhorst}, {Liu}, {van Dokkum}, {Conroy}, \&
  {Greco}}]{Gilhuly:2020}
{Gilhuly}, C., {Hendel}, D., {Merritt}, A., {et~al.} 2020, \apj, 897, 108

\bibitem[{{G{\'o}rski} {et~al.}(2005){G{\'o}rski}, {Hivon}, {Banday},
  {Wandelt}, {Hansen}, {Reinecke}, \& {Bartelmann}}]{Gorski:2005}
{G{\'o}rski}, K.~M., {Hivon}, E., {Banday}, A.~J., {et~al.} 2005, \apj, 622,
  759

\bibitem[{{Graham} \& {Driver}(2005)}]{Graham:2005}
{Graham}, A.~W., \& {Driver}, S.~P. 2005, \pasa, 22, 118

\bibitem[{{Greco} {et~al.}(2018){Greco}, {Greene}, {Strauss}, {Macarthur},
  {Flowers}, {Goulding}, {Huang}, {Kim}, {Komiyama}, {Leauthaud}, {Leisman},
  {Lupton}, {Sif{\'o}n}, \& {Wang}}]{Greco:2018}
{Greco}, J.~P., {Greene}, J.~E., {Strauss}, M.~A., {et~al.} 2018, \apj, 857,
  104

\bibitem[{Hastie {et~al.}(2001)Hastie, Tibshirani, \& Friedman}]{Elements}
Hastie, T., Tibshirani, R., \& Friedman, J. 2001, The Elements of Statistical
  Learning, Springer Series in Statistics (New York, NY, USA: Springer New York
  Inc.)

\bibitem[{{H{\"a}u{\ss}ler} {et~al.}(2013){H{\"a}u{\ss}ler}, {Bamford}, {Vika},
  {Rojas}, {Barden}, {Kelvin}, {Alpaslan}, {Robotham}, {Driver}, {Baldry},
  {Brough}, {Hopkins}, {Liske}, {Nichol}, {Popescu}, \&
  {Tuffs}}]{Haussler:2013}
{H{\"a}u{\ss}ler}, B., {Bamford}, S.~P., {Vika}, M., {et~al.} 2013, \mnras,
  430, 330

\bibitem[{{Hayward} {et~al.}(2005){Hayward}, {Irwin}, \&
  {Bregman}}]{Hayward:2005}
{Hayward}, C.~C., {Irwin}, J.~A., \& {Bregman}, J.~N. 2005, \apj, 635, 827

\bibitem[{{Hilker} {et~al.}(1999){Hilker}, {Kissler-Patig}, {Richtler},
  {Infante}, \& {Quintana}}]{Hilker:1999}
{Hilker}, M., {Kissler-Patig}, M., {Richtler}, T., {Infante}, L., \&
  {Quintana}, H. 1999, \aaps, 134, 59

\bibitem[{{Hogg} {et~al.}(2003){Hogg}, {Blanton}, {Eisenstein}, {Gunn},
  {Schlegel}, {Zehavi}, {Bahcall}, {Brinkmann}, {Csabai}, {Schneider},
  {Weinberg}, \& {York}}]{Hogg:2003}
{Hogg}, D.~W., {Blanton}, M.~R., {Eisenstein}, D.~J., {et~al.} 2003, \apjl,
  585, L5

\bibitem[{{Hoyle} {et~al.}(2018){Hoyle}, {Gruen}, {Bernstein}, {Rau}, {De
  Vicente}, {Hartley}, {Gaztanaga}, {DeRose}, {Troxel}, {Davis}, {Alarcon},
  {MacCrann}, {Prat}, {S{\'a}nchez}, {Sheldon}, {Wechsler}, {Asorey}, {Becker},
  {Bonnett}, {Carnero Rosell}, {Carollo}, {Carrasco Kind}, {Castander},
  {Cawthon}, {Chang}, {Childress}, {Davis}, {Drlica-Wagner}, {Gatti},
  {Glazebrook}, {Gschwend}, {Hinton}, {Hoormann}, {Kim}, {King}, {Kuehn},
  {Lewis}, {Lidman}, {Lin}, {Macaulay}, {Maia}, {Martini}, {Mudd},
  {M{\"o}ller}, {Nichol}, {Ogando}, {Rollins}, {Roodman}, {Ross}, {Rozo},
  {Rykoff}, {Samuroff}, {Sevilla-Noarbe}, {Sharp}, {Sommer}, {Tucker}, {Uddin},
  {Varga}, {Vielzeuf}, {Yuan}, {Zhang}, {Abbott}, {Abdalla}, {Allam}, {Annis},
  {Bechtol}, {Benoit-L{\'e}vy}, {Bertin}, {Brooks}, {Buckley-Geer}, {Burke},
  {Busha}, {Capozzi}, {Carretero}, {Crocce}, {D'Andrea}, {da Costa}, {DePoy},
  {Desai}, {Diehl}, {Doel}, {Eifler}, {Estrada}, {Evrard}, {Fernandez},
  {Flaugher}, {Fosalba}, {Frieman}, {Garc{\'\i}a-Bellido}, {Gerdes},
  {Giannantonio}, {Goldstein}, {Gruendl}, {Gutierrez}, {Honscheid}, {James},
  {Jarvis}, {Jeltema}, {Johnson}, {Johnson}, {Kirk}, {Krause}, {Kuhlmann},
  {Kuropatkin}, {Lahav}, {Li}, {Lima}, {March}, {Marshall}, {Melchior},
  {Menanteau}, {Miquel}, {Nord}, {O'Neill}, {Plazas}, {Romer}, {Sako},
  {Sanchez}, {Santiago}, {Scarpine}, {Schindler}, {Schubnell}, {Smith},
  {Smith}, {Soares-Santos}, {Sobreira}, {Suchyta}, {Swanson}, {Tarle},
  {Thomas}, {Tucker}, {Vikram}, {Walker}, {Weller}, {Wester}, {Wolf}, {Yanny},
  {Zuntz}, \& {DES Collaboration}}]{Hoyle:2018}
{Hoyle}, B., {Gruen}, D., {Bernstein}, G.~M., {et~al.} 2018, \mnras, 478, 592

\bibitem[{Hunter(2007)}]{Hunter:2007}
Hunter, J.~D. 2007, Computing In Science \& Engineering, 9, 90

\bibitem[{{Janssens} {et~al.}(2017){Janssens}, {Abraham}, {Brodie}, {Forbes},
  {Romanowsky}, \& {van Dokkum}}]{Janssens:2017}
{Janssens}, S., {Abraham}, R., {Brodie}, J., {et~al.} 2017, \apjl, 839, L17

\bibitem[{{Jarvis}(2015)}]{Jarvis:2015}
{Jarvis}, M. 2015, {TreeCorr: Two-point correlation functions}, ascl:1508.007

\bibitem[{Jones {et~al.}(2001)Jones, Oliphant, Peterson, {et~al.}}]{scipy:2001}
Jones, E., Oliphant, T., Peterson, P., {et~al.} 2001, {SciPy}: Open source
  scientific tools for {Python}

\bibitem[{{Koda} {et~al.}(2015){Koda}, {Yagi}, {Yamanoi}, \&
  {Komiyama}}]{Koda:2015}
{Koda}, J., {Yagi}, M., {Yamanoi}, H., \& {Komiyama}, Y. 2015, \apjl, 807, L2

\bibitem[{{Landy} \& {Szalay}(1993)}]{Landy:1993}
{Landy}, S.~D., \& {Szalay}, A.~S. 1993, \apj, 412, 64

\bibitem[{{Larson} {et~al.}(1980){Larson}, {Tinsley}, \&
  {Caldwell}}]{Larson:1980}
{Larson}, R.~B., {Tinsley}, B.~M., \& {Caldwell}, C.~N. 1980, \apj, 237, 692

\bibitem[{{Law-Smith} \& {Eisenstein}(2017)}]{Law-Smith:2017}
{Law-Smith}, J., \& {Eisenstein}, D.~J. 2017, \apj, 836, 87

\bibitem[{{Lintott} {et~al.}(2011){Lintott}, {Schawinski}, {Bamford}, {Slosar},
  {Land}, {Thomas}, {Edmondson}, {Masters}, {Nichol}, {Raddick}, {Szalay},
  {Andreescu}, {Murray}, \& {Vandenberg}}]{Lintott:2011}
{Lintott}, C., {Schawinski}, K., {Bamford}, S., {et~al.} 2011, \mnras, 410, 166

\bibitem[{{Maller} {et~al.}(2005){Maller}, {McIntosh}, {Katz}, \&
  {Weinberg}}]{Maller:2005}
{Maller}, A.~H., {McIntosh}, D.~H., {Katz}, N., \& {Weinberg}, M.~D. 2005,
  \apj, 619, 147

\bibitem[{{Marigo} {et~al.}(2017){Marigo}, {Girardi}, {Bressan}, {Rosenfield},
  {Aringer}, {Chen}, {Dussin}, {Nanni}, {Pastorelli}, {Rodrigues}, {Trabucchi},
  {Bladh}, {Dalcanton}, {Groenewegen}, {Montalb{\'a}n}, \&
  {Wood}}]{Marigo:2017}
{Marigo}, P., {Girardi}, L., {Bressan}, A., {et~al.} 2017, \apj, 835, 77

\bibitem[{{Martin} {et~al.}(2019){Martin}, {Kaviraj}, {Laigle}, {Devriendt},
  {Jackson}, {Peirani}, {Dubois}, {Pichon}, \& {Slyz}}]{Martin:2019}
{Martin}, G., {Kaviraj}, S., {Laigle}, C., {et~al.} 2019, \mnras, 485, 796

\bibitem[{{Martin} {et~al.}(2013){Martin}, {Ibata}, {McConnachie}, {Mackey},
  {Ferguson}, {Irwin}, {Lewis}, \& {Fardal}}]{Martin:2013}
{Martin}, N.~F., {Ibata}, R.~A., {McConnachie}, A.~W., {et~al.} 2013, \apj,
  776, 80

\bibitem[{{Martin} {et~al.}(2016){Martin}, {Ibata}, {Lewis}, {McConnachie},
  {Babul}, {Bate}, {Bernard}, {Chapman}, {Collins}, {Conn}, {Crnojevi{\'c}},
  {Fardal}, {Ferguson}, {Irwin}, {Mackey}, {McMonigal}, {Navarro}, \&
  {Rich}}]{Martin:2016}
{Martin}, N.~F., {Ibata}, R.~A., {Lewis}, G.~F., {et~al.} 2016, \apj, 833, 167

\bibitem[{{McConnachie}(2012)}]{McConnachie:2012}
{McConnachie}, A.~W. 2012, \aj, 144, 4

\bibitem[{{McGaugh} {et~al.}(1995){McGaugh}, {Bothun}, \&
  {Schombert}}]{McGaugh:1995}
{McGaugh}, S.~S., {Bothun}, G.~D., \& {Schombert}, J.~M. 1995, \aj, 110, 573

\bibitem[{{Merritt} {et~al.}(2016){Merritt}, {van Dokkum}, {Danieli},
  {Abraham}, {Zhang}, {Karachentsev}, \& {Makarova}}]{Merritt:2016}
{Merritt}, A., {van Dokkum}, P., {Danieli}, S., {et~al.} 2016, \apj, 833, 168

\bibitem[{{Mihos} {et~al.}(2017){Mihos}, {Harding}, {Feldmeier}, {Rudick},
  {Janowiecki}, {Morrison}, {Slater}, \& {Watkins}}]{Mihos:2017}
{Mihos}, J.~C., {Harding}, P., {Feldmeier}, J.~J., {et~al.} 2017, \apj, 834, 16

\bibitem[{{Mihos} {et~al.}(2015){Mihos}, {Durrell}, {Ferrarese}, {Feldmeier},
  {C{\^o}t{\'e}}, {Peng}, {Harding}, {Liu}, {Gwyn}, \&
  {Cuillandre}}]{Mihos:2015}
{Mihos}, J.~C., {Durrell}, P.~R., {Ferrarese}, L., {et~al.} 2015, \apjl, 809,
  L21

\bibitem[{{Minchin} {et~al.}(2004){Minchin}, {Disney}, {Parker}, {Boyce}, {de
  Blok}, {Banks}, {Ekers}, {Freeman}, {Garcia}, {Gibson}, {Grossi}, {Haynes},
  {Knezek}, {Lang}, {Malin}, {Price}, {Putman}, {Stewart}, \&
  {Wright}}]{Minchin:2004}
{Minchin}, R.~F., {Disney}, M.~J., {Parker}, Q.~A., {et~al.} 2004, \mnras, 355,
  1303

\bibitem[{{Morganson} {et~al.}(2018){Morganson}, {Gruendl}, {Menanteau},
  {Carrasco Kind}, {Chen}, {Daues}, {Drlica-Wagner}, {Friedel}, {Gower},
  {Johnson}, {Johnson}, {Kessler}, {Paz-Chinch{\'o}n}, {Petravick}, {Pond},
  {Yanny}, {Allam}, {Armstrong}, {Barkhouse}, {Bechtol}, {Benoit-L{\'e}vy},
  {Bernstein}, {Bertin}, {Buckley-Geer}, {Covarrubias}, {Desai}, {Diehl},
  {Goldstein}, {Gruen}, {Li}, {Lin}, {Marriner}, {Mohr}, {Neilsen}, {Ngeow},
  {Paech}, {Rykoff}, {Sako}, {Sevilla-Noarbe}, {Sheldon}, {Sobreira}, {Tucker},
  {Wester}, \& {DES Collaboration}}]{Morganson:2018}
{Morganson}, E., {Gruendl}, R.~A., {Menanteau}, F., {et~al.} 2018, \pasp, 130,
  074501

\bibitem[{{Moster} {et~al.}(2013){Moster}, {Naab}, \& {White}}]{Moster:2013}
{Moster}, B.~P., {Naab}, T., \& {White}, S. D.~M. 2013, \mnras, 428, 3121

\bibitem[{{Mu{\~n}oz} {et~al.}(2015){Mu{\~n}oz}, {Eigenthaler}, {Puzia},
  {Taylor}, {Ordenes-Brice{\~n}o}, {Alamo-Mart{\'\i}nez}, {Ribbeck},
  {{\'A}ngel}, {Capaccioli}, {C{\^o}t{\'e}}, {Ferrarese}, {Galaz}, {Hempel},
  {Hilker}, {Jord{\'a}n}, {Lan{\c{c}}on}, {Mieske}, {Paolillo}, {Richtler},
  {S{\'a}nchez-Janssen}, \& {Zhang}}]{Munoz:2015}
{Mu{\~n}oz}, R.~P., {Eigenthaler}, P., {Puzia}, T.~H., {et~al.} 2015, \apjl,
  813, L15

\bibitem[{{Neilsen} {et~al.}(2019){Neilsen}, {Annis}, {Diehl}, {Swanson},
  {D'Andrea}, {Kent}, \& {Drlica-Wagner}}]{Neilsen:2019}
{Neilsen}, Eric~H., J., {Annis}, J.~T., {Diehl}, H.~T., {et~al.} 2019, arXiv
  e-prints, arXiv:1912.06254

\bibitem[{{Norberg} {et~al.}(2002){Norberg}, {Baugh}, {Hawkins}, {Maddox},
  {Madgwick}, {Lahav}, {Cole}, {Frenk}, {Baldry}, {Bland -Hawthorn}, {Bridges},
  {Cannon}, {Colless}, {Collins}, {Couch}, {Dalton}, {De Propris}, {Driver},
  {Efstathiou}, {Ellis}, {Glazebrook}, {Jackson}, {Lewis}, {Lumsden},
  {Peacock}, {Peterson}, {Sutherland}, \& {Taylor}}]{Norberg:2002}
{Norberg}, P., {Baugh}, C.~M., {Hawkins}, E., {et~al.} 2002, \mnras, 332, 827

\bibitem[{{O'Neil} {et~al.}(1997){O'Neil}, {Bothun}, \& {Cornell}}]{ONeil:1997}
{O'Neil}, K., {Bothun}, G.~D., \& {Cornell}, M.~E. 1997, \aj, 113, 1212

\bibitem[{{O'Neil} {et~al.}(2000){O'Neil}, {Bothun}, \&
  {Schombert}}]{ONeil:2000}
{O'Neil}, K., {Bothun}, G.~D., \& {Schombert}, J. 2000, \aj, 119, 136

\bibitem[{Ordenes-Brice{\~n}o {et~al.}(2018)Ordenes-Brice{\~n}o, Eigenthaler,
  Taylor, Puzia, Alamo-Martínez, Ribbeck, P.~Muñoz, Zhang, Grebel, Ángel, \&
  et~al.}]{Ordenes:2018}
Ordenes-Brice{\~n}o, Y., Eigenthaler, P., Taylor, M.~A., {et~al.} 2018, The
  Astrophysical Journal, 859, 52

\bibitem[{{Papastergis} {et~al.}(2017){Papastergis}, {Adams}, \&
  {Romanowsky}}]{Papastergis:2017}
{Papastergis}, E., {Adams}, E.~A.~K., \& {Romanowsky}, A.~J. 2017, \aap, 601,
  L10

\bibitem[{{Papastergis} {et~al.}(2015){Papastergis}, {Giovanelli}, {Haynes}, \&
  {Shankar}}]{Papastergis:2015}
{Papastergis}, E., {Giovanelli}, R., {Haynes}, M.~P., \& {Shankar}, F. 2015,
  \aap, 574, A113

\bibitem[{Pedregosa {et~al.}(2011)Pedregosa, Varoquaux, Gramfort, Michel,
  Thirion, Grisel, Blondel, Prettenhofer, Weiss, Dubourg, Vanderplas, Passos,
  Cournapeau, Brucher, Perrot, \& Duchesnay}]{Pedregosa:2011}
Pedregosa, F., Varoquaux, G., Gramfort, A., {et~al.} 2011, Journal of Machine
  Learning Research, 12, 2825

\bibitem[{{Peebles}(1980)}]{Peebles:1980}
{Peebles}, P.~J.~E. 1980, {The large-scale structure of the universe}
  (Princeton University Press)

\bibitem[{{Peng} {et~al.}(2002){Peng}, {Ho}, {Impey}, \& {Rix}}]{Peng:2002}
{Peng}, C.~Y., {Ho}, L.~C., {Impey}, C.~D., \& {Rix}, H.-W. 2002, \aj, 124, 266

\bibitem[{{Rom{\'a}n} \& {Trujillo}(2017)}]{Roman:2017b}
{Rom{\'a}n}, J., \& {Trujillo}, I. 2017, \mnras, 468, 4039

\bibitem[{{Rosenbaum} {et~al.}(2009){Rosenbaum}, {Krusch}, {Bomans}, \&
  {Dettmar}}]{Rosenbaum:2009}
{Rosenbaum}, S.~D., {Krusch}, E., {Bomans}, D.~J., \& {Dettmar}, R.~J. 2009,
  \aap, 504, 807

\bibitem[{{Sabatini} {et~al.}(2005){Sabatini}, {Davies}, {van Driel}, {Baes},
  {Roberts}, {Smith}, {Linder}, \& {O'Neil}}]{Sabatini:2005}
{Sabatini}, S., {Davies}, J., {van Driel}, W., {et~al.} 2005, \mnras, 357, 819

\bibitem[{{Sales} {et~al.}(2019){Sales}, {Navarro}, {Penafiel}, {Peng}, {Lim},
  \& {Hernquist}}]{Sales:2019}
{Sales}, L.~V., {Navarro}, J.~F., {Penafiel}, L., {et~al.} 2019, arXiv
  e-prints, arXiv:1909.01347

\bibitem[{{Schlafly} \& {Finkbeiner}(2011)}]{Schlafly:2011}
{Schlafly}, E.~F., \& {Finkbeiner}, D.~P. 2011, \apj, 737, 103

\bibitem[{{Schlegel} {et~al.}(1998){Schlegel}, {Finkbeiner}, \&
  {Davis}}]{Schlegel1998}
{Schlegel}, D.~J., {Finkbeiner}, D.~P., \& {Davis}, M. 1998, \apj, 500, 525

\bibitem[{{Sevilla-Noarbe} {et~al.}(2020){Sevilla-Noarbe}, {Bechtol},
  {et~al.}}]{Sevilla:2020}
{Sevilla-Noarbe}, I., {Bechtol}, K., {et~al.} 2020, in prep

\bibitem[{{Shen} {et~al.}(2003){Shen}, {Mo}, {White}, {Blanton}, {Kauffmann},
  {Voges}, {Brinkmann}, \& {Csabai}}]{Shen:2003}
{Shen}, S., {Mo}, H.~J., {White}, S. D.~M., {et~al.} 2003, \mnras, 343, 978

\bibitem[{{Sif{\'o}n} {et~al.}(2018){Sif{\'o}n}, {van der Burg}, {Hoekstra},
  {Muzzin}, \& {Herbonnet}}]{Sifon:2018}
{Sif{\'o}n}, C., {van der Burg}, R. F.~J., {Hoekstra}, H., {Muzzin}, A., \&
  {Herbonnet}, R. 2018, \mnras, 473, 3747

\bibitem[{{Simon}(2019)}]{Simon:2019}
{Simon}, J.~D. 2019, \araa, 57, 375

\bibitem[{{Strateva} {et~al.}(2001){Strateva}, {Ivezi{\'c}}, {Knapp},
  {Narayanan}, {Strauss}, {Gunn}, {Lupton}, {Schlegel}, {Bahcall}, {Brinkmann},
  {Brunner}, {Budav{\'a}ri}, {Csabai}, {Castander}, {Doi}, {Fukugita},
  {Gy{\H{o}}ry}, {Hamabe}, {Hennessy}, {Ichikawa}, {Kunszt}, {Lamb}, {McKay},
  {Okamura}, {Racusin}, {Sekiguchi}, {Schneider}, {Shimasaku}, \&
  {York}}]{Strateva:2001}
{Strateva}, I., {Ivezi{\'c}}, {\v{Z}}., {Knapp}, G.~R., {et~al.} 2001, \aj,
  122, 1861

\bibitem[{{Suchyta} {et~al.}(2016){Suchyta}, {Huff}, {Aleksi{\'c}}, {Melchior},
  {Jouvel}, {MacCrann}, {Ross}, {Crocce}, {Gaztanaga}, {Honscheid}, {Leistedt},
  {Peiris}, {Rykoff}, {Sheldon}, {Abbott}, {Abdalla}, {Allam}, {Banerji},
  {Benoit-L{\'e}vy}, {Bertin}, {Brooks}, {Burke}, {Carnero Rosell}, {Carrasco
  Kind}, {Carretero}, {Cunha}, {D'Andrea}, {da Costa}, {DePoy}, {Desai},
  {Diehl}, {Dietrich}, {Doel}, {Eifler}, {Estrada}, {Evrard}, {Flaugher},
  {Fosalba}, {Frieman}, {Gerdes}, {Gruen}, {Gruendl}, {James}, {Jarvis},
  {Kuehn}, {Kuropatkin}, {Lahav}, {Lima}, {Maia}, {March}, {Marshall},
  {Miller}, {Miquel}, {Neilsen}, {Nichol}, {Nord}, {Ogando}, {Percival},
  {Reil}, {Roodman}, {Sako}, {Sanchez}, {Scarpine}, {Sevilla-Noarbe}, {Smith},
  {Soares-Santos}, {Sobreira}, {Swanson}, {Tarle}, {Thaler}, {Thomas},
  {Vikram}, {Walker}, {Wechsler}, {Zhang}, \& {DES
  Collaboration}}]{Suchyta:2016}
{Suchyta}, E., {Huff}, E.~M., {Aleksi{\'c}}, J., {et~al.} 2016, \mnras, 457,
  786

\bibitem[{{Sulentic} \& {Tifft}(1999)}]{NGC:1999}
{Sulentic}, J.~W., \& {Tifft}, W.~G. 1999, VizieR Online Data Catalog, 7001

\bibitem[{{Swanson} {et~al.}(2008){Swanson}, {Tegmark}, {Hamilton}, \&
  {Hill}}]{Swanson:2008}
{Swanson}, M.~E.~C., {Tegmark}, M., {Hamilton}, A. J.~S., \& {Hill}, J.~C.
  2008, \mnras, 387, 1391

\bibitem[{{Tremmel} {et~al.}(2019){Tremmel}, {Wright}, {Brooks}, {Munshi},
  {Nagai}, \& {Quinn}}]{Tremmel:2019}
{Tremmel}, M., {Wright}, A.~C., {Brooks}, A.~M., {et~al.} 2019, arXiv e-prints,
  arXiv:1908.05684

\bibitem[{{Tully}(2015)}]{Tully:2015}
{Tully}, R.~B. 2015, \aj, 149, 171

\bibitem[{{van der Burg} {et~al.}(2016){van der Burg}, {Muzzin}, \&
  {Hoekstra}}]{vanderBurg:2016}
{van der Burg}, R. F.~J., {Muzzin}, A., \& {Hoekstra}, H. 2016, \aap, 590, A20

\bibitem[{{Van Der Walt} {et~al.}(2011){Van Der Walt}, {Colbert}, \&
  {Varoquaux}}]{numpy:2011}
{Van Der Walt}, S., {Colbert}, S.~C., \& {Varoquaux}, G. 2011, Computing in
  Science \& Engineering, 13, 22

\bibitem[{{van Dokkum} {et~al.}(2019){van Dokkum}, {Danieli}, {Abraham},
  {Conroy}, \& {Romanowsky}}]{vanDokkum:2019}
{van Dokkum}, P., {Danieli}, S., {Abraham}, R., {Conroy}, C., \& {Romanowsky},
  A.~J. 2019, \apjl, 874, L5

\bibitem[{{van Dokkum} {et~al.}(2018){van Dokkum}, {Danieli}, {Cohen},
  {Merritt}, {Romanowsky}, {Abraham}, {Brodie}, {Conroy}, {Lokhorst}, {Mowla},
  {O'Sullivan}, \& {Zhang}}]{vanDokkum:2018}
{van Dokkum}, P., {Danieli}, S., {Cohen}, Y., {et~al.} 2018, \nat, 555, 629

\bibitem[{{van Dokkum} {et~al.}(2020){van Dokkum}, {Lokhorst}, {Danieli}, {Li},
  {Merritt}, {Abraham}, {Gilhuly}, {Greco}, \& {Liu}}]{vanDokkum:2020}
{van Dokkum}, P., {Lokhorst}, D., {Danieli}, S., {et~al.} 2020, \pasp, 132,
  074503

\bibitem[{{van Dokkum} {et~al.}(2015){van Dokkum}, {Abraham}, {Merritt},
  {Zhang}, {Geha}, \& {Conroy}}]{vanDokkum:2015a}
{van Dokkum}, P.~G., {Abraham}, R., {Merritt}, A., {et~al.} 2015, \apjl, 798,
  L45

\bibitem[{{Venhola} {et~al.}(2017){Venhola}, {Peletier}, {Laurikainen}, {Salo},
  {Lisker}, {Iodice}, {Capaccioli}, {Verdois Kleijn}, {Valentijn}, {Mieske},
  {Hilker}, {Wittmann}, {van de Ven}, {Grado}, {Spavone}, {Cantiello},
  {Napolitano}, {Paolillo}, \& {Falc{\'o}n-Barroso}}]{Venhola:2017}
{Venhola}, A., {Peletier}, R., {Laurikainen}, E., {et~al.} 2017, \aap, 608,
  A142

\bibitem[{{Venhola} {et~al.}(2018){Venhola}, {Peletier}, {Laurikainen}, {Salo},
  {Iodice}, {Mieske}, {Hilker}, {Wittmann}, {Lisker}, {Paolillo}, {Cantiello},
  {Janz}, {Spavone}, {D'Abrusco}, {Ven}, {Napolitano}, {Kleijn}, {Maddox},
  {Capaccioli}, {Grado}, {Valentijn}, {Falc{\'o}n-Barroso}, \&
  {Limatola}}]{Venhola2018}
---. 2018, \aap, 620, A165

\bibitem[{{Wang} {et~al.}(2013){Wang}, {Brunner}, \& {Dolence}}]{Wang:2013}
{Wang}, Y., {Brunner}, R.~J., \& {Dolence}, J.~C. 2013, \mnras, 432, 1961

\bibitem[{{Wang} {et~al.}(2009){Wang}, {Yang}, {Mo}, {van den Bosch}, {Katz},
  {Pasquali}, {McIntosh}, \& {Weinmann}}]{Wang:2009}
{Wang}, Y., {Yang}, X., {Mo}, H.~J., {et~al.} 2009, \apj, 697, 247

\bibitem[{{Wechsler} \& {Tinker}(2018)}]{Wechsler:2018}
{Wechsler}, R.~H., \& {Tinker}, J.~L. 2018, \araa, 56, 435

\bibitem[{{White} \& {Frenk}(1991)}]{White:1991}
{White}, S. D.~M., \& {Frenk}, C.~S. 1991, \apj, 379, 52

\bibitem[{{Wittmann} {et~al.}(2017){Wittmann}, {Lisker}, {Ambachew Tilahun},
  {Grebel}, {Conselice}, {Penny}, {Janz}, {Gallagher}, {Kotulla}, \&
  {McCormac}}]{Wittmann:2017}
{Wittmann}, C., {Lisker}, T., {Ambachew Tilahun}, L., {et~al.} 2017, \mnras,
  470, 1512

\bibitem[{{Zehavi} {et~al.}(2002){Zehavi}, {Blanton}, {Frieman}, {Weinberg},
  {Mo}, {Strauss}, {Anderson}, {Annis}, {Bahcall}, {Bernardi}, {Briggs},
  {Brinkmann}, {Burles}, {Carey}, {Castander}, {Connolly}, {Csabai},
  {Dalcanton}, {Dodelson}, {Doi}, {Eisenstein}, {Evans}, {Finkbeiner},
  {Friedman}, {Fukugita}, {Gunn}, {Hennessy}, {Hindsley}, {Ivezi{\'c}}, {Kent},
  {Knapp}, {Kron}, {Kunszt}, {Lamb}, {Leger}, {Long}, {Loveday}, {Lupton},
  {McKay}, {Meiksin}, {Merrelli}, {Munn}, {Narayanan}, {Newcomb}, {Nichol},
  {Owen}, {Peoples}, {Pope}, {Rockosi}, {Schlegel}, {Schneider}, {Scoccimarro},
  {Sheth}, {Siegmund}, {Smee}, {Snir}, {Stebbins}, {Stoughton}, {SubbaRao},
  {Szalay}, {Szapudi}, {Tegmark}, {Tucker}, {Uomoto}, {Vanden Berk}, {Vogeley},
  {Waddell}, {Yanny}, \& {York}}]{Zehavi:2002}
{Zehavi}, I., {Blanton}, M.~R., {Frieman}, J.~A., {et~al.} 2002, \apj, 571, 172

\bibitem[{{Zehavi} {et~al.}(2005){Zehavi}, {Zheng}, {Weinberg}, {Frieman},
  {Berlind}, {Blanton}, {Scoccimarro}, {Sheth}, {Strauss}, {Kayo}, {Suto},
  {Fukugita}, {Nakamura}, {Bahcall}, {Brinkmann}, {Gunn}, {Hennessy},
  {Ivezi{\'c}}, {Knapp}, {Loveday}, {Meiksin}, {Schlegel}, {Schneider},
  {Szapudi}, {Tegmark}, {Vogeley}, {York}, \& {SDSS
  Collaboration}}]{Zehavi:2005}
{Zehavi}, I., {Zheng}, Z., {Weinberg}, D.~H., {et~al.} 2005, \apj, 630, 1

\bibitem[{{Zehavi} {et~al.}(2011){Zehavi}, {Zheng}, {Weinberg}, {Blanton},
  {Bahcall}, {Berlind}, {Brinkmann}, {Frieman}, {Gunn}, {Lupton}, {Nichol},
  {Percival}, {Schneider}, {Skibba}, {Strauss}, {Tegmark}, \&
  {York}}]{Zehavi:2011}
---. 2011, \apj, 736, 59

\bibitem[{{Zhong} {et~al.}(2008){Zhong}, {Liang}, {Liu}, {Hammer}, {Hu},
  {Chen}, {Deng}, \& {Zhang}}]{Zhong:2008}
{Zhong}, G.~H., {Liang}, Y.~C., {Liu}, F.~S., {et~al.} 2008, \mnras, 391, 986

\end{thebibliography}

\appendix

\section{Surface-Brightness Limits} \label{app:sbcontrast}

\begin{figure}[ht!]
\centering
\includegraphics[width=0.9\textwidth]{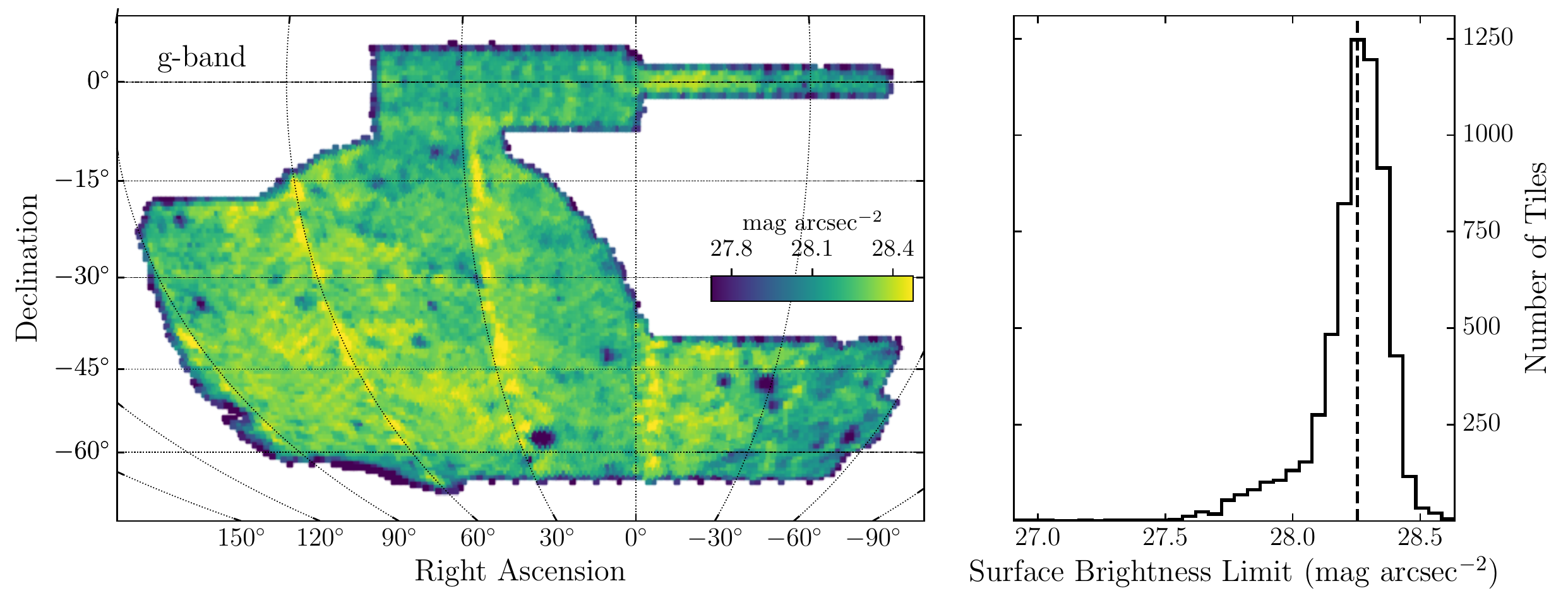}
\includegraphics[width=0.9\textwidth]{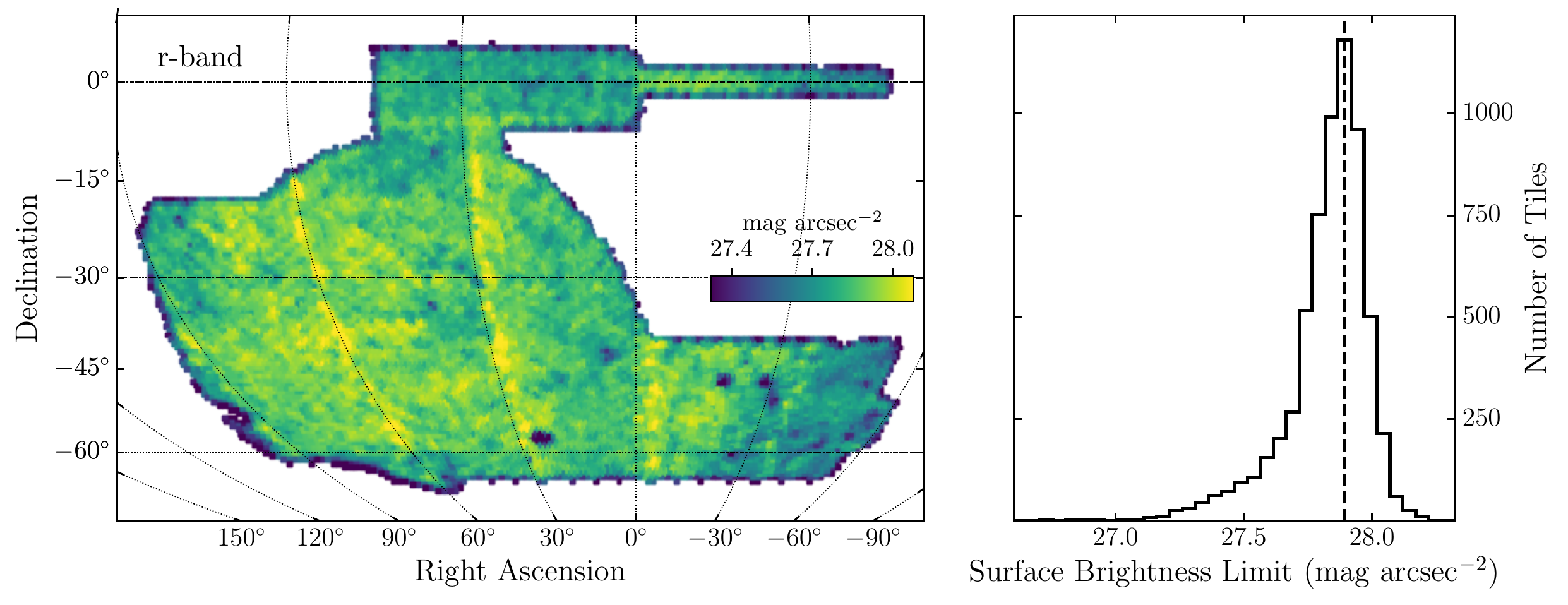}
\includegraphics[width=0.9\textwidth]{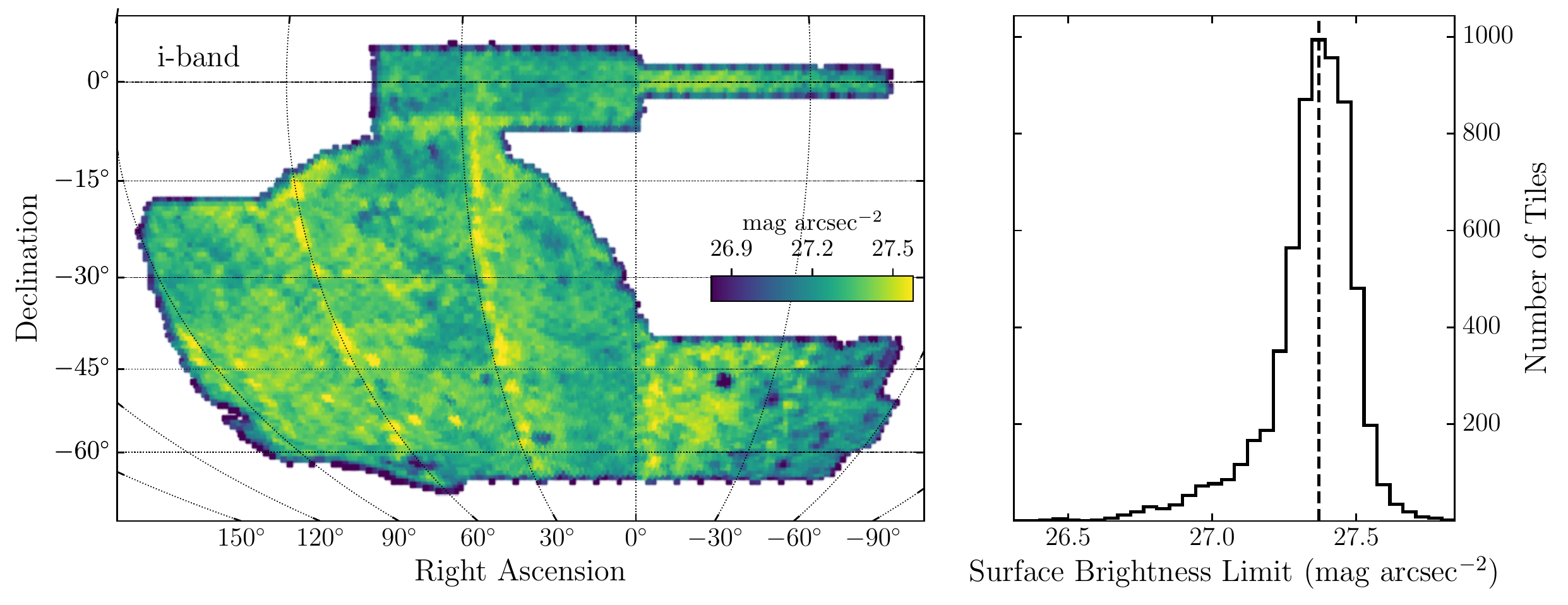}
\caption{Surface-brightness limits at $3\sigma$ estimated from the surface-brightness contrast in $10'' \times 10''$ regions over the DES coadd tiles in the $g$ band (top), $r$ band (middle), and $i$ band (bottom).}
\label{fig:sbcontrast}
\end{figure}

We estimate the surface-brightness limit of the DES data by applying the \code{sbcontrast} module from Multi-Resolution Filtering packaged developed for the Dragonfly Telephoto Array \citep{vanDokkum:2020}.\footnote{\url{https://github.com/AstroJacobLi/mrf}}
This procedure bins each coadd image into $10'' \times 10''$ regions, subtracts a local background from each binned pixel based on the surrounding 8 pixels, and calculates the variation among the binned and background-subtracted pixels.  
We applied this procedure to each DES coadd tile after masking bad pixels and sources detected by \SExtractor.
The resulting maps and 1-D distributions of $3\sigma$ surface-brightness limits are shown in \figref{sbcontrast}.
The tail to lower surface-brightness limits comes dominantly from tiles around the survey boarder, which have fewer tilings and less homogenous coverage.

\section{Selection Criteria} \label{app:select}

Removal of point sources (star--galaxy separation):
\begin{verbatim}
(EXTENDED_CLASS_COADD != 0) &
(SPREAD_MODEL_I + 5/3*SPREADERR_MODEL_I > 0.007)
\end{verbatim}

Selection of LSBG candidates:\\

$\bullet$ Surface-brightness and  radius cuts:
\begin{verbatim}
(FLUX_RADIUS_G > 2.5) & (FLUX_RADIUS_G < 20)
(MU_MEAN_MODEL_G > 24.2) & (MU_MEAN_MODEL_G < 28.8)
\end{verbatim}

$\bullet$ Ellipticity cut:
\begin{verbatim}
(1 - B_IMAGE/A_IMAGE) < 0.7
\end{verbatim}

$\bullet$ Color cuts:
\begin{verbatim}
-0.1 < (MAG_AUTO_G-MAG_AUTO_I) < 1.4
(MAG_AUTO_G - MAG_AUTO_R) > 0.7*(MAG_AUTO_G - MAG_AUTO_I) - 0.4
(MAG_AUTO_G - MAG_AUTO_R) < 0.7*(MAG_AUTO_G - MAG_AUTO_I) + 0.4
\end{verbatim}

\section{Magnitude Distributions} \label{app:magnitude_dists}

This appendix presents supplemental plots characterizing the magnitude distribution of our LSBG sample and associated external 2MPZ sample.

\begin{figure}[ht!]
\centering
\epsscale{0.8}
\plotone{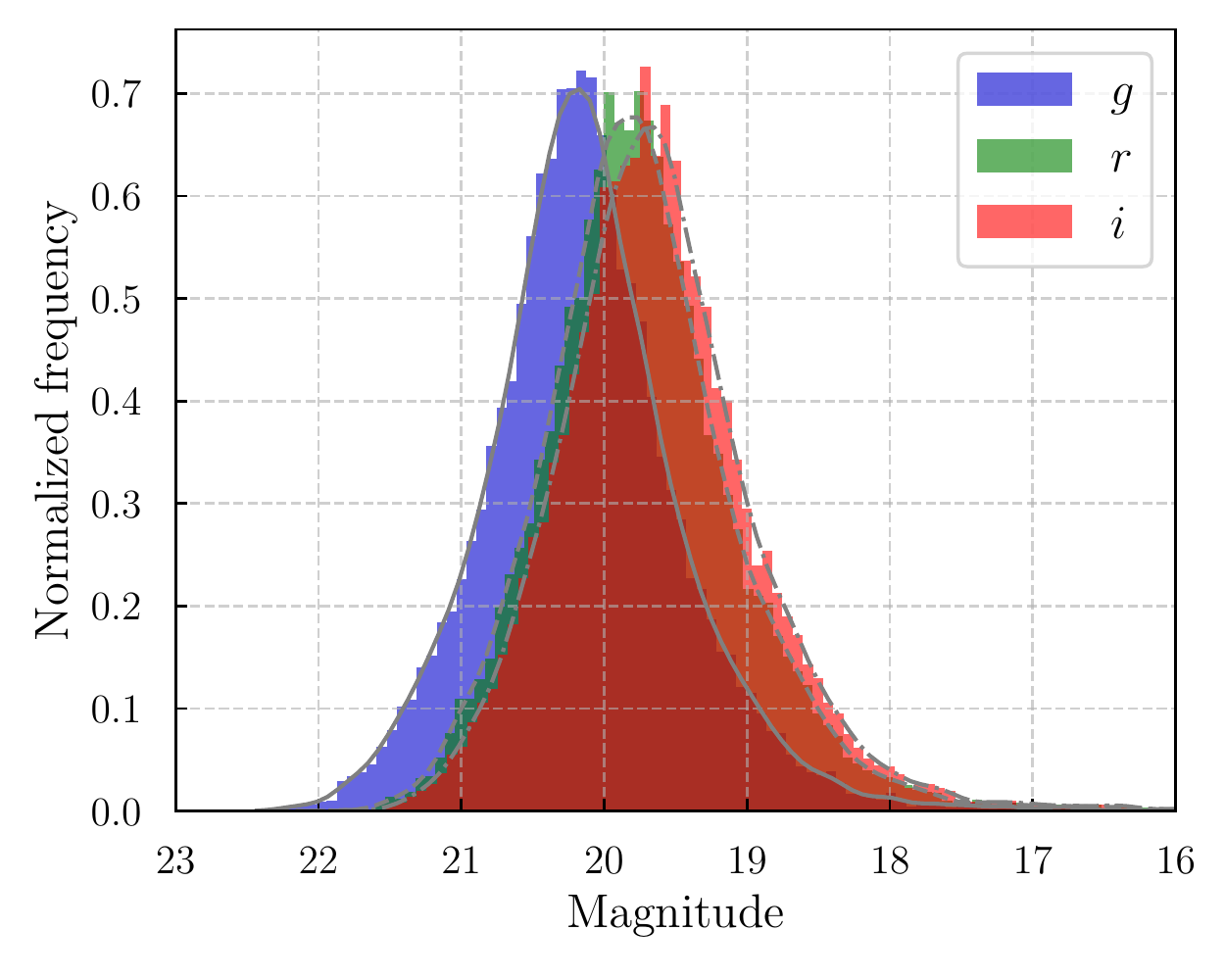}
\vspace{0.1cm}
\caption{Normalized distribution of the $g$-, $r$-, and $i$-band magnitudes of our LSBG sample.}
\label{fig:Mag_dists}
\end{figure}

\begin{figure*}[t]
\centering
\subfigure[]{\includegraphics[width=0.45\textwidth]{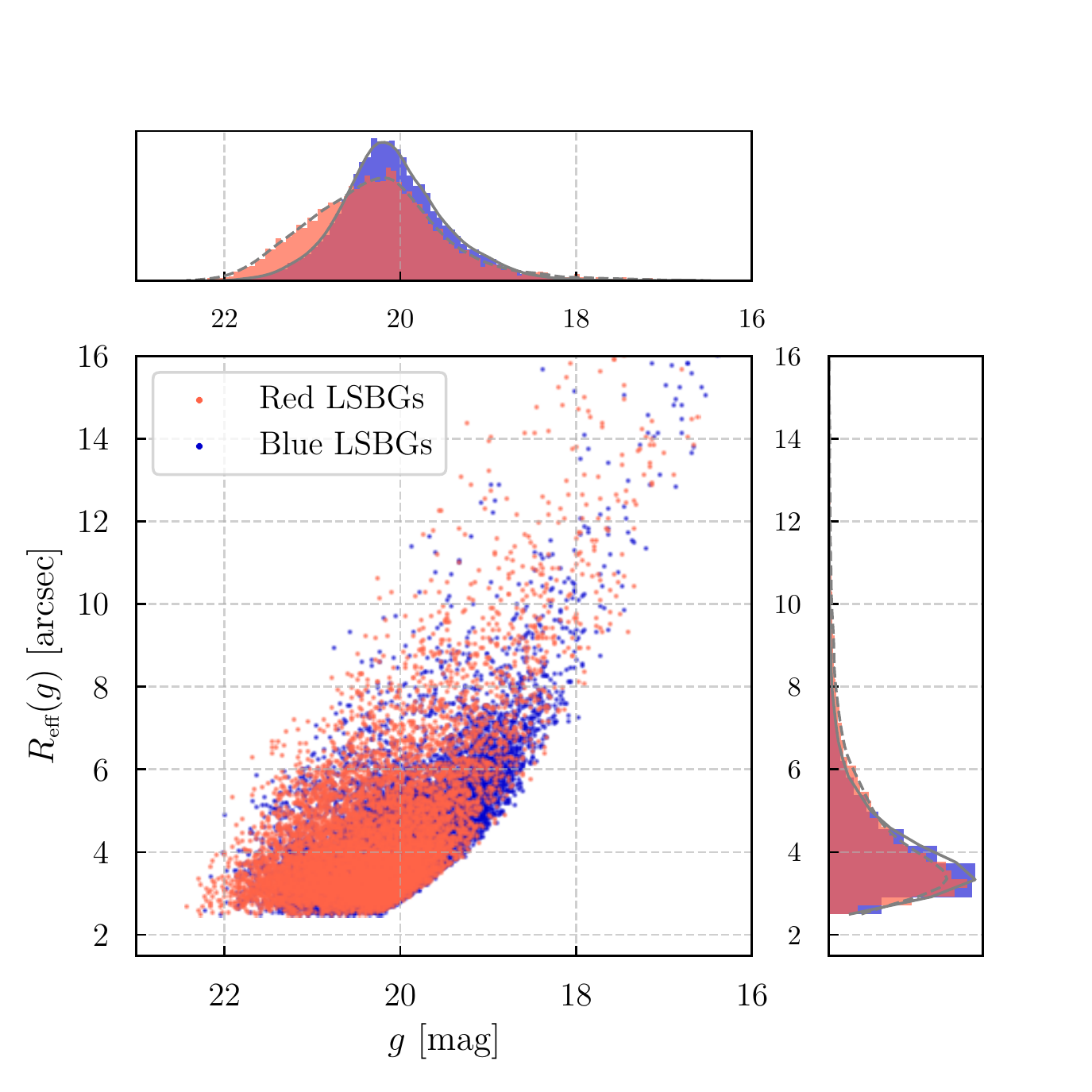}}
\subfigure[]{\includegraphics[width=0.45\textwidth]{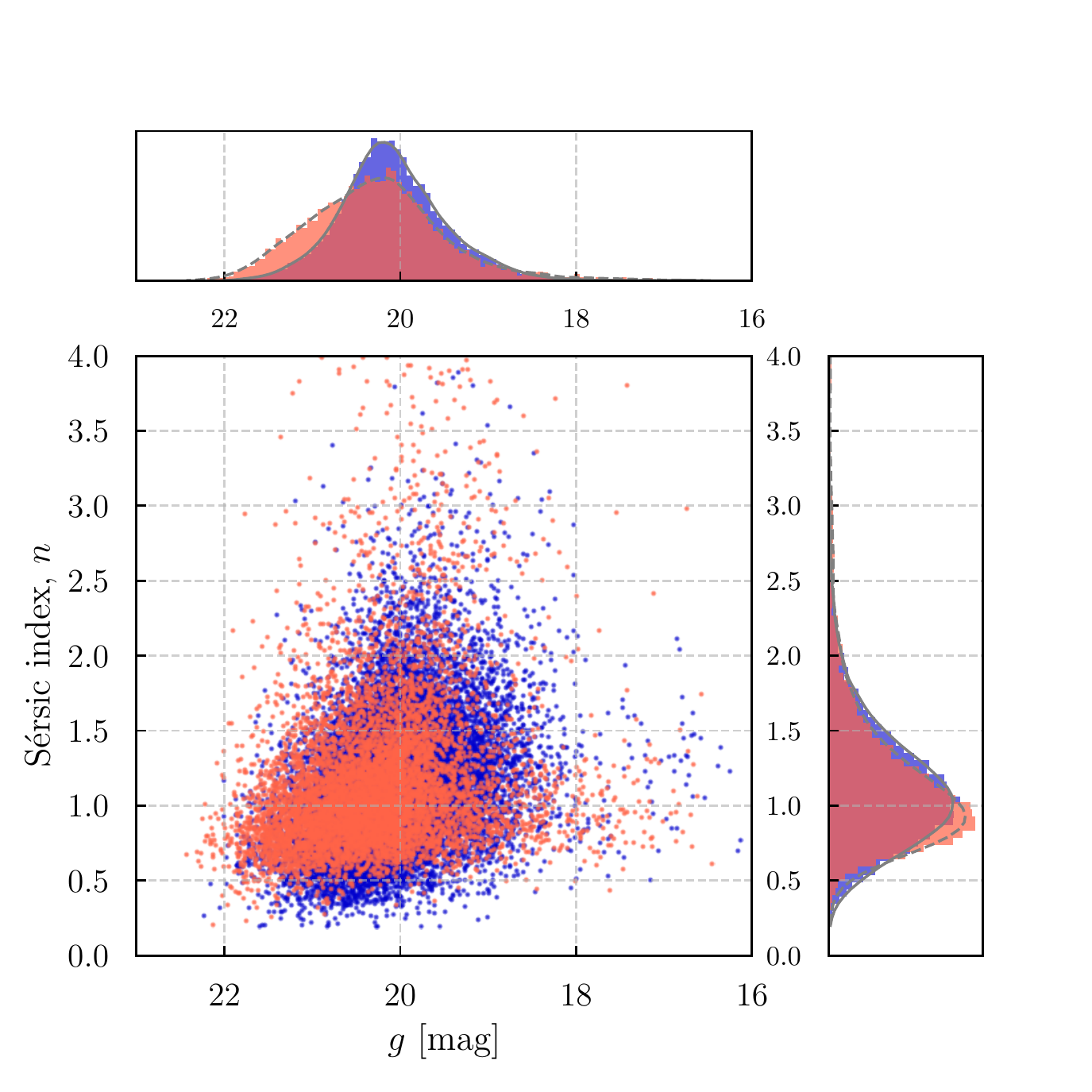}}
\caption{Joint distributions of the red and blue LSBGs in the space of $g$-band magnitude vs (a) effective radius, \Reff, and (b) S\'ersic index, $n$, both in $g$-band.}
\label{fig:mag_rad_ser}
\end{figure*}

In \figref{Mag_dists} we present the $g$, $r$, and $i$-band magnitude distributions of our LSBG sample. 
The magnitudes come from the \code{galfitm} S\'ersic  model fitting of the sample. 
The median magnitudes in each band are $g = 20.2$, $r = 19.8$, and $i = 19.7$.

Similar to \figref{eff_rad}, in \figref{mag_rad_ser} we present joint distributions of the blue and red LSBG subsamples in the space of (a) effective radius, \Reff, and (b) S\'ersic index vs the $g$-band magnitude this time. We note that there is no strong color dependence of the $g$-magnitude distribution.

\begin{figure}[ht!]
\centering
\epsscale{0.8}
\plotone{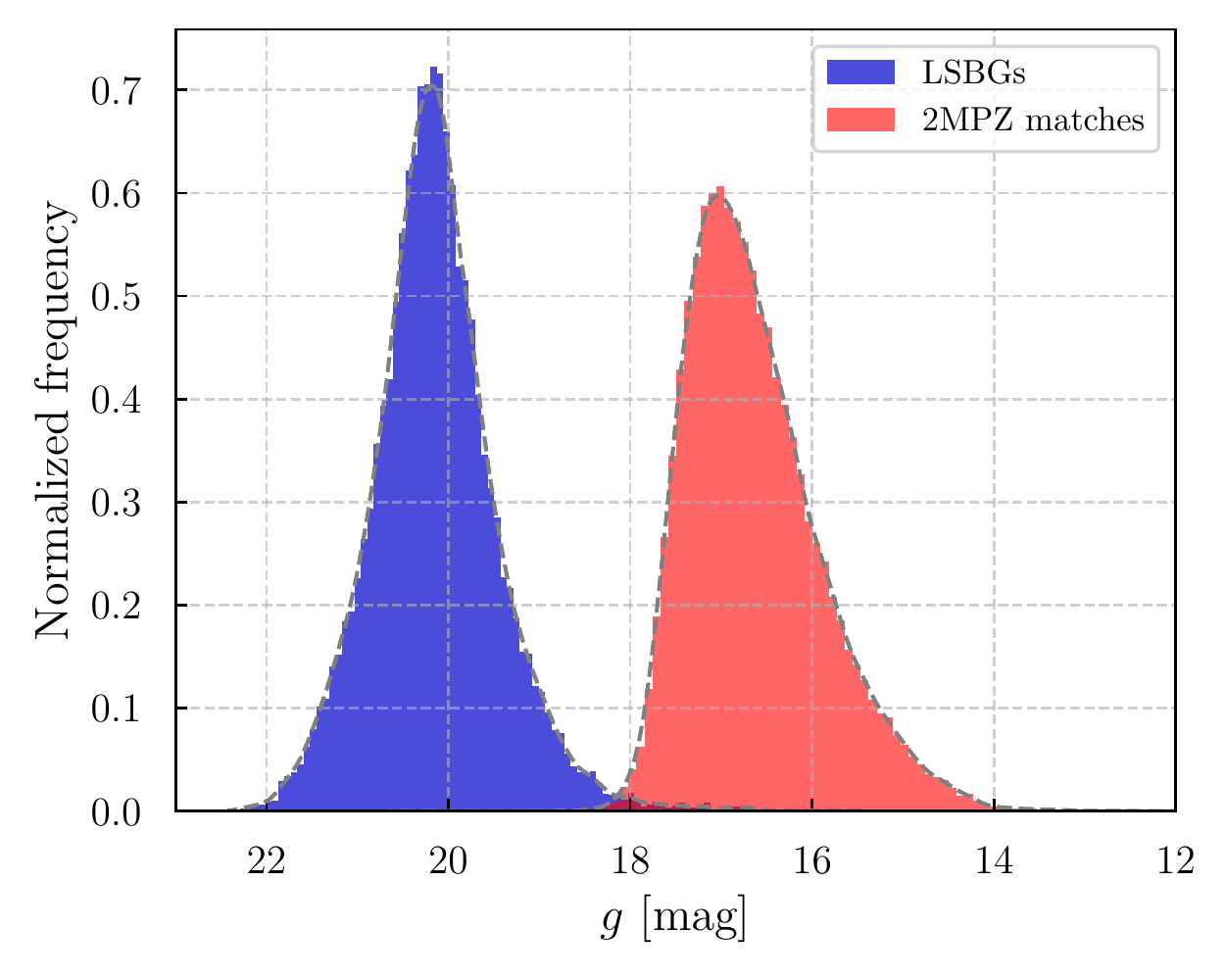}
\vspace{0.1cm}
\caption{$g$-band magnitude distributions of the LSBG sample and the DES catalog matches on the 2MPZ sample.}
\label{fig:g_band_mags}
\end{figure}

Finally, in \figref{g_band_mags}, we compare the $g$-band magnitude distributions of the LSBG sample and the 2MPZ galaxy sample that we used in the main text. Because the 2MPZ catalog did not provide such magnitudes, we matched the 2MPZ catalog with the DES Y3 GOLD catalog. The distribution presented here is derived from the \SExtractor's \code{MAG\_AUTO} magnitudes of these matches. That sample is significantly brighter than the LSBGs, with a median magnitude $g \sim 16.8$.

Note that we do not consider the HSBG sample separately in this section, as by construction it has the same magnitude distributions as the LSBG sample.

\end{document}